\documentclass[12pt]{article}
\newif\iffigs\figstrue
\usepackage{latexsym}
\iffigs
  \input{epsf}
\else
\fi
\font\tenmsbm=msbm10 scaled 1200
\font\sevenmsbm=msbm9
\newfam\msbmfam
\textfont\msbmfam=\tenmsbm
\scriptfont\msbmfam=\sevenmsbm

\makeatletter
\@addtoreset{equation}{section}
\makeatother

\newcommand{\eqn}[1]{(\ref{#1})}
\newcommand{\ft}[2]{{\textstyle\frac{#1}{#2}}}
\newsavebox{\uuunit}
\sbox{\uuunit}
    {\setlength{\unitlength}{0.825em}
     \begin{picture}(0.6,0.7)
        \thinlines
        \put(0,0){\line(1,0){0.5}}
        \put(0.15,0){\line(0,1){0.7}}
        \put(0.35,0){\line(0,1){0.8}}
       \multiput(0.3,0.8)(-0.04,-0.02){10}{\rule{0.5pt}{0.5pt}}
     \end {picture}}

\newsavebox{\bobox}
\sbox{\bobox}
    {\setlength{\unitlength}{0.825em}
     \begin{picture}(0.6,0.7)
        \thinlines
        \put(0,0){\line(1,0){0.7}}
        \put(0,0){\line(0,1){0.7}}
        \put(0.7,0){\line(0,1){0.7}}
        \put(0,0.7){\line(1,0){0.7}}
       \multiput(0,0)(0.05,0.05){14}{\rule{0.4pt}{0.4pt}}
       \multiput(0,0.65)(0.05,-0.05){13}{\rule{0.4pt}{0.4pt}}
     \end {picture}}
\newcommand {\xbox}{\mathord{\!\usebox{\bobox}}\,}

\def\IP{\relax{\rm I\kern-.18em P}}

\font\cmss=cmss10 \font\cmsss=cmss10 at 7pt

\def\inbar{\vrule height1.5ex width.4pt depth0pt}
\def\IC{\relax\,\hbox{$\inbar\kern-.3em{\rm C}$}}
\def\IG{\relax\,\hbox{$\inbar\kern-.3em{\rm G}$}}
\def\IB{\relax{\rm I\kern-.18em B}}
\def\ID{\relax{\rm I\kern-.18em D}}
\def\IL{\relax{\rm I\kern-.18em L}}
\def\IF{\relax{\rm I\kern-.18em F}}
\def\IH{\relax{\rm I\kern-.18em H}}
\def\II{\relax{\rm I\kern-.17em I}}
\def\IN{\relax{\rm I\kern-.18em N}}
\def\IM{\relax{\rm I\kern-.18em M}}
\def\IP{\relax{\rm I\kern-.18em P}}
\def\IQ{\relax\,\hbox{$\inbar\kern-.3em{\rm Q}$}}
\def\bfzero{\relax\,\hbox{$\inbar\kern-.3em{\rm 0}$}}
\def\IK{\relax{\rm I\kern-.18em K}}
\def\IG{\relax\,\hbox{$\inbar\kern-.3em{\rm G}$}}
 \font\cmss=cmss10 \font\cmsss=cmss10 at 7pt
\def\IR{\relax{\rm I\kern-.18em R}}
\def\ZZ{\relax\ifmmode\mathchoice
{\hbox{\cmss Z\kern-.4em Z}}{\hbox{\cmss Z\kern-.4em Z}}
{\lower.9pt\hbox{\cmsss Z\kern-.4em Z}}
{\lower1.2pt\hbox{\cmsss Z\kern-.4em Z}}\else{\cmss Z\kern-.4em
Z}\fi}
\def\bfone{\relax{\rm 1\kern-.35em 1}}

 \def\cB{{\cal B}}
\def\cC{{\cal C}} 
 
\def\cH{{\cal H}} 
 
\def\cL{{\cal L}} \def\cM{{\cal M}}

\def\cR{{\cal R}} 

\def\tilde{\widetilde}

\def\IE{\relax{{\rm I\kern-.18em E}}}

\def\IGam{\relax{{\rm I}\kern-.18em \Gamma}}

\def\bet{\begin{tabular}}
\def\eet{\end{tabular}}

\def\a{\alpha}
\def\b{\beta}
\def\l{\lambda}

\def\g{\gamma}
\def\d{\delta}

\def\r{\rho}
\def\s{\sigma}

\def\e{\epsilon}

\begin{document}
\begin{titlepage}
\begin{flushright}
Preprint DFTT 99/11\\
hep-th/9903036 \\
March 1999\\
\end{flushright}
\vskip 2cm
\begin{center}
{{\Large \bf    M -theory on $AdS_4\times M^{111}$:\\}
\vskip 0.2cm
{\Large \bf  the  complete $Osp(2\vert 4)\times SU(3)\times SU(2)$ spectrum \\
\vskip 0.2cm
from harmonic analysis$^*$}}\\
\vfill
{\large  Davide Fabbri , Pietro Fr\'e  ,
  Leonardo Gualtieri  and Piet Termonia } \\
\vfill
{ \sl  Dipartimento di Fisica Teorica, Universit\'a di Torino, via P.
Giuria 1,
I-10125 Torino, \\
 Istituto Nazionale di Fisica Nucleare (INFN) - Sezione di Torino,
Italy \\}
\end{center}
\vfill
\begin{abstract}
We reconsider
the Kaluza Klein compactifications
of $D=11$ supergravity on 
$AdS_4 \times (G/H)_7$ manifolds
that were classified in the eighties,
in the modern perspective of 
$AdS_4/CFT_3$ correspondence. 
We focus on one of the three ${\cal N}=2$ cases:
$(G/H)_7 = M^{111}=SU(3) \times SU(2) \times U(1) / SU(2)
\times U(1)^\prime\times U(1)^{\prime\prime}$.
Relying on the
systematic use of the harmonic analysis techniques developed in the eighties 
by one of us (P. Fr\'e) with R. D'Auria, 
we derive the complete spectrum of long, short
and massless $Osp(2 \vert 4) \times SU(3) \times SU(2)$ unitary
irreducible representations obtained in this compactification. 
Our result also provides a general scheme for the other ${\cal N}=2$
compactifications. Furthermore
it  is a necessary comparison term in the $AdS_4/CFT_3$
correspondence:
the complete $AdS/CFT$ match of the spectra that we obtain
will provide  a much more
stringent proof of the $AdS/CFT$ correspondence than in the $S^7$ case,
since the structure of the superconformal field
theory on the $M2$--brane world volume  must be such as to reproduce,
at the level of composite operators, the flavor group
representations, the conformal dimensions and the hypercharges 
that we obtain in the present article. 
The investigation of this match
is left to future publications. Here we provide an exhaustive
construction of the Kaluza Klein side of our spectroscopy.
\end{abstract}
\vspace{2mm} \vfill \hrule width 3.cm
{\footnotesize
 $^*$ Supported in part by   EEC  under TMR contract
 ERBFMRX-CT96-0045}
\end{titlepage}
\section{Introduction to $M^{111}$: old and new viewpoints}
In this paper we present the complete Kaluza Klein spectrum of M-theory
compactified on the  seven--manifold, $M^{111}$
that has ${\cal N}=2$ supersymmetry and $SU(3)\times SU(2) \times
U(1)$ as isometry group. This is part of a wider project \cite{stiefel},
\cite{n010} that aims at constructing the Kaluza Klein spectra for all the supersymmetric
homogeneous seven--manifolds $G/H$ classified in 1984 by Castellani  Romans
and Warner \cite{castromwar}.
\par 
As we illustrate in this introductory section, this is the final
completion  of a research programme that was
actively ongoing sixteen years ago \cite{noi321,spectfer,univer,spec321,multanna}.
Actually  the results    presented here   supersede all previous partial
results  and provide, in our opinion, a comprehensive and complete understanding of the involved
problem. Sixteen years ago the project  was suddenly stopped
for two reasons: a) failure to obtain the goals that in the 1984
perspective were considered the main motivations, b) occurrence of the
first string revolution that provided alternative perspectives to
pursue the same goals, that is the derivation of the standard model from
a higher--dimensional, unified, locally supersymmetric theory.
\par
The motivation to resume the project is provided by the current
interest in the $AdS/CFT$ correspondence and by the occurrence of the
second string revolution. This latter has shown that all ten-dimensional string
theories are, together with D=11 Supergravity, perturbative limits of
the same quantum theory, named M--theory. Hence all compactified
solutions of D=11 supergravity are relevant and highlight different aspects of
M--theory. In particular, as shown in \cite{g/hm2},  each
supersymmetric $AdS_4\times (G/H)_7$ vacuum of the Castellani et al
classification has a  corresponding supersymmetric $M2$--brane
solution which interpolates between such a vacuum at the horizon and
a vacuum $M_3 \times {\cal C}_8$ at infinity, $M_3$ being the
three--dimensional minkowskian world volume of the M2--brane and
${\cal C}_8$ being a suitable K\"ahlerian cone constructed over the
Sasakian manifold $(G/H)_7$.
Following the ideas explained in \cite{noim2} and the
suggestions coming from the recent literature \cite{tatar,witkleb}
the final goal of our investigation is to construct the
three-dimensional superconformal field theory that is dual to each
$(G/H)_7$ compactification of M-theory. In this direction an
essential constructive step and comparison term is the derivation of
the corresponding Kaluza Klein spectrum, which is our present aim.
\par
In order to introduce the specific example of Kaluza Klein spectrum we shall
calculate and the harmonic analysis techniques we shall employ we
feel it necessary to make a step back and provide a brief historical
perspective for our problem.
\subsection{Old Kaluza Klein supergravity: the search for a realistic model based on $M^{pqr}$ spaces and
harmonic analysis}
In the beginning of the eighties, after $D=11$ supergravity had been
discovered \cite{cre} and also reformulated geometrically
\cite{ricpie11}, a lot of interest was attracted by the revival of
the Kaluza Klein  idea \cite{Kkidea}.
It was conceived that the gauge symmetries of
fundamental particle interactions might be interpreted as isometries
of the seven compactified  extra dimensions. In particular it was realized
by Freund and Rubin \cite{freurub} that
$D=11$ supergravity admits classical vacua  where the eleven--dimensional space is
the product of four--dimensional anti de Sitter space with a compact
seven-dimensional Einstein space:
\begin{equation}
{\cal M}_{11} = AdS_4 \times M_7
\label{11=4+7}
\end{equation}
The Killing vectors $k^m_I(y)$ ($I=1,\dots,\mbox{dim} \, {\cal G}$)
admitted by the Einstein metric on $M_7$ close the Lie algebra of
the gauge group ${\cal G}$, while the number ${\cal N}$ of Killing spinors
$\eta^A(y)$ ($A=1,\dots, {\cal N}$)
existing on such a geometry   coincides with the number of
supersymmetries preserved by the four--dimensional compactified
theory.
Initially, Kaluza Klein supergravity \cite{duffrev} focused on the case where the compact
manifold $M_7$ is topologically a sphere, endowed either with the  $SO(8)$ invariant Einstein
metric ({\it round $S^7$}) \cite{round7a} or with a second Einstein metric which is
$SO(5)\times SO(3)$ invariant ({\it squashed $S^7$}) and   preserves ${\cal N}=1$ rather than
${\cal N}=8$ supersymmetry, as it was shown by Awada, Duff and Pope \cite{squash7a}.
Although the two gauge groups ${\cal G}$ emerging from such compactifications are not realistic
and cannot embed the standard model, an in--depth analysis \cite{englert}, \cite{biran},
\cite{casher}, \cite{dewit2} of these
cases  revealed many subtle features of Kaluza Klein supergravity and
of its relation with the gauged ${\cal N}=8$ supergravity of de Wit and Nicolai
\cite{dewit1}.
Furthermore after D'Auria and Fr\'e proved \cite{osp48}  that $Osp(8 \vert 4)  $
is the  isometry superalgebra of the round $S^7$  background,  then
the mass spectrum of such a compactification could be organized
by G\"unaydin and Warner \cite{gunawar} into unitary irreducible representations
(UIR) of this non--compact superalgebra.
This result of 1985, later extended by G\"unaydin and Marcus to the  spectrum of
$SU(2,2\vert 4)$
supermultiplets that arise in the compactification of type IIB
superstring on $AdS_5 \times S^5$ \cite{gunay2}, is one of the
cornerstones for the recent reinterpretation of Kaluza Klein supergravity
within the framework of an $AdS/CFT$ correspondence. Indeed as a
consequence of the Maldacena conjecture \cite{maldapasto} that
classical supergravity on $AdS_{p+2}\times M_{D-p-2}$ is dual to a
quantum superconformal  theory on the $(p\!+\!1)$--dimensional boundary of
$AdS_{p+2}$ it follows that the Kaluza Klein spectrum of $AdS_{p+2}$
supermultiplets corresponds to the spectrum of composite primary
operators in the conformal field theory \cite{minwalla},\cite{serfro1},\cite{serfro2},
\cite{serfro3},\cite{serlau},\cite{serlau2}.
Yet from the  Kaluza Klein   point of view the seven
sphere  was not very satisfactory since, as already observed, neither $SO(8)$  nor $SO(5)
\times SO(3)$ contain the gauge group of the standard model:
\begin{equation}
  {\cal G}_s = SU(3)^c \times SU(2)^w \times U(1)^Y
\label{gstand}
\end{equation}
However, already in 1981 Witten had observed \cite{kkwitten} that seven is the smallest
number of dimensions where the group (\ref{gstand}) can be realized
as a regular isometry group. Indeed one can introduce the
seven--dimensional homogeneous spaces:
\begin{equation}
  M^{pqr} =  \frac{{\cal G}_s}{H}\equiv \frac{SU(3)^c
  \times SU(2)^w \times U(1)^Y}{SU(2)^c \times U(1)' \times U(1)''}
\label{mpqrdef}
\end{equation}
where the integer numbers $p,q,r \in \ZZ$  characterize the
topology of the manifold and are introduced through the embedding of the
isotropy subgroup $H=SU(2)^c \times U(1)' \times U(1)''$ into ${\cal G}_s$
as explained in section \ref{geometry}.
In order to show that   $AdS_4 \times M^{pqr}$ are exact vacua of
$D=11$ supergravity that produce the standard model gauge group through
the Kaluza Klein mechanism it was necessary to construct an $SU(3)\times
SU(2)\times U(1)$ invariant Einstein metric on each of these spaces.
This was done by Castellani, D'Auria and  Fr\'e in \cite{noi321}. In
the same paper the properties of these spaces with respect to
supersymmetry breaking were discussed. It was established that for
$p\ne q $ there are no Killing spinors while in the case
$p=q$ the resulting four--dimensional theory has ${\cal N}=2$
supersymmetry. Since the spectrum does not depend on the actual
value of $p=q$ or of $r\ne 0$, we can just focus on the space
$M^{111}$
Hence an early conclusion reached in \cite{noi321} was
that by compactifying $D=11$ supergravity on $AdS_4 \times M^{111}$
one obtains a low energy effective action containing the
${\cal N}=2$ supergravity multiplet plus, at least, the vector multiplets
of the group $SU(3)\times SU(2)$. The factor $U(1)^Y$ in the isometry group
is the R--symmetry  of ${\cal N}=2$ supergravity and the associated
gauge boson  is the graviphoton.
\par
In order to establish the precise content of the theory, both at the
massless and the massive level more information on the spectrum was
necessary.
\par In this respect the case of $M^{111}$ was (and it is)
quite different with respect to the case of the round $S^7$.
In the latter case it suffices to know that the isometry superalgebra is $Osp(8\vert 4)$
to deduce the entire spectrum from  a purely algebraic construction.
The reason is the following. As shown by G\"unaydin and Warner
\cite{gunawar} the complete Kaluza Klein spectrum of the round $S^7$
compactification is composed of {\it short $Osp(8\vert 4)$
supermultiplets} characterized by a quantization of masses (or better
anti de Sitter energy labels) in terms of the $SO(8)$ R-symmetry
representation in which their Clifford vacuum falls. In modern
language all the Kaluza Klein states are BPS states and their
spectrum can be derived by constructing the {\it short} UIR  of
$Osp(8\vert 4)$ with highest spin limited to be two.
From the  perspective of the three dimensional superconformal theory
this means that all the composite primary operators have conformal
weights equal to their naive dimensions, no anomalous dimensions being generated.
\par The same results can also be derived from harmonic analysis on the
compact space $S^7$ but no new information is obtained with respect
to the algebraic construction.
\par
On the contrary in the case of anti de Sitter vacua with ${\cal N}=1,2,3$
supersymmetry the Kaluza Klein states have to be organized in
supermultiplets of $Osp({\cal N}\vert 4) \times G^\prime$ where
$G^\prime$ is the factor in the isometry group of the seven--manifold
that commutes with supersymmetry and with the $R$--symmetry factor
$O(\cal N )$. In this case the Kaluza Klein states do not necessarily
fall into {\it short multiplets} of $Osp({\cal N}\vert 4)$ and do not
necessarily have quantized energy labels (or masses). Indeed their
masses depend not only on the $R$--symmetry representation but also
on the $G^\prime$ representation. In the modern perspective of the
dual superconformal field theory this means that anomalous dimensions
are generated, yet if the $AdS/CFT$ correspondence is really true
then these anomalous dimensions can be exactly calculated from the
supergravity side by harmonic analysis on the compact seven--manifold.
Because of the  interest in the $D=11$ compactification on $M^{111}$
that displayed the standard model group (\ref{gstand}) as gauge
group, the oldest of us (P.Fr\'e), in
collaboration with R. D'Auria, developed  in the years 1983-1984 a systematic approach
to the harmonic analysis on coset manifolds with Killing spinors and in
particular on $M^{pqr}$ spaces \cite{spectfer,univer,spec321}.
The formalism and techniques developed in those papers are the
basis for the present investigation where we provide the completion
of the classification programme started fifteen years ago
in \cite{multanna}.
The results obtained in \cite{spectfer,univer,spec321,multanna}
showed that the massless sector of M-theory on $AdS_4 \times
M^{111}$ is given by the ${\cal N}=2$ graviton multiplet plus the vector
multiplets of $SU(3)\times SU(2) \times U(1)$. Indeed, although
$U(1)^Y$ is the R--symmetry and its gauge-boson is the graviphoton, an
additional massless vector multiplet,  at the time  named the
Betti multiplet, comes from the three--form $A_{\mu ij}$ with a
space--time index and two internal indices. This happens because the
$M^{111}$ space has a closed non trivial two--form. At the classical
level there are no states charged with respect to such a
$U(1)$ but in non perturbative string theory there might be. Indeed
this is like a Ramond-Ramond multiplet. It is still
an open problem to determine the  twelve--dimensional
special K\"ahler manifold:
\begin{equation}
  {\cal SK}_{12}^{111}
\label{sk111}
\end{equation}
whose geometry determines the  ${\cal N}=2$ low energy
supergravity action for the $AdS_4\times M^{111} $ compactification.
\subsection{The results of this paper in the modern perspective of
$AdS/CFT$ correspondence}
The programme of deriving the complete $Osp(2\vert 4) \times SU(3)
\times SU(2))$ spectrum was left unachieved in 1984 since the hopes to
obtain a realistic Kaluza Klein model were doomed. Indeed not only
was the theory plagued by a too large cosmological constant (anti de
Sitter space) and by an intrinsic non chiral nature, but also the
quark and lepton representations could not be found in the
hypermultiplet spectrum \cite{fretrst}. The harmonic analysis had been carried
through to the point of determining the spectrum of graviton and
gravitino multiplets (see \cite{multanna}) but in order to derive the
spectrum of vector multiplets and hypermultiplets a major effort was
still necessary. Indeed one needed to find the eigenvalues of the
Laplace Beltrami operators:
\begin{equation}
  M_{(1)(0)^2} \quad , \quad M_{(1)^2(0)} \quad , \quad M_{(1)^3}
\label{operatori}
\end{equation}
acting respectively on 1--forms, 2--forms and 3--forms.
\par
The spectra of these eigenvalues is precisely what we determine in
the present paper and this information combined with the old results
allows us to organize all the Kaluza Klein states in $Osp(2\vert 4)$
supermultiplets. Indeed as a by product of our analysis we also find
the structure of {\it short and long} UIR representations of the
superalgebra $Osp(2\vert 4)$ limited by highest spin less or equal to
two. In other words the path we follow here is somehow the reciprocal
path to that followed by G\"unaydin and Warner in the case of the
round $S^7$. There representation theory of the non compact
superalgebra $Osp(8\vert 4)$ could be used to retrieve all the
data inherent to  harmonic analysis on $S^7$. Here full fledged
harmonic analysis on $M^{111}$, besides giving detailed information on
the energy labels of the supermultiplets, determines also the structure of
the UIR of the non--compact superalgebra $Osp(2\vert 4)$. Obviously
such a structure is universal and must be the same in the other
${\cal N}=2$ compactifications. What will be different in the other 
cases is the $G^\prime$ group and the $G^\prime$ representation assignment of the
supermultiplets. Similarly, for long multiplets, but not for short multiplets
we will have different  values of the energy labels (anomalous conformal
dimensions in the three dimensional SCFT interpretation). According
to the Castellani et al classification (see \cite{g/hm2} for a summary
in modern perspective) there are just three ${\cal N}=2$ theories in
this setup and they correspond  to the cases,
\begin{eqnarray}
M^{ppr} & ; & G^\prime = SU(3) \times SU(2) \nonumber\\
V_{5,2} & ; & G^\prime = SO(5)\nonumber\\
Q^{ppp} & ; & G^\prime = SU(2)^3
\label{casidevita}
\end{eqnarray}
The first manifold is treated in the present paper. The spectrum of
the Stiefel manifold $V_{5,2}$ is presently under construction and will
appear in a forthcoming publication \cite{stiefel}. As for the last
case $Q^{ppp}$, originally introduced by R. D'Auria, P. Fr\'e and Van Nieuwenhuizen
\cite{dafrepvn}, in this case there already exists a proposal for the
dual superconformal field theory \cite{tatar}. If the $Osp(2\vert 4)\times SU(2)^3$
spectrum had been determined the conjectured $AdS/CFT$ correspondence
could be tested. Unfortunately the harmonic analysis on $Q^{111}$ is
not yet available and has still to be planned.
\par
Hence the present paper is a first essential step along a path that
we outline in the outlook at the end of the paper.
\subsection{Structure of the paper}
Our paper is divided in two parts.
In the first we give a summary of our result, namely
we present the full spectrum of $Osp(2|4)\times SU(3)\times SU(2)$
supermultiplets while in the second part we give all the details
of its derivation from systematic harmonic analysis on $M^{111}$.
The first part is written in such a way as to be self-consistent.
The reader interested only in the spectrum but not in harmonic analysis
can completely skip part two.
More specifically the content of the sections is as follows.
\par
Section two describes the structure of $Osp(2|4)$ supermultiplets from
a general stand point.
\par
Section three presents the spectrum of $Osp(2|4)$ supermultiplets that
occur in the $M^{111}$ compactifications giving their
$SU(3)\times SU(2)$-quantum numbers.
The multiplets are divided in three classes: long, short and massless
and for each class we have graviton, gravitino and vector multiplets.
The hypermultiplets are always short.
There are no massless gravitino multiplets because this would mean that
supersymmetry is enhanced from ${\cal N}=2$ to higher ${\cal N}$, which
is not the case.
\par
Section four describes the geometry of the $M^{111}$ space.
\par
Section five introduces the basic items of harmonic analysis and applies
them to the space under consideration.
\par
Section six explains how to derive the eigenvalues of the Laplace-Beltrami
operators relevant to our problem from harmonic analysis.
The section is divided in subsections dealing with the Laplace-Beltrami
operators acting on zero-forms, one-forms, two-forms, three-forms and
spinors respectively.
\par
Section seven is devoted to filling the various states found by harmonic
analysis into $Osp(2|4)$ supermultiplets.
\par
Section eight gives an outlook on the continuation of our research
project and future developments.
\par
Appendix A contains our conventions while appendix B summarizes, for
the reader convenience, the main formulas on harmonic analysis and
mass relations that we use throughout the paper.
\newpage
\vskip 1cm
\centerline{\LARGE
{\bf PART ONE: {\sl  THE RESULT}}}
\vskip .7cm
In this part of the article we present our result while in the second part
we give all the details of its derivation.
\par
Specifically, in part two we shall give a summary of $M^{111}$ geometry and
a full fledged discussion of harmonic analysis on such a manifold. In that
part the interested reader can find all the details concerning the
calculation of the eigenvalue spectra for the operators $M_{(1)(0)^2},~
M_{(1)^2(0)}~$ and $M_{(1)^3}$ \cite{univer,castdauriafre} which are the
key tool to obtain the complete spectrum of supermultiplets.
Yet the reader whose interest is limited to the actual form of this
spectrum can just confine his reading to part one.
\par
Here we give the result in two steps. In the first step we present the
structure of long and short multiplets of the $Osp\left(2|4\right)$
superalgebra that are limited by highest spin less or equal to two.
Such a structure could have been derived either by means of the
Freedman Nicolai method of vanishing norms \cite{frenico},
or by the G\"unaydin et Warner oscillator method \cite{gunawar}.
We bypass such calculations obtaining the result as a byproduct of
harmonic analysis on the specific manifold $M^{111}$. Yet  we emphasize
(as we did in the introduction) that this part of the result is universal
and applies to all other ${\cal N}=2$ compactifications on $AdS_4$.
In the second step we exactly specify how many long and short
$Osp\left(2|4\right)$ multiplets occur in the $M^{111}$ compactification
and for each of them we give both the $SU\left(3\right)\times SU\left(2
\right)$ representation assignment and the value of the energy  and
hypercharge labels $\left(E_0,y_0\right)$ of the corresponding Clifford vacuum.
\section{Structure of the $Osp\left(2|4\right)$ multiplets}
The structure of the relevant ${\cal N}=2$ supermultiplets in $AdS_4$
superspace is summarized in tables \ref{longgraviton},\ref{longgravitino},
\ref{longvector},\ref{shortgraviton},
\ref{shortgravitino},\ref{shortvector},\ref{hyper},\ref{masslessgraviton},
\ref{masslessvector} whose content we discuss in the present section.
To describe the organization of these tables we recall the basic facts
about $Osp\left(2|4\right)$ unitary irreducible representations (UIR).
\par
The even subalgebra of the superalgebra:
\begin{equation}
\label{geven}
G_{\rm even}=Sp\left(4,\IR\right)\oplus SO\left(2\right)\subset
Osp\left(2|4\right)
\end{equation}
is a direct sum of subalgebras, where $Sp\left(4,\IR\right)\sim SO\left(2,3\right)$
is the isometry algebra of $AdS_4$ while the compact  subalgebra
$SO\left(2\right)$ generates $R$--symmetry. The maximally compact
subalgebra of $G_{\rm even}$ is
\begin{equation}
\label{gcompact}
G_{\rm compact}=SO\left(2\right)_E\oplus SO\left(3\right)_S\oplus SO\left(
2\right)_R\subset G_{\rm even}.
\end{equation}
The generator of $SO\left(2\right)_E$  is interpreted as the
hamiltonian of the system when $Osp\left(2|4\right)$ acts as the isometry
group of anti de Sitter superspace.
Consequently its eigenvalues $E$  are the energy levels of possible
states for the system. The group $SO\left(3\right)_S$ is the ordinary
rotation group and similarly its representation labels $s$ describe
the possible spin states of the system.
Finally the eigenvalue $y$ of the generator of $SO\left(2\right)_R$ is the
hypercharge of a state.
\par
A supermultiplet, namely a UIR of the superalgebra $Osp\left(2|4\right)$
is composed of a {\sl finite} number of UIR of the even subalgebra
$G_{\rm even}$ (\ref{geven}), each of them being what in physical language
we call a particle state, characterized by a spin ``$s$'', a mass ``$m$''
and a hypercharge ``$y$''. From a mathematical viewpoint each UIR
representation of the non-compact even subalgebra $G_{\rm even}$ is an
{\sl infinite} tower of finite dimensional UIR representations
of the compact subalgebra $G_{\rm compact}$ (\ref{gcompact}).
The lowest lying representation of such a tower $|E,s,y>$ is the Clifford
vacuum,
The mass, spin and hypercharge of the corresponding particle are read
from the labels of the Clifford vacuum by use of the relations between mass
and energy \cite{multanna} that we have recalled in appendix B eq.
(\ref{massenergy}).
\par
In the same way there is a Clifford vacuum $|E_0,s_0,y_0>$ for the entire
supermultiplet out of which we not only construct the corresponding
particle state but also, through the action of  the $SUSY$ charges, we construct
the Clifford vacua $|E_0+\dots,s_0+\dots,y_0+\dots>$ of the other members
of the same supermultiplet. Hence the structure of a supermultiplet
is conveniently described by listing the energy $E$, the spin $s$ and the
hypercharge $y$ of all the Clifford vacua of the multiplet.
\par
In tables \ref{longgraviton},\ref{longgravitino},
\ref{longvector},\ref{shortgraviton},
\ref{shortgravitino},\ref{shortvector},\ref{hyper},\ref{masslessgraviton},
\ref{masslessvector} we provide such information in the first three columns.
The remaining columns provide additional information concerning the way
the abstract $Osp\left(2|4\right)$ supermultiplets are actually realized
in Kaluza Klein supergravity. Indeed for each particle state appearing in
the supermultiplet we write the name of the corresponding field in the
Kaluza Klein expansion of $D=11$ supergravity to which such a particle state
contributes. The standard expansion of linearized $D=11$ supergravity and
the conventions for the names of the $D=4$ fields are given in appendix B,
eq. (\ref{kkexpansion}).
\par
Given these preliminaries let us discuss our result for the structure
of the $Osp\left(2|4\right)$ supermultiplets with $s\le 2$.
\par
In $\cal N$--extended $AdS_4$ superspace there are three kinds of supermultiplets:
\begin{itemize}
\item the long multiplets
\item the short multiplets
\item the massless multiplets
\end{itemize}
For ${\cal N}=2$ the long multiplets satisfy the following unitarity relation, without saturation:
\begin{equation}
E_0>|y_0|+s_0+1.
\label{longcond} 
\end{equation}
Furthermore we distinguish three kinds of long multiplets depending on the highest
spin state they contain: $s_{\rm max}=2,~{3\over2},~1$.
These multiplets are respectively named {\sl long graviton}, {\sl long gravitino} and
{\sl long vector multiplets}.
\par
The {\sl long graviton multiplet}, satisfying $E_0>|y_0|+2$, has the structure
displayed in table \ref{longgraviton}.
\par
\begin{table}
\centering
\begin{tabular}{||c|c|c|c|c||}
\hline
Spin &
Energy &
Hypercharge &
Mass ($^2$)&
Name \\
\hline
\hline
$2$      & $E_0+1    $  & $y_0$     & $16(E_0+1)(E_0-2)$  & $h$ \\
$\ft32$  & $E_0+\ft32$  & $y_0-1$   & $-4E_0-4$  & $\chi^-$ \\
$\ft32$  & $E_0+\ft32$  & $y_0+1$   & $-4E_0-4$  & $\chi^-$ \\
$\ft32$  & $E_0+\ft12$  & $y_0-1$   & $4E_0-8 $  & $\chi^+$ \\
$\ft32$  & $E_0+\ft12$  & $y_0+1$   & $4E_0-8 $  & $\chi^+$ \\
$1$      & $E_0+2    $  & $y_0$     & $16E_0(E_0+1)$    & $W$ \\
$1$      & $E_0+1    $  & $y_0-2$   & $16E_0(E_0-1)$    & $Z$ \\
$1$      & $E_0+1    $  & $y_0+2$   & $16E_0(E_0-1)$    & $Z$ \\
$1$      & $E_0+1    $  & $y_0$     & $16E_0(E_0-1)$    & $Z$ \\
$1$      & $E_0+1    $  & $y_0$     & $16E_0(E_0-1)$    & $Z$ \\
$1$      & $E_0      $  & $y_0$     & $16(E_0-1)(E_0-2)$& $A$  \\
$\ft12$  & $E_0+\ft32$  & $y_0-1$   & $4E_0$     &  $\lambda_T$  \\
$\ft12$  & $E_0+\ft32$  & $y_0+1$   & $4E_0$     &  $\lambda_T$ \\
$\ft12$  & $E_0+\ft12$  & $y_0-1$   & $-4E_0+4$  &  $\lambda_T$ \\
$\ft12$  & $E_0+\ft12$  & $y_0+1$   & $-4E_0+4$  &  $\lambda_T$ \\
$0$      & $E_0+1    $  & $y_0$     & $16E_0(E_0-1)  $  & $\phi$ \\
\hline
\end{tabular}\\[.13in]
\caption{${\cal N}=2$ long graviton multiplet}
\label{longgraviton}
\end{table}
The {\sl long gravitino multiplet}, satisfying $E_0>|y_0|+\ft32$, has the structure
displayed in table \ref{longgravitino}.
\begin{table}
\centering
\begin{small}
\begin{tabular}{||c|c|c|c|c|c|c||}
\hline
Spin &
Energy &
Hypercharge &
Mass ($^2$)&
Name &
Mass ($^2$)&
name \\
\hline
\hline
$\ft32$  & $E_0+1$      & $y_0$   & $4E_0-6$  & $\chi^+$
                & $-4E_0-2$ & $\chi^-$  \\
$1$      & $E_0+\ft32$  & $y_0-1$ & $16(E_0-\ft12)(E_0+\ft12)$ & $Z$
                & $16(E_0-\ft12)(E_0+\ft12)$ & $W$ \\
$1$      & $E_0+\ft32$  & $y_0+1$ & $16(E_0-\ft12)(E_0+\ft12)$ & $Z$
                & $16(E_0-\ft12)(E_0+\ft12)$ & $W$ \\
$1$      & $E_0+\ft12$  & $y_0-1$ & $16(E_0-\ft32)(E_0-\ft12)$ & $A$
                & $16(E_0-\ft32)(E_0-\ft12)$ & $Z$ \\
$1$      & $E_0+\ft12$  & $y_0+1$ & $16(E_0-\ft32)(E_0-\ft12)$ & $A$
                & $16(E_0-\ft32)(E_0-\ft12)$ & $Z$ \\
$\ft12$  & $E_0+2$      & $y_0$   & $4E_0+2$  &  $\lambda_T$
                & $-4E_0-2$ &  $\lambda_L$ \\
$\ft12$  & $E_0+1$      & $y_0-2$ & $-4E_0+2$ &  $\lambda_T$
                & $-4E_0-2$ &  $\lambda_T$ \\
$\ft12$  & $E_0+1$      & $y_0$   & $-4E_0+2$ &  $\lambda_T$
                & $4E_0-2$  &  $\lambda_T$  \\
$\ft12$  & $E_0+1$      & $y_0+2$ & $-4E_0+2$ &  $\lambda_T$
                & $4E_0-2$  &  $\lambda_T$ \\
$\ft12$  & $E_0+1$      & $y_0$   & $-4E_0+2$ &  $\lambda_T$
                & $4E_0-2$  &  $\lambda_T$  \\
$\ft12$  & $E_0$        & $y_0$   & $4E_0-6$  &  $\lambda_L$
                & $-4E_0+6$ &  $\lambda_T$ \\
$0$      & $E_0+\ft32$  & $y_0-1$ & $16(E_0-\ft12)(E_0+\ft12)$  & $\phi$
                & $16(E_0-\ft12)(E_0+\ft12)$  & $\pi$ \\
$0$      & $E_0+\ft32$  & $y_0+1$ & $16(E_0-\ft12)(E_0+\ft12)$  & $\phi$
                & $16(E_0-\ft12)(E_0+\ft12)$  & $\pi$ \\
$0$      & $E_0+\ft12$  & $y_0-1$ & $16(E_0-\ft32)(E_0-\ft12)$  & $\pi$
                & $16(E_0-\ft32)(E_0-\ft12)$  & $\phi$ \\
$0$      & $E_0+\ft12$  & $y_0+1$ & $16(E_0-\ft32)(E_0-\ft12)$  & $\pi$
                & $16(E_0-\ft32)(E_0-\ft12)$  & $\phi$ \\
\hline
\end{tabular}\\[.13in]
\end{small}
\caption{${\cal N}=2$ long gravitino multiplets $\chi^+$ and $\chi^-$}
\label{longgravitino}
\end{table}
As we see this type of multiplet is realized in Kaluza Klein supergravity
in two different ways. The two ways correspond to a resolution of the
ambiguity inherent to the quadratic form of the mass/energy relations
(\ref{massenergy}).
One realization of the multiplets chooses one branch of the relation,
the second realization chooses the second branch. The two realizations
of the multiplets are implemented by different fields of the Kaluza Klein
expansion. For instance in one realization   the vector
fields ($Z$) arise from the expansion of the three--form $A_{\mu ij}$
(with one space--time index and two internal indices), in the other
realization the vector fields ($A/W$) arise both from the metric $g_{
\mu i}$ and from the three--form $A_{\mu\nu i}$ with two legs on
space--time and one internal leg.
\par
The {\sl long vector multiplet}, satisfying $E_0>|y_0|+1$, has the structure
displayed in table \ref{longvector}.
\begin{table}
\centering
\begin{footnotesize}
\begin{tabular}{||c|c|c|c|c|c|c|c||}
\hline
Spin &
Energy &
Hypercharge &
Mass ($^2$)&
Name &
Name &
Mass ($^2$)&
Name \\
\hline
\hline
$1$      & $E_0+1$      & $y_0$   & $16E_0(E_0-1)$  & $A$
                 & $W$     & $16E_0(E_0-1)$  & $Z$      \\
$\ft12$  & $E_0+\ft32$  & $y_0-1$ & $-4E_0$  &   $\lambda_T$
                 &   $\lambda_L$  & $4E_0$   &   $\lambda_T$   \\
$\ft12$  & $E_0+\ft32$  & $y_0+1$ & $-4E_0$  &   $\lambda_T$
                  &   $\lambda_L$  & $4E_0$   &   $\lambda_T$   \\
$\ft12$  & $E_0+\ft12$  & $y_0-1$ & $4E_0-4$ &   $\lambda_L$
                 &   $\lambda_T$   & $-4E_0+4$&   $\lambda_T$ \\
$\ft12$  & $E_0+\ft12$  & $y_0+1$ & $4E_0-4$ &   $\lambda_L$
                 &   $\lambda_T$   & $-4E_0+4$&   $\lambda_T$  \\
$0$      & $E_0+2$      & $y_0$   & $16E_0(E_0+1)$  & $\phi$
                 & $\Sigma$  & $16E_0(E_0+1)$  & $\pi$  \\
$0$      & $E_0+1$      & $y_0-2$ & $16E_0(E_0-1)$  & $\pi$
                  & $\pi$   & $16E_0(E_0-1)$  & $\phi$   \\
$0$      & $E_0+1$      & $y_0+2$ & $16E_0(E_0-1)$  & $\pi$
                  & $\pi$   & $16E_0(E_0-1)$  & $\phi$   \\
$0$      & $E_0+1$      & $y_0$   & $16E_0(E_0-1)$  & $\pi$
                  & $\pi$    & $16E_0(E_0-1)$  & $\phi$   \\
$0$      & $E_0$        & $y_0$   & $16(E_0-2)(E_0-1)$  & $S$
                  & $\phi$  & $16(E_0-2)(E_0-1)$  & $\pi$  \\
\hline
\end{tabular}\\[.13in]
\end{footnotesize}
\caption{${\cal N}=2$ long vector multiplets $A$,$W$ and $Z$}
\label{longvector}
\end{table}
Also the long vector multiplet has different realizations from the
Kaluza Klein viewpoint.
\par
The short multiplets are of two kinds: the {\sl short graviton,
gravitino} and {\sl vector  multiplets}, that saturate the bound
\begin{equation}
E_0=|y_0|+s_0+1
\label{shortcond}
\end{equation}
and the {\sl hypermultiplets} (spin $1/2$ multiplets), that saturate
the other bound
\begin{equation}
E_0=|y_0|~~{\rm with}~~|y_0|\ge{1\over 2}.
\label{masslesscond}
\end{equation}
\par
The {\sl short graviton multiplet}, satisfying $E_0=|y_0|+2$, has the structure
displayed in table \ref{shortgraviton}.
\begin{table}
\centering
\begin{footnotesize}
\begin{tabular}{||c|c|c|c|c||}
\hline
Spin &
Energy &
Hypercharge &
Mass ($^2$)&
Name \\
\hline
\hline
$2$      & $y_0+3$        & $y_0$     & $16y_0(y_0+3)$  & $h$        \\
$\ft32$  & $y_0+\ft72$    & $y_0-1$& $-4y_0-12$  & $\chi^-$      \\
$\ft32$  & $y_0+\ft52$    & $y_0+1$   & $4y_0 $  & $\chi^+$             \\
$\ft32$  & $y_0+\ft52$    & $y_0-1$   & $4y_0 $  & $\chi^+$             \\
$1$      & $y_0+3    $    & $y_0-2$& $16(y_0+2)(y_0+1)$    & $Z$        \\
$1$      & $y_0+3    $    & $y_0$     & $16(y_0+2)(y_0+1)$    & $Z$     \\
$1$      & $y_0+2      $  & $y_0$     & $16y_0(y_0+1)$& $A$         \\
$\ft12$  & $y_0+\ft52$  & $y_0-1$& $-4y_0-4$  &  $\lambda_T$    \\
\hline
\end{tabular}\\[.13in]
\end{footnotesize}
\caption{${\cal N}=2$ short graviton multiplet with positive hypercharge $y_0>0$}
\label{shortgraviton}
\end{table}
\par
The {\sl short gravitino multiplet}, satisfying $E_0=|y_0|+\ft32$, has the structure
displayed in table \ref{shortgravitino}
\footnote{in this table there is a $\pm$. The actual sign can be calculated
looking at the norms of a three--creation operator state}.
\begin{table}
\centering
\begin{footnotesize}
\begin{tabular}{||c|c|c|c|c||}
\hline
Spin &
Energy &
Hypercharge &
Mass ($^2$)&
Name \\
\hline
\hline
$\ft32$  & $y_0+\ft52$    & $y_0$         & $4y_0$                  & $\chi^+$  \\
$1$      & $y_0+3$        & $y_0-1$  & $16(y_0+1)(y_0+2)$      & $Z$    \\
$1$      & $y_0+2$        & $y_0+1$       & $16y_0(y_0+1)$       & $A$  \\
$1$      & $y_0+2$        & $y_0-1$       & $16y_0(y_0+1)$       & $A$  \\
$\ft12$   & $y_0+\ft52$   & $y_0$         & $-4y_0-4$            &  $\lambda_T$ \\
$\ft12$  & $y_0+\ft52$    & $y_0-2$         & $-4y_0-4$            &  $\lambda_T$   \\
$\ft12$  & $y_0+\ft32$    & $y_0$         & $4y_0$                  &  $\lambda_L$  \\
$0$       &  $y_0+3$   & $y_0\pm 1$     & $16(y_0+1)(y_0+2)$ &  $\phi$  \\
\hline
\end{tabular}\\[.13in]
\end{footnotesize}
\caption{${\cal N}=2$ short gravitino multiplet $\chi^+$ with positive hypercharge $y_0 >0$}
\label{shortgravitino}
\end{table}
\par
The {\sl short vector multiplet}, satisfying $E_0=|y_0|+1$, has the structure
displayed in table \ref{shortvector} $^1$.
\par
We must stress that the multiplets displayed in tables
\ref{shortgraviton}, \ref{shortgravitino}, \ref{shortvector} are only
half of the story, since they can be viewed as the BPS states where
$E_0 = y_0 +s_0 +1$ and $y_0 >0 $. In addition one has also the anti
BPS states. These are the short multiplets where $E_0 = -y_0 +s_0 +1$
with $y_0 <0$.  The structure of these anti short multiplets
can be easily read off  from tables \ref{shortgraviton}, \ref{shortgravitino}, \ref{shortvector}
by reversing the sign of all hypercharges.
\par
The {\sl hypermultiplet}, satisfying $E_0=|y_0|\ge{1\over 2}$, has the structure
displayed in table \ref{hyper}.
\begin{table}
\centering
\begin{footnotesize}
\begin{tabular}{||c|c|c|c|c||}
\hline
Spin &
Energy &
Hypercharge &
Mass ($^2$)&
Name  \\
\hline
\hline
$1$      & $y_0+2$      & $y_0$      & $16y_0(y_0+1)$   & $A$       \\
$\ft12$  & $y_0+\ft52$  & $y_0\pm 1$ & $-4y_0-4$    & $\lambda_T$   \\
$\ft12$  & $y_0+\ft32$  & $y_0+1$    & $4y_0$       & $\lambda_L$   \\
$\ft12$  & $y_0+\ft32$  & $y_0-1$    & $4y_0$       & $\lambda_L$   \\
$0$      & $y_0+2$      & $y_0-2$ & $16y_0(y_0+1)$  & $\pi$     \\
$0$      & $y_0+2$      & $y_0$      & $16y_0(y_0+1)$   & $\pi$         \\
$0$      & $y_0+1$      & $y_0$      & $16y_0(y_0-1)$   & $S$       \\
\hline
\end{tabular}\\[.13in]
\end{footnotesize}
\caption{${\cal N}=2$ short vector multiplet $A$ with positive hypercharge $y_0 >0$}
\label{shortvector}
\end{table}
\begin{table}
\centering
\begin{footnotesize}
\begin{tabular}{||c|c|c|c|c||}
\hline
Spin &
Energy &
Hypercharge &
Mass ($^2$)&
Name  \\
\hline
\hline
$\ft12$  & $y_0+\ft12$  & $y_0-1$    & $4y_0-4$       & $\lambda_L$   \\
$0$      & $y_0+1$      & $y_0-2$    & $16y_0(y_0-1)$   & $\pi$     \\
$0$      & $y_0$        & $y_0$      & $16(y_0-2)(y_0-1)$   & $S$       \\
\hline
\hline
$\ft12$  & $y_0+\ft12$  & $-y_0+1$    & $4y_0-4$      & $\lambda_L$   \\
$0$      & $y_0+1$      & $-y_0+2$    & $16y_0(y_0-1)$   & $\pi$     \\
$0$      & $y_0$        & $-y_0$      & $16(y_0-1)(y_0-2)$   & $S$       \\
\hline
\end{tabular}\\[.13in]
\end{footnotesize}
\caption{${\cal N}=2$ hypermultiplet, $y_0>0$}
\label{hyper}
\end{table}
This structure is different from the others because this multiplet is
complex. This means that for each field there is another field with same
energy and spin but opposite hypercharge. So it is built with two ${\cal N}=1$
Wess-Zumino multiplets. The four real scalar field can be arranged into a
quaternionic complex form.
\par
The massless multiplets are either  short graviton or short vector
multiplets satisfying the further condition
\begin{equation}
E_0=s_0+1~~~{\rm equivalent~to}~~~y_0=0.
\end{equation}
\par
The {\sl massless graviton multiplet}, satisfying $E_0=2~~y_0=0$, has the structure
displayed in table \ref{masslessgraviton}.
\begin{table}
\centering
\begin{footnotesize}
\begin{tabular}{||c|c|c|c|c||}
\hline
Spin &
Energy &
Hypercharge &
Mass ($^2$)&
Name \\
\hline
\hline
$2$      & $3$        & $0$     & $0$  & $h$     \\
$\ft32$  & $\ft52$    & $-1$   & $0$  & $\chi^+$\\
$\ft32$  & $\ft52$    & $+1$   & $0$  & $\chi^+$\\
$1$      & $2$        & $0$     & $0$  & $A$     \\
\hline
\end{tabular}\\[.13in]
\end{footnotesize}
\caption{${\cal N}=2$ massless graviton multiplet}
\label{masslessgraviton}
\end{table}
\par
The {\sl massless vector multiplet}, satisfying $E_0=1~~y_0=0$, has the structure
displayed in table \ref{masslessvector}.
\begin{table}
\centering
\begin{footnotesize}
\begin{tabular}{||c|c|c|c|c|c|c||}
\hline
Spin &
Energy &
Hypercharge &
Mass ($^2$)&
Name &
Mass ($^2$)&
Name \\
\hline
\hline
$1$      & $2$      & $0$      & $0$  & $A$
                       & $0$  & $Z$         \\
$\ft12$  & $\ft32$  & $-1$    & $0$  & $\lambda_L$
                   & $0$  & $\lambda_T$ \\
$\ft12$  & $\ft32$  & $+1$    & $0$  & $\lambda_L$
                   & $0$  & $\lambda_T$ \\
$0$      & $2$      & $0$ & $0$  & $\pi$
                   & $0$ & $\phi$      \\
$0$      & $1$        & $0$      & $0$  & $S$
                   & $0$  & $\pi$   \\
\hline
\end{tabular}\\[.13in]
\end{footnotesize} 
\caption{${\cal N}=2$ massless vector multiplets $A$ and $Z$}
\label{masslessvector}
\end{table}
These are the ${\cal N}=2$ supermultiplets in anti de Sitter space that can occur  in
Kaluza Klein supergravity.
\par
The structure  of the long multiplets was derived in the eighties (see \cite{multanna}),
whereas the structure of the short and massless multiplets  we have
given here, is, to the best of our knowledge, a new result that we have
obtained  as a byproduct of harmonic analysis
on $M^{111}$.
In establishing our result we have also used as a tool the necessary decomposition
of the ${\cal N}=2$ multiplets into ${\cal N}=1$ multiplets (see Figures \ref{N1longgraviton},
\ref{N1longgravitino}, \ref{N1longvector}, \ref{N1shortgraviton}, \ref{N1shortgravitino},
\ref{N1shortvector}, \ref{N1hyper}) for long and short multiplets. Finally we note that
the structure of massless multiplets is identical in the anti de Sitter  and in the Poincar\'e
case, as it is well known.
\par
\begin{figure}
\centering
\begin{picture}(230,170)
\put (20,40){\shortstack{\large{$1\over 2$}}}
\put (40,0){\shortstack{\large{$0$}}}
\put (60,40){\shortstack{\large{$1\over 2$}}}
\put (40,80){\shortstack{\large{$1$}}}
\put (28,53){\line (1,2){12}}
\put (28,35){\line (1,-2){12}}
\put (59,53){\line (-1,2){12}}
\put (59,35){\line (-1,-2){12}}
\put (60,120){\shortstack{\large{$3\over 2$}}}
\put (80,80){\shortstack{\large{$1$}}}
\put (100,120){\shortstack{\large{$3\over 2$}}}
\put (80,160){\shortstack{\large{$2$}}}
\put (68,133){\line (1,2){12}}
\put (68,115){\line (1,-2){12}}
\put (99,133){\line (-1,2){12}}
\put (99,115){\line (-1,-2){12}}
\put (140,120){\shortstack{\large{$3\over 2$}}}
\put (120,80){\shortstack{\large{$1$}}}
\put (140,40){\shortstack{\large{$1\over 2$}}}
\put (160,80){\shortstack{\large{$1$}}}
\put (128,93){\line (1,2){12}}
\put (128,75){\line (1,-2){12}}
\put (159,93){\line (-1,2){12}}
\put (159,75){\line (-1,-2){12}}
\put (180,80){\shortstack{\large{$1$}}}
\put (200,120){\shortstack{\large{$3\over 2$}}}
\put (220,80){\shortstack{\large{$1$}}}
\put (200,40){\shortstack{\large{$1\over 2$}}}
\put (188,93){\line (1,2){12}}
\put (188,75){\line (1,-2){12}}
\put (219,93){\line (-1,2){12}}
\put (219,75){\line (-1,-2){12}}
\end{picture}
\caption{${\cal N}=2\rightarrow {\cal N}=1$ decomposition of the long graviton multiplet}
\label{N1longgraviton}
\end{figure}
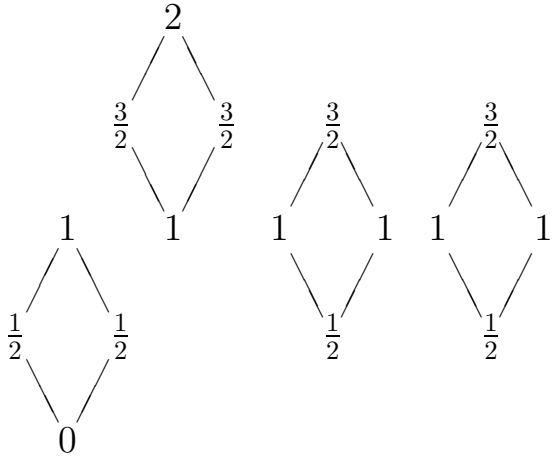
\begin{figure}
\centering
\begin{picture}(230,130)
\put (20,0){\shortstack{\large{$0$}}}
\put (60,0){\shortstack{\large{$0$}}}
\put (40,40){\shortstack{\large{$1\over 2$}}}
\put (28,13){\line (1,2){12}}
\put (59,13){\line (-1,2){12}}
\put (60,80){\shortstack{\large{$1$}}}
\put (80,40){\shortstack{\large{$1\over 2$}}}
\put (100,80){\shortstack{\large{$1$}}}
\put (80,120){\shortstack{\large{$3\over 2$}}}
\put (68,93){\line (1,2){12}}
\put (68,75){\line (1,-2){12}}
\put (99,93){\line (-1,2){12}}
\put (99,75){\line (-1,-2){12}}
\put (140,80){\shortstack{\large{$1$}}}
\put (120,40){\shortstack{\large{$1\over 2$}}}
\put (140,0){\shortstack{\large{$0$}}}
\put (160,40){\shortstack{\large{$1\over 2$}}}
\put (128,53){\line (1,2){12}}
\put (128,35){\line (1,-2){12}}
\put (159,53){\line (-1,2){12}}
\put (159,35){\line (-1,-2){12}}
\put (180,40){\shortstack{\large{$1\over 2$}}}
\put (200,80){\shortstack{\large{$1$}}}
\put (220,40){\shortstack{\large{$1\over 2$}}}
\put (200,0){\shortstack{\large{$0$}}}
\put (188,53){\line (1,2){12}}
\put (188,35){\line (1,-2){12}}
\put (219,53){\line (-1,2){12}}
\put (219,35){\line (-1,-2){12}}
\end{picture}
\caption{${\cal N}=2\rightarrow {\cal N}=1$ decomposition of the long gravitino multiplet}
\label{N1longgravitino}
\end{figure}
\begin{figure}
\centering
\begin{picture}(230,90)
\put (60,40){\shortstack{\large{$1\over 2$}}}
\put (80,0){\shortstack{\large{$0$}}}
\put (100,40){\shortstack{\large{$1\over 2$}}}
\put (80,80){\shortstack{\large{$1$}}}
\put (68,53){\line (1,2){12}}
\put (68,35){\line (1,-2){12}}
\put (99,53){\line (-1,2){12}}
\put (99,35){\line (-1,-2){12}}
\put (140,40){\shortstack{\large{$1\over 2$}}}
\put (120,0){\shortstack{\large{$0$}}}
\put (160,0){\shortstack{\large{$0$}}}
\put (128,13){\line (1,2){12}}
\put (159,13){\line (-1,2){12}}
\put (180,0){\shortstack{\large{$0$}}}
\put (200,40){\shortstack{\large{$1\over 2$}}}
\put (220,0){\shortstack{\large{$0$}}}
\put (188,13){\line (1,2){12}}
\put (219,13){\line (-1,2){12}}
\end{picture}
\caption{${\cal N}=2\rightarrow{\cal N}=1$ decomposition of the long vector multiplet}
\label{N1longvector}
\end{figure}
\begin{figure}
\centering
\begin{picture}(230,130)
\put (60,80){\shortstack{\large{$3\over 2$}}}
\put (80,40){\shortstack{\large{$1$}}}
\put (100,80){\shortstack{\large{$3\over 2$}}}
\put (80,120){\shortstack{\large{$2$}}}
\put (68,93){\line (1,2){12}}
\put (68,75){\line (1,-2){12}}
\put (99,93){\line (-1,2){12}}
\put (99,75){\line (-1,-2){12}}
\put (140,80){\shortstack{\large{$3\over 2$}}}
\put (120,40){\shortstack{\large{$1$}}}
\put (140,0){\shortstack{\large{$1\over 2$}}}
\put (160,40){\shortstack{\large{$1$}}}
\put (128,53){\line (1,2){12}}
\put (128,35){\line (1,-2){12}}
\put (159,53){\line (-1,2){12}}
\put (159,35){\line (-1,-2){12}}
\end{picture}
\caption{${\cal N}=2\rightarrow{\cal N}=1$ decomposition of the short graviton multiplet}
\label{N1shortgraviton}
\end{figure}
\begin{figure}
\centering
\begin{picture}(230,130)
\put (60,80){\shortstack{\large{$1$}}}
\put (80,40){\shortstack{\large{$1\over 2$}}}
\put (100,80){\shortstack{\large{$1$}}}
\put (80,120){\shortstack{\large{$3\over 2$}}}
\put (68,93){\line (1,2){12}}
\put (68,75){\line (1,-2){12}}
\put (99,93){\line (-1,2){12}}
\put (99,75){\line (-1,-2){12}}
\put (140,80){\shortstack{\large{$1$}}}
\put (120,40){\shortstack{\large{$1\over 2$}}}
\put (140,0){\shortstack{\large{$0$}}}
\put (160,40){\shortstack{\large{$1\over 2$}}}
\put (128,53){\line (1,2){12}}
\put (128,35){\line (1,-2){12}}
\put (159,53){\line (-1,2){12}}
\put (159,35){\line (-1,-2){12}}
\end{picture}
\caption{${\cal N}=2\rightarrow{\cal N}=1$ decomposition of the short gravitino multiplet}
\label{N1shortgravitino}
\end{figure}
\begin{figure}
\centering
\begin{picture}(230,90)
\put (60,40){\shortstack{\large{$1\over 2$}}}
\put (80,0){\shortstack{\large{$0$}}}
\put (100,40){\shortstack{\large{$1\over 2$}}}
\put (80,80){\shortstack{\large{$1$}}}
\put (68,53){\line (1,2){12}}
\put (68,35){\line (1,-2){12}}
\put (99,53){\line (-1,2){12}}
\put (99,35){\line (-1,-2){12}}
\put (140,40){\shortstack{\large{$1\over 2$}}}
\put (120,0){\shortstack{\large{$0$}}}
\put (160,0){\shortstack{\large{$0$}}}
\put (128,13){\line (1,2){12}}
\put (159,13){\line (-1,2){12}}
\end{picture}
\caption{${\cal N}=2\rightarrow{\cal N}=1$ decomposition of the short vector multiplet}
\label{N1shortvector}
\end{figure}
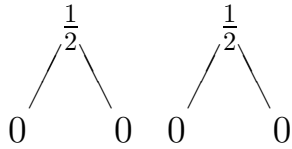
\begin{figure}
\centering
\begin{picture}(230,90)
\put (80,40){\shortstack{\large{$1\over 2$}}}
\put (60,0){\shortstack{\large{$0$}}}
\put (100,0){\shortstack{\large{$0$}}}
\put (68,13){\line (1,2){12}}
\put (99,13){\line (-1,2){12}}
\put (120,0){\shortstack{\large{$0$}}}
\put (140,40){\shortstack{\large{$1\over 2$}}}
\put (160,0){\shortstack{\large{$0$}}}
\put (128,13){\line (1,2){12}}
\put (159,13){\line (-1,2){12}}
\end{picture}
\caption{${\cal N}=2\rightarrow{\cal N}=1$ decomposition of the hypermultiplet}
\label{N1hyper}
\end{figure}
\par
This is all the information that  we can have on the multiplets from
superalgebra theory. The determination of the actual spectrum by harmonic analysis
teaches us a lot more: indeed we find out which multiplets are present for each representation
of $G'$, how many they are, and the exact value of the $E_0$ label for each multiplet.
So, at the end  we know the energy of every field in the multiplets. In the dual superconformal
field theory on the two--brane these are the conformal dimensions of the primary operators.
In addition we find also the hypercharge of each multiplet in the spectrum.
\section{The complete spectrum of $Osp(2\vert 4) \times SU\left(3\right)\times SU\left(2\right)$ supermultiplets}
As stressed in the introduction the irreducible supermultiplets of $Osp\left(2|4\right)$
occur into irreducible representations
of the bosonic group
\[
G'=SU\left(3\right)\times SU\left(2\right) \,,
\]
whose interpretation is the {\sl flavor group} in the conformal field
theory side of the correspondence and it is the {\sl gauge group} in the
supergravity side. In any case the crucial information one needs to
extract from harmonic analysis is precisely the $G^\prime$
representation assignment of the supermultiplets and the actual value of $E_0$ and
$y_0$.
\par
To present our result, we first need to fix our conventions for
labelling the irreducible representations of $G^\prime$.
It has rank three, so that its irreducible representations are labeled
by three integer numbers. A representation of $SU\left(3\right)$ can be identified by a Young
diagram of the following type
\begin{eqnarray*}
\begin{array}{l}
\begin{array}{|c|c|c|c|c|c|}
\hline
             \hskip .3 cm & \cdots & \hskip .3 cm &
             \hskip .3 cm & \cdots & \hskip .3 cm \\
\hline
\end{array}\,,\\
\begin{array}{|c|c|c|}
             \hskip .3 cm & \cdots & \hskip .3 cm \\
\hline
\end{array}
\end{array}\\
\underbrace{\hskip 2.2 cm}_{M_2}
\underbrace{\hskip 2.2 cm}_{M_1}
\end{eqnarray*}
while an irreducible representation of $SU\left(2\right)$ can be described by a Young diagram
as follows
\begin{eqnarray*}
\begin{array}{|c|c|c|}
\hline
             \hskip .3 cm & \cdots & \hskip .3 cm \\
\hline
\end{array}\,.\\
\underbrace{\hskip 2.2 cm}_{2J}
\end{eqnarray*}
Hence we can take the  nonnegative integers $M_1,~M_2,~2J$, 
as the labels of a $G'$ irreducible representation.
\par
Relying on the procedures explained in the following sections and originally
introduced in \cite{spectfer}, we have found the  following result.
\par
Not every $G'$ representation  is actually present, but only
those representations that satisfy the following relations
\begin{equation}
M_2-M_1 \in 3 \, \ZZ \quad ; \quad J \in \IN.
\label{labcond}
\end{equation}
In the following pages,  for each type  of ${\cal N}=2$ multiplet
we list the  $G'$ representations through which it occurs in the
spectrum. We do this by writing bounds on the range of values for
the $M_1,~M_2,~2J~$ labels. The reader  should take into account
that, case by case, in addition to the specific bounds we write, also the
general restriction (\ref{labcond}) ) is to be imposed.
\par
Furthermore for every multiplet, we give   the energy and hypercharge
values $E_0$ and $y_0$  of the Clifford vacuum. From the tables \ref{longgraviton},
\ref{longgravitino}, \ref{longvector}, \ref{shortgraviton},
\ref{shortgravitino}, \ref{shortvector}, \ref{hyper}, \ref{masslessgraviton},
\ref{masslessvector} it is straightforward to get the energies and the
hypercharges of all other  fields in each  multiplet.
\par
As a short--hand notation let us name $H_0$ the following quadratic form in the representation
labels:
\begin{equation}
\label{defH0}
H_0\equiv {64\over 3}\left(M_2+M_1+M_2M_1\right)+32J\left(J+1\right)+{32\over 9}
\left(M_2-M_1\right)^2.
\end{equation}
Up to multiplicative constants, the first two addends $M_2 + M_1 + M_2\,
M_1$ and  $J(J+1)$   are the Casimirs of  $G^\prime = SU(3)
\times SU(2)$. The last addendum is contributed by the square of the hypercharge through
its relation with the $SU(3)$ representation implied by the geometry of the space.
\vskip 0.6cm
\centerline{LONG MULTIPLETS}
\begin{enumerate}
  \item {{\underline {\bf Long graviton multiplets}} 
$~~\left(1\left(2\right),4\left(3\over 2\right),6\left(1\right),4\left(1\over 2\right),1\right)$\par
One long graviton multiplet (table \ref{longgraviton}) in each representation of the series
\begin{eqnarray}
&&\left\{M_1\ge 0,~M_2\ge 0,~J>\left|{1\over 3}\left(M_2-M_1\right)\right|\right\}\cup\nonumber\\
&&\left\{M_1>0,~M_2>0,~J=\left|{1\over 3}\left(M_2-M_1\right)\right|\right\}
\end{eqnarray}
with
\begin{equation}
h:~E_0={1\over 2}+{1\over 4}\sqrt{H_0+36},~y_0={2\over 3}\left(M_2-M_1\right)
\end{equation}}
  \item {{\underline {\bf Long gravitino multiplets}}
$~~\left(1\left({3\over 2}\right),4\left(1\right),6\left(1\over 2\right),4\left(0\right)\right)$\par
  \begin{itemize}
\item Four long gravitino multiplets(two $\chi^+$ and two $\chi^-$, table
\ref{longgravitino}) in each representation of the series
\[
\left\{M_2>0,~M_1>0,~J>\left|{1\over 3}\left(M_2-M_1\right)\right|+1\right\}\cup
\]
\[
\left\{M_2\ge M_1>1,~J={1\over 3}\left(M_2-M_1\right)+1\right\}\cup
\]
\begin{equation}
\left\{M_1\ge M_2>1,~J=-{1\over 3}\left(M_2-M_1\right)+1\right\}
\end{equation}
with
\begin{eqnarray}
\chi^+:~&E_0=-{1\over 2}+{1\over 4}\sqrt{H_0+{32\over 3}\left(M_2-M_1\right)+16},
&y_0={2\over 3}\left(M_2-M_1\right)-1\\
\chi^+:~&E_0=-{1\over 2}+{1\over 4}\sqrt{H_0-{32\over 3}\left(M_2-M_1\right)+16},
&y_0={2\over 3}\left(M_2-M_1\right)+1\\
\chi^-:~&E_0={3\over 2}+{1\over 4}\sqrt{H_0+{32\over 3}\left(M_2-M_1\right)+16},
&y_0={2\over 3}\left(M_2-M_1\right)-1\\
\chi^-:~&E_0={3\over 2}+{1\over 4}\sqrt{H_0-{32\over 3}\left(M_2-M_1\right)+16},
&y_0={2\over 3}\left(M_2-M_1\right)+1
\end{eqnarray}
\item Three long gravitino multiplets (one $\chi^+$ and two $\chi^-$, table
\ref{longgravitino}), in each representation of the series
\begin{equation}
\left\{M_2\ge M_1=1,~J={1\over 3}\left(M_2-M_1\right)+1\right\}
\end{equation}
with
\begin{eqnarray}
\chi^+:~&E_0=-{1\over 2}+{1\over 4}\sqrt{H_0+{32\over 3}\left(M_2-M_1\right)+16},
&y_0={2\over 3}\left(M_2-M_1\right)-1\\
\chi^-:~&E_0={3\over 2}+{1\over 4}\sqrt{H_0-{32\over 3}\left(M_2-M_1\right)+16},
&y_0={2\over 3}\left(M_2-M_1\right)+1\\
\chi^-:~&E_0={3\over 2}+{1\over 4}\sqrt{H_0+{32\over 3}\left(M_2-M_1\right)+16},
&y_0={2\over 3}\left(M_2-M_1\right)-1
\end{eqnarray}
and the conjugate multiplets in the complex conjugate representations of $G'$.
\item Two long gravitino multiplets (one $\chi^+$ and one $\chi^-$, table
\ref{longgravitino}) in each representation of the series
\[
\left\{M_2>M_1> 0,~J={1\over 3}\left(M_2-M_1\right)
\right\}\cup
\]
\[
\left\{M_2>M_1> 0,~J={1\over 3}\left(M_2-M_1\right)-1
\right\}\cup
\]
\begin{equation}
\left\{M_2>M_1=0,~J\ge {1\over 3}\left(M_2-M_1\right)\right\}
\end{equation}
with
\begin{eqnarray}
\chi^+:~&E_0=-{1\over 2}+{1\over 4}\sqrt{H_0+{32\over 3}\left(M_2-M_1\right)+16},
&y_0={2\over 3}\left(M_2-M_1\right)-1\\
\chi^-:~&E_0={3\over 2}+{1\over 4}\sqrt{H_0+{32\over 3}\left(M_2-M_1\right)+16},
&y_0={2\over 3}\left(M_2-M_1\right)-1
\end{eqnarray}
and the conjugate multiplets in the complex conjugate representations of $G'$.
\item One long gravitino multiplet (a $\chi^-$, table \ref{longgravitino}),
in each representation of the series
\begin{equation}
\left\{M_2>M_1=0,~J={1\over 3}\left(M_2-M_1\right)-1\right\}
\end{equation}
with
\begin{equation}
\chi^-:~E_0={3\over 2}+{1\over 4}\sqrt{H_0+{32\over 3}\left(M_2-M_1\right)+16},
~y_0={2\over 3}\left(M_2-M_1\right)-1
\end{equation}
and the conjugate multiplet in the complex conjugate representations of
$G'$.
\end{itemize}
}
\item {{\underline {\bf Long vector multiplets}}
$~~\left(1\left(1\right),4\left(1\over 2\right),5\left(0\right)\right)$\par
As already stressed there are different realizations of the long vector
 multiplet
  arising   from different fields of the $D=11$ theory. We have the $W$ vector
multiplets, the $A$ vector multiplets, and   the  $Z$ vector
multiplets.
\begin{itemize}
  \item One $W$ long vector multiplet (table \ref{longvector})
in each representation of the series
\begin{equation}
\left\{M_2\ge 0,~M_1\ge 0,~J\ge\left|{1\over 3}\left(M_2-M_1\right)\right|\right\}
\end{equation}
with
\begin{equation}
W:~E_0={5\over 2}+{1\over 4}\sqrt{H_0+36},~y_0={2\over 3}\left(M_2-M_1\right)
\end{equation}
\item One $A$ long vector multiplet (table \ref{longvector})
in each representation of the series
\[
\left\{M_2\ge M_1=0,~J>{1\over 3}\left(M_2-M_1\right)+1\right\}\cup
\]
\[
\left\{M_2\ge M_1=1,~J>{1\over 3}\left(M_2-M_1\right)\right\}\cup
\]
\begin{equation}
\left\{M_2\ge M_1>1,~J\ge{1\over 3}\left(M_2-M_1\right)\right\}
\end{equation}
with
\begin{equation}
A:~E_0=-{3\over 2}+{1\over 4}\sqrt{H_0+36},~y_0={2\over 3}\left(M_2-M_1\right)
\end{equation}
and the conjugate multiplet in the complex conjugate representations of $G'$.
\item One $Z$ long vector multiplet (table \ref{longvector})
in each representation of the series
\begin{equation}
\left\{M_2>0,~M_1>0,~J\ge\left|{1\over 3}\left(M_2-M_1\right)\right|\right\}
\end{equation}
with
\begin{equation}
Z:~E_0={1\over 2}+{1\over 4}\sqrt{H_0+4},~y_0={2\over 3}\left(M_2-M_1\right)
\end{equation}
\item One $Z$ long vector multiplet (table \ref{longvector})
in each representation of the series
\[
\left\{M_2>M_1+3,~J\ge {1\over 3}\left(M_2-M_1\right)-2\right\}\cup
\]
\begin{equation}
\left\{M_1+3\ge M_2>1,~J>-{1\over 3}\left(M_2-M_1\right)+1\right\}
\end{equation}
with
\begin{equation}
Z:~E_0={1\over 2}+{1\over 4}\sqrt{H_0+{64\over 3}\left(M_2-M_1\right)-28},
~y_0={2\over 3}\left(M_2-M_1\right)-2
\end{equation}
and the conjugate multiplet in the complex conjugate representations of $G'$.
\end{itemize}
}
\end{enumerate}
\vskip 0.6cm
\centerline{SHORT MULTIPLETS}
\begin{enumerate}
  \item {{\underline {\bf Short graviton multiplets}}
  $~~\left(1\left(2\right),3\left(3\over 2\right),3\left(1\right),{1\over 2}\right)$\par
\begin{itemize}
  \item One short graviton multiplet (table \ref{shortgraviton})
in each representation of the series
\begin{equation}
\cases{
M_2=3k\cr
M_1=0\cr
J=k\cr}~~k>0~{\rm integer}
\end{equation}
with
\begin{equation}
E_0=2k+2,~y_0=2k
\end{equation}
and the conjugate multiplet in the complex conjugate representations of $G'$.
\end{itemize} } 
  \item {{\underline {\bf Short gravitino multiplets}}
   $~~\left(1\left({3\over 2}\right),3\left(1\right),3\left(1\over 2\right),0\right)$\par
  \begin{itemize}
\item One short gravitino multiplet ($\chi^+$, table \ref{shortgravitino}) in
each representation of the series
\begin{equation}
\cases{
M_2=3k+1\cr
M_1=1\cr
J=k+1\cr}~~
k\ge 0~{\rm integer}
\end{equation}
with
\begin{equation}
E_0=2k+{1\over 2},~y_0=2k+1
\end{equation}
and the conjugate multiplet in the complex conjugate representations of $G'$.
\item One short gravitino multiplet ($\chi^+$, table \ref{shortgravitino}) in
each representation of the series
\begin{equation}
\cases{
M_2=3k\cr
M_1=0\cr
J=k-1\cr}~~
k>0~{\rm integer}
\end{equation}
with
\begin{equation}
E_0=2k+{1\over 2},~y_0=2k-1
\end{equation}
and the conjugate multiplet in the complex conjugate representations of $G'$.
\end{itemize}
}
  \item {{\underline {\bf  Short vector multiplets}}
 $~~\left(1\left(1\right),3\left(1\over 2\right),3\left(0\right)\right)$\par
  \begin{itemize}
\item One short vector multiplet ($A$, table \ref{shortvector}), in each
representation of the series
\begin{equation}
\cases{
M_2=3k+1\cr
M_1=1\cr
J=k\cr}~~k>0~{\rm integer}
\end{equation}
\begin{equation}
\cases{
M_2=3k\cr
M_1=0\cr
J=k+1\cr}~~k>0~{\rm integer}
\end{equation}
with
\begin{equation}
E_0=2k+1,~y_0=2k
\end{equation}
and the conjugate multiplet in the complex conjugate representations of $G'$.
\end{itemize}
}
\item {{\underline {\bf  Hypermultiplets}}
 $~~\left(2\left(1\over 2\right),4\left(0\right)\right)$\par
\begin{itemize}
\item One hypermultiplet (table \ref{hyper}) in each representation of the series
\begin{equation}
\label{hyp1}
\cases{
M_2=3k\cr
M_1=0\cr
J=k\cr}~~k>0~{\rm integer}
\end{equation}
\begin{equation}
E_0=\left|y_0\right|=2k
\end{equation}
We mean that a half of the hypermultiplet is in a representation among the (\ref{hyp1}),
the other half is in a representation among their conjugates. Indeed   the hypermultiplet
is complex, so half of it transforms in a representation $\rho$ of $G'$, half
of it in the conjugate representation $\rho^*$.
\end{itemize}
}
\end{enumerate}
\vskip 0.6cm
\centerline{MASSLESS MULTIPLETS}
\vskip 0.6cm
\begin{enumerate}
\item {\underline{\bf The massless graviton multiplet}} (table \ref{masslessgraviton})
$~~\left(1\left(2\right),2\left(3\right),1\right)$\\
in the singlet representation
\begin{equation}
M_2=M_1=J=0
\end{equation}
with
\begin{equation}
E_0=2,~y_0=0.
\end{equation}
In this multiplet the graviphoton is associated with the Killing vector of the $R$--symmetry group
$U\left(1\right)_R$.
\item {\bf The massless vector multiplet} (table \ref{masslessvector})
$~~\left(1\left(1\right),2\left(1\right),2\left(0\right)\right)$\\
in  the {\em adjoint representation} of the $G^\prime$ group
\begin{eqnarray}
M_2=M_1=1,~J=0&&\label{rkilsu3}\\
M_2=M_1=0,~J=1&&\label{rkilsu2}
\end{eqnarray}
with
\begin{equation}
  E_0=1,~y_0=0.
\label{labellini}
\end{equation}
\item {\bf An additional massless vector multiplet} in the singlet
representation of the gauge group
\begin{equation}
M_2=M_1=J=0\label{rbetti}
\end{equation}
with the same energy and hypercharges as in (\ref{labellini}) that
arises from the three--form $A_{\mu ij}$ and is due to the existence
of one closed cohomology two--form on the $M^{111}$ manifold. This
multiplet is named the Betti multiplet.
\end{enumerate}
\par
Summarizing, the massless spectrum, besides the supergravity multiplet
contains  twelve vector multiplets: so the total number
of  massless gauge bosons is thirteen, one of them being the graviphoton.
In the low energy effective lagrangian we just couple to supergravity these twelve vector
multiplets. However we expect the gauging of
a thirteen--parameter group:
\begin{equation}
SU\left(3\right)\times SU\left(2\right)\times U(1)_R \times U\left(1\right)'
\end{equation}
the further $U\left(1\right)'$  being  associated with the Betti
multiplet. All Kaluza Klein states are neutral under
  $U\left(1\right)'$ yet non perturbative states might carry
  $U(1)^\prime$ charges. This is a completely open problem.
\vskip 1cm
\centerline{\LARGE
{\bf PART TWO: {\sl  THE DERIVATION}}}
\vskip .7cm
In the second part of the paper we give the detailed derivation
of the results presented in the first part. Our main tools
in this derivation are:
\begin{itemize}
\item harmonic analysis on $G\over H$ seven--dimensional coset manifolds
as developed in the eighties by D'Auria and Fr\`e 
\cite{spectfer,univer}
\item the general diagonalization of linearized $D=11$ supergravity
field equations and the resulting mass--formulae derived by D'Auria and
Fr\`e \cite{spectfer} and by Castellani et al \cite{bosmass} also in the
eighties 
\item the mass relations following from the existence of Killing spinors 
on $G\over H$ derived by D'Auria and Fr\`e in \cite{univer}
\item The structure of ${\cal N}=2$ long multiplets that were obtained
in \cite{frenico,multanna}.
\end{itemize}
We begin with a review of $M^{111}$ differential geometry.
\par
\section{The differential geometry of $M^{111}$}\label{geometry}
Let us briefly review the essential features of $M^{111}$ geometry.
For notations and basic lore we mainly refer to 
\cite{noi321,spectfer,castdauriafre}.
\par
The homogeneous space $M^{111}$ is the quotient of
$G=SU(3) \times SU(2) \times U(1)$ by the action of its
subgroup $H=SU(2)^c \times U(1)' \times U(1)''$, where the
embedding of $H$ in $G$ is defined as follows.
$SU(2)^c$
\footnote{The superscript $^c$ stands for \emph{color}.
This nomenclature is due to the original interest in the
group $G$ as the gauge group of the GSW standard model.
Even if this perspective has faded, the terminology is
still useful.
So we will label quantities referring to $SU(3)$ with the
index $^c$, and quantities referring to the \emph{weak isospin}
$SU(2)\not\subset SU(3)$ with $^w$.}
is the isospin $SU\left(2\right)$ subgroup of $SU\left(3\right)$ 
with respect to which the fundamental triplet representation
branches as follows:
\begin{equation}
{\bf 3}\rightarrow{\bf 2}\oplus{\bf 1}.
\end{equation}
The additional $U(1)'$ and $U(1)''$ factors of the $H$ subgroup
are generated by $Z'$ and $Z''$ being two independent linear
combinations of the three remaining abelian generators of
$SU\left(3\right)\times SU\left(2\right)\times U\left(1\right)$
commuting with $SU\left(2\right)^c$. Using the Gell--Mann
matrices for $SU\left(3\right)$ and the Pauli matrices for
$SU\left(2\right)$ (see appendix A for conventions) these three
generators are ${\sqrt{3}\over 2}i\lambda_8,~{1\over 2}i\sigma_{\ddot{3}},~$
and $iY_{\stackrel{.\!.\!.}{3}}$. The two linear combinations  
$Z'$ and $Z''$ are defined as a basis for the orthogonal complement
of the generator
\begin{equation} \label{Z}
Z=p\ft{\sqrt{3}}{2}i\l_8+\ft{1}{2}qi\s_{\ddot{3}}+riY_{\stackrel{.\!.\!.}{3}}
\end{equation}
\[
{\rm with}~~~p=q=r=1
\]
which belongs to $\IK$ rather than $\IH$ in the decomposition of the
$SU\left(3\right)\times SU\left(2\right)\times U\left(1\right)$
Lie algebra
\begin{equation}
\cases{
\IG=\IH\oplus\IK\cr
\IG=SU\left(3\right)\times SU\left(2\right)\times U\left(1\right)\cr
\IH=SU\left(2\right)\times U\left(1\right)'\times U\left(1\right)''\cr}
\end{equation}
\par
In the above equations $i\sqrt{3}\l_8$ is the hypercharge generator of
$SU(3)$, $\ft{1}{2}i\s_{\ddot{3}}$ is the third component of the weak
isospin and $iY_{\stackrel{.\!.\!.}{3}}$ is the weak $U(1)$ generator.
\par
An explicit representation of the group $G$ is given by the following $6\times 6$ 
block--diagonal matrix:
\begin{equation}
G\ni g=
\begin{array}{c}
\left(\begin{array}{c|c|c}
SU\left(3\right)&0&0\\
\hline
0&SU\left(2\right)&0\\
\hline
0&0&U\left(1\right)
\end{array}
\right)\,,\\
\underbrace{\hspace{1.3 cm}}_{3}
\underbrace{\hspace{1.3 cm}}_{2}
\underbrace{\hspace{1.15 cm}}_{1}
\end{array}
\end{equation}
where the diagonal blocks contain the fundamental representation of
$SU(3)$, $SU(2)$ and $U(1)$ respectively.
The whole set of generators of $G$ is given by:
\begin{equation}
T_{\Lambda}\equiv\left(\ft{1}{2}i\l_A,\,\ft{1}{2}i\l_{\dot m},\,
\ft{1}{2}i\l_8,\,\ft{1}{2}i\s_{\ddot n},\,\ft{1}{2}i\s_{\ddot 3},\,
iY_{\stackrel{.\!.\!.}{3}}\right),
\end{equation}
where $\l_i$ stands for the $i$-th Gell-Mann matrix trivially extended
to a $6\times6$ matrix:
\begin{equation}
\l_i\to\left(\begin{array}{ccc}
\l_i&0&0\\
0&0&0\\
0&0&0
\end{array}
\right)\,.
\end{equation}
Similarly $\s_i$ denotes the following extension of the Pauli matrices:
\begin{equation}
\s_i\to\left(\begin{array}{ccc}
0&0&0\\
0&\s_i&0\\
0&0&0
\end{array}
\right)\,,
\end{equation}
and $Y_{\stackrel{.\!.\!.}{3}}$ is given by:
\begin{equation}
Y_{\stackrel{.\!.\!.}{3}}=\left(\begin{array}{ccc}
0&0&0\\
0&0&0\\
0&0&1
\end{array}
\right)\,.
\end{equation}
A basis for the two abelian generators of $H$ can is given by
\begin{eqnarray}
Z'=\sqrt{3}i\l_8+i\s_{\ddot 3}-4iY_{\stackrel{.\!.\!.}{3}}\,,\label{z1}\\
Z''=-\ft{\sqrt{3}}{2}i\l_8+\ft{3}{2}i\s_{\ddot 3}\,,
\label{z2}
\end{eqnarray}
which are orthogonal among themselves and with $Z$:
\begin{equation}
Tr(ZZ')=Tr(ZZ'')=Tr(Z'Z'')=0\,.
\end{equation}
An explicit parametrization of the coset $G/H$ is given by the
seven coordinates $(y^A,y^m,y^3)$:
\begin{equation}
L(y^A,y^m,y^3)=\exp(\ft{1}{2}i\l_Ay^A)\exp(\ft{1}{2}i\s_my^m)\exp(Zy^3)\ .
\end{equation}
From this we can construct the left-invariant one-forms on $G/H$ as:
\begin{equation}
\Omega(y)=L^{-1}(y){\rm d}L(y)=\Omega^{\Lambda}(y)T_{\Lambda}\,,
\end{equation}
which satisfy the Maurer-Cartan equations
\begin{equation}\label{Omega}
{\rm d}\Omega^{\Lambda}+\ft{1}{2}\cC^{\Lambda}_{\ \Sigma\Pi}
\Omega^{\Sigma}\wedge\Omega^{\Pi}=0
\end{equation}
with the structure constants of $G$:
\begin{equation}
\left[T_{\Sigma},T_{\Pi}\right]=\cC^{\Lambda}_{\ \Sigma\Pi}T_{\Lambda}\,.
\end{equation}
The one-forms $\Omega^{\Lambda}$ can be separated into a set
$\{\Omega^H\}$ corresponding to the generators of the subalgebra $\IH$
and a set $\{\Omega^{\a}\}$ corresponding to the coset generators.
These latter can be identified with the $SU(3)\times SU(2)\times U(1)$
invariant seven-vielbeins on $G/H$:
\begin{eqnarray}
\cB^{\a} \equiv (\cB^A, \cB^m, \cB^3),\nonumber\\
\left\{\begin{array}{ccl}\label{B}
\cB^A & = & \ft{\sqrt{3}}{8}\Omega^A,\\
\cB^m & = & \ft{\sqrt{2}}{8}\Omega^m,\\
\cB^3 & = & \ft{1}{8}(\sqrt{3}\Omega^8+\Omega^{\ddot 3}+
2\Omega^{\stackrel{.\!.\!.}{3}}) = \ft{3}{2}\Omega^Z,
\end{array}\right.
\end{eqnarray}
where the multiplicative coefficients in front of the vielbeins
have been properly chosen to let the metric on $G/H$ be Einstein.
The invariant forms $\Omega^H$ are:
\begin{equation}
\left\{\begin{array}{ccl}
\Omega^{\dot m}, &&\\
\Omega^{Z'} & = & \ft{1}{24}(\sqrt{3}\Omega^8+\Omega^{\ddot 3}
 -4\Omega^{\stackrel{.\!.\!.}{3}}),\\
\Omega^{Z''} & = & \ft{1}{12}(3\Omega^{\ddot 3}-\sqrt{3}\Omega^8).
\end{array}\right.
\end{equation}
\par
The spin-connection $\cB^{\a}_{\ \b}$ is easily determined from
the vielbeins $\cB^{\a}$ by imposing vanishing torsion:
\begin{equation}
{\rm d}\cB^{\a}-\cB^{\a}_{\ \b}\wedge\cB^{\b}=0,
\end{equation}
\begin{equation}
\left\{
\begin{array}{ccl}
\cB^{mn} &=& \e^{mn}\left(\Omega^{\ddot 3}-2\cB^3\right),\\
\cB^{3m} &=& -2\e^{mn}\cB_n,\\
\cB^{mA} &=& 0,\\
\cB^{3A} &=& -\ft{4}{\sqrt{3}}f^{8AB}\cB_B,\\
\cB^{AB} &=& f^{{\dot m}AB}\Omega_{\dot m}+f^{8AB}\Omega_8-
\ft{4}{\sqrt{3}}f^{8AB}\cB^3.
\end{array}\right.
\end{equation}
\section{Harmonic analysis on $M^{111}$}
In this section we summarize the essential ideas concerning the 
techniques of harmonic analysis on homogeneous seven--manifolds
originally developed in \cite{spectfer,univer}. These techniques are the 
basic ingredient of our calculations and in the present summary we 
present them already applied to the specific case  of the $M^{111}$
manifold.
\par
The essential goal of harmonic analysis, is that of translating
a differential equation problem into a linear algebraic one,
by means of group theory.
In the present case, the differential equations to solve are
the linearized field equations of Kaluza Klein supergravity, whose
typical form is:
\begin{equation}
\left( \Box_x^{[J_1 J_2]}+\xbox_y^{[\l_1\l_2\l_3]}\right)
\Phi_{[\l_1\l_2\l_3]}^{[J_1 J_2]}(x,y)=0,
\end{equation}
where $\Phi_{[\l_1\l_2\l_3]}^{[J_1 J_2]}(x,y)$ is a field transforming
in the irreducible representations $[J_1 J_2]$ of $SO(3,2)$
and  $[\l_1\l_2\l_3]$ of $SO(7)$, and depends both on the
coordinates $x$ of Anti-de Sitter space and on the coordinates $y$
of $G/H$.
$\Box_x^{[J_1 J_2]}$ is the kinetic operator for a field of spin
$[J_1 J_2]$ in four dimension while $\xbox_y^{[\l_1\l_2\l_3]}$ is
the kinetic operator for a field of spin $[\l_1\l_2\l_3]$ in
seven dimensions.
\par
Now, the harmonics constitute a complete set of functions for
the expansion of any $SO(7)$-irreducible field over $G/H$,
${\cal Y}^i_{[\l_1\l_2\l_3]}(y)$.
But their most important property is that they transform irreducibly
under $G$, the group of isometries of the coset space.
This group acts on ${\cal Y}^i_{[\l_1\l_2\l_3]}(y)$ through the
so-called covariant Lie derivative (see eq.(2.25) of
\cite{spectfer}):
\begin{equation}
\d_{\Lambda}{\cal Y}^i_{[\l_1\l_2\l_3]}(y):=
\cL_{\Lambda}{\cal Y}^i_{[\l_1\l_2\l_3]}(y),
\end{equation}
which satisfy the Lie algebra of the group $G$:
\begin{equation}
\left[\cL_{\Lambda},\cL_{\Sigma}\right]=
\cC^{\Pi}_{\ \Lambda\Sigma}\cL_{\Pi}.
\end{equation}
Moreover, the operators $\cL_{\Lambda}$ commute with the $SO(7)$
covariant derivative:
\begin{equation}\label{dercomm}
\cL_{\Lambda}D{\cal Y}^i=D\cL_{\Lambda}{\cal Y}^i\,,
\end{equation}
where $D$ is defined by
\begin{equation}\label{covder}
D=d+\cB^{\a\b}t_{\a\b},
\end{equation}
and $(t_{\a\b})^i_{\ j}$ are the generators of the $SO(7)$ irreducible 
representation $[\l_1\l_2\l_3]$ of ${\cal Y}^i$ (e.g. $(t_{\a\b})^{\g\d}=-\d_{\a\b}^{\g\d}$
for the vector representation).
\par
An important thing to note is that $H=SU(2)\times U(1)\times U(1)$
is necessarily a subgroup of $SO(7)$, whose natural embedding is given in
terms of the following embedding of the algebra $\IH$ into the adjoint
representation of $SO(7)$:
\begin{eqnarray}
(T_H)^{\a}_{\ \b}=\cC^{\a}_{H\,\b},\nonumber\\
(T_{Z'})^{\a}_{\ \b}=\left(
\begin{array}{ccc}
2\sqrt{3}f^{8AB}&0&0\\
0&2\e^{mn}&0\\
0&0&0
\end{array}\right),\\
(T_{Z''})^{\a}_{\ \b}=\left(
\begin{array}{ccc}
-\sqrt{3}f^{8AB}&0&0\\
0&3\e^{mn}&0\\
0&0&0
\end{array}\right),\\
(T_{\dot m})^{\a}_{\ \b}=\left(
\begin{array}{ccc}
f^{{\dot m}AB}&0&0\\
0&0&0\\
0&0&0
\end{array}\right).
\end{eqnarray}
This means that the $SO(7)$-indices of the various $n$-forms can be
split in the following subsets, each one transforming into an irreducible
representation of $H$:
\begin{equation}
\begin{array}{ccl}
{\cal Y}^{\a} &=& \{{\cal Y}^A,{\cal Y}^m,{\cal Y}^3\}\\
{\cal Y}^{[\a\b]} &=& \{{\cal Y}^{AB},{\cal Y}^{Am},{\cal Y}^{mn},
{\cal Y}^{A3},{\cal Y}^{m3}\}\\
{\cal Y}^{[\a\b\g]} &=& \{{\cal Y}^{ABC},{\cal Y}^{ABm},{\cal Y}^{AB3},
{\cal Y}^{Amn},{\cal Y}^{Am3},{\cal Y}^{mn3}\}
\end{array}
\end{equation}
and the $SO(7)$ irreducible representations $[\l_1\l_2\l_3]$ break into the direct
sum of $H$ irreducible representations.
Let us then introduce some notation for the harmonics of $M^{111}$,
which will be denoted by $\cH$.
\par
A generic $SO(3,2)\!\!\times\!\!H$-irreducible field can be
expanded as follows:
\begin{equation} \label{expansion}
\Phi_{[J^cZ'Z'']i_1\cdots i_{2J^c}}^{[J_1 J_2]}(x,y) =
{\sum}'_{[M_1M_2J\,Y]}
\sum_{\mu}\ \sum_m
\cH_{[J^c Z' Z'']i_1\cdots i_{2J^c}}^{[M_1M_2J\,Y]m\mu}(y)\cdot
\varphi^{[J_1J_2]}_{[M_1M_2J\,Y]m}(x)
\end{equation}
The coefficients $\varphi(x)$ of the expansion will become
the space--time fields of the theory in $AdS_4$.
The first sum is over all the $G$ irreducible representations $[M_1M_2J\,Y]$ which
break into the given $H$-one.
 We call ${\sum}'$ the sum over this 
subset of the possible representations of $G$.
The subscripts $_{i_1,\cdots,i_{2J^c}}$ span the representation
space of $[J^cZ'Z'']$, while $m$ is a collective index which
spans the representation space of $[M_1M_2J\,Y]$.
Finally $\mu$ accounts for the fact that the same $H$ irreducible representation
can be embedded in $G$ in different ways.
The cases of interest for us will be the following:
\begin{equation}
\label{muab}
J^c=\frac{1}{2}:\ \left\{
\begin{array}{ccc}
\mu=(a) & \to &
\begin{array}{l}
\begin{array}{|c|c|c|c|c|c|c|}
\hline
            1  & \cdots & 1 & 3 & \cdots & 3 & i\\
\hline
\end{array}\\
\begin{array}{|c|c|c|}
            2 & \cdots & 2 \\
\hline
\end{array}
\end{array}\\
 & & \\
\mu=(b) & \to &
\begin{array}{l}
\begin{array}{|c|c|c|c|c|c|c|}
\hline
             \hskip .03 cm i \hskip .04 cm & 1
             & \cdots & 1 & 3 & \cdots & 3\\
\hline
\end{array}\\
\begin{array}{|c|c|c|c|c|}
            3 & 2 & \cdots & 2 \\
\hline
\end{array}
\end{array}
\end{array}\right.
\end{equation}
\par
\begin{equation}
\label{mucde}
J^c=1:\ \left\{
\begin{array}{ccc}
\mu=(c) & \to & \frac{1}{2}\left(
\begin{array}{l}
\begin{array}{|c|c|c|c|c|c|c|c|}
\hline
            1 & \cdots & 1 & \hskip .03 cm i \hskip .03 cm
            & j & 3 & \cdots & 3\\
\hline
\end{array}\\
\begin{array}{|c|c|c|c|c|}
            2 & \cdots & 2 & 3\\
\hline
\end{array}
\end{array} + (i\leftrightarrow j)\right)\\
 && \\
\mu=(d) & \to &
\begin{array}{l}
\begin{array}{|c|c|c|c|c|c|c|c|}
\hline
             \hskip .03 cm i \hskip .04 cm & j & 1
             & \cdots & 1 & 3 & \cdots & 3\\
\hline
\end{array}\\
\begin{array}{|c|c|c|c|c|c|}
            3 & 3 & 2 & \cdots & 2 \\
\hline
\end{array}
\end{array}\\
 && \\
\mu=(e) & \to &
\begin{array}{l}
\begin{array}{|c|c|c|c|c|c|c|c|}
\hline
            1 & \cdots & 1 & 3 & \cdots & 3 &
            \hskip .03 cm i \hskip .03 cm & j\\
\hline
\end{array}\\
\begin{array}{|c|c|c|c|}
            2 & \cdots & 2 \\
\hline
\end{array}
\end{array}
\end{array}\right.
\end{equation}
where a double-row Yang tableau diagrammatically represents the
$SU(3)$ irreducible representation labeled by $M_1$ and $M_2$, the number of boxes in
the two rows of the diagram:

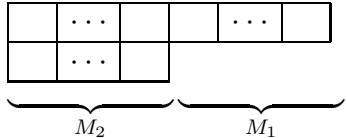
\begin{eqnarray*}
\begin{array}{l}
\begin{array}{|c|c|c|c|c|c|}
\hline
             \hskip .3 cm & \cdots & \hskip .3 cm &
             \hskip .3 cm & \cdots & \hskip .3 cm \\
\hline
\end{array}\\
\begin{array}{|c|c|c|}
             \hskip .3 cm & \cdots & \hskip .3 cm \\
\hline
\end{array}
\end{array}\\
\underbrace{\hskip 2.2 cm}_{M_2}
\underbrace{\hskip 2.2 cm}_{M_1}
\end{eqnarray*}
$J$ is the \emph{flavor isospin} characterizing the representation
of $SU(2)^w$, and $Y$ is the \emph{hypercharge} of
the generator $U(1)\not\subset SU(3)^c\times SU(2)^w$.
\subsection{The constraints on the irreducible representations}
\label{irrepconstraint}
As we have seen in eq. \eqn{expansion}, the expansion of a
generic field contains only the harmonics whose $H$- and
$G$-quantum numbers are such that the $G$ representation, 
decomposed under $H$, contain the $H$ representation of 
the field. This fact poses some constraints on the 
$G$-quantum numbers.
\par
Depending on which constraints are satisfied by a certain $G$ representation,
only part of the harmonics is  present,
and only their corresponding four--dimensional fields 
appear in the spectrum. Then, in the $G$ representations in which
such field  disappear, there is  {\sl multiplet shortening}.
In the modern perspective of Kaluza Klein theory, the exact spectrum
of the short multiplets is crucial.
Hence the importance of analyzing this disappearance of harmonics with care.
\par
Every harmonic is defined by its $SU\left(2\right)\times U\left(1\right)'
\times U\left(1\right)''$ representation, identified by the labels
$[J^c~Z'~Z'']$. Substituting these values in equations (\ref{z1}), (\ref{z2}),
(\ref{muab}), (\ref{mucde}), we can determine the constraints on
the $G$ representations.
\par
The first constraint gives the value of $Y$ in terms of $M_1$ 
and $M_2$. We have five possible expressions of $Y$, identifying five families of
$G$ representations which we will denote with the superscripts
$^0$, $^+$, $^-$, $^{++}$ and $^{--}$:
\begin{eqnarray}
\left.
\begin{array}{rccl}
^0:&\quad Y & = & 2/3(M_2-M_1)\\
^{++}:&\quad Y & = & 2/3(M_2-M_1)-2\\
^{--}:&\quad Y & = & 2/3(M_2-M_1)+2
\end{array}
\right\}
&&\begin{array}{c}
{\rm for\ \ bosonic}\\
{\rm fields}
\end{array}\nonumber\\
\left.
\begin{array}{rccl}
^+:&\quad Y & = & 2/3(M_2-M_1)-1\\
^-:&\quad Y & = & 2/3(M_2-M_1)+1
\end{array}
\right\}
&&\begin{array}{c}
{\rm for\ \ fermionic}\\
{\rm fields}
\end{array}
\label{constraintY}
\end{eqnarray}
It is worth noting that the value of $Y$ identifies a $U\left(1\right)_R$ 
representation, so these five families of representations correspond
to the five possible representations of $U\left(1\right)_R$.
\par
A second constraint is the lower bound on the quantum number $J$,
since the third component of the \emph{weak isospin}, $J_3$,
is linked to $Y$.
We have three possibilities:
\begin{equation}
\label{constraintJ}
J \geq \left\{
\begin{array}{c}
 \left|Y/2\right|\\
\left|Y/2+1\right|\\
\left|Y/2-1\right|
\end{array}\right.
\end{equation}
\par
The last kind of constraint refers to $M_1$ and $M_2$:
\begin{equation}
\label{constraintM}
\begin{array}{|c||c|c|}\hline
{\rm constraints} & J^c & \mu\\
\hline
\begin{array}{l}
M_1 \geq 0\\
M_2 \geq 0
\end{array}&0&-\\
\hline
\begin{array}{l}
M_1 \geq 1\\
M_2 \geq 0
\end{array}&\ft{1}{2}&(a)\\
\hline
\begin{array}{l}
M_1 \geq 0\\
M_2 \geq 1
\end{array}&\ft{1}{2}&(b)\\
\hline
\begin{array}{l}
M_1 \geq 1\\
M_2 \geq 1
\end{array}&1&(c)\\
\hline
\begin{array}{l}
M_1 \geq 0\\
M_2 \geq 2
\end{array}&1&(d)\\
\hline
\begin{array}{l}
M_1 \geq 2\\
M_2 \geq 0
\end{array}&1&(e)\\
\hline
\end{array}
\end{equation}
When some of the (\ref{constraintM}) constraints are not 
satisfied, the Young tableau (see (\ref{muab}), (\ref{mucde}) ) 
corresponding to $\mu$ in  (\ref{constraintM}) does not exist.
\par
We organize the series of the $G=G'\times U\left(1\right)_R$ 
representations in the following way.
The constraints  (\ref{constraintJ}) and (\ref{constraintM}),  with
the five  values of $Y$ in terms of $M_1,~M_2$ given by 
(\ref{constraintY}), define  the series of $G'$ representations that we list in
table \ref{defseries}. Every $G'$ representation, together with a 
superscript $^0,\ ^+,\ ^-,\ ^{++}$ or $^{--}$ that define the value of $Y$,
is  a $G$ representation. So the series of $G'$ representations
defined in table \ref{defseries} with such a superscript are 
series of  representations of the whole $G$ group.
\par
For each family of representations ($^0,\ ^+,\ ^-,\ ^{++}, $ and
$^{--}$) we will call a series {\sl regular} if it contains the
maximum number of harmonics.
The regular series cover all the representations with $M_1$, $M_2$
and $J$ sufficiently high to satisfy all the inequality constraints.
When some of these inequalities are not satisfied instead, some of
the harmonics may be absent in the expansion.
\par
In tables \ref{0series}, \ref{+series}, \ref{-series}, \ref{++series}, \ref{--series},
we show which harmonics are present for the different series of $G$ representations.
The first column contains the name of each series.
The other columns contain the possible harmonics, each labeled
by its $H$-quantum numbers.
An asterisk denotes the presence of a given harmonic.
 \
To obtain the constraints on the conjugate series 
it suffices to exchange $M_1$ and $M_2$, as explained in \cite{spectfer,castdauriafre}.
\begin{table}
\begin{footnotesize}
\[
\begin{array}{||c|c|c||}\hline
G'{\rm -name} & M_1,\ M_2 & J\ \ {\rm constraints}\\
\hline \hline
A_R & M_2>0,\ M_1>0 & J>\left|(M_2-M_1)/3\right|\\
\hline
A_1 & M_2>M_1>0 & J=(M_2-M_1)/3\\
\hline
A_2 & M_2>M_1>0 & J=(M_2-M_1)/3-1\\
\hline
A_3 & M_2>M_1=0 & J>(M_2-M_1)/3\\
\hline
A_4 & M_2>M_1=0 & J=(M_2-M_1)/3\\
\hline
A_5 & M_2>M_1=0 & J=(M_2-M_1)/3-1\\
\hline
A_6 & M_2=M_1>0 & J=0\\
\hline
A_7 & M_2=M_1=0 & J>0\\
\hline
A_8 & M_2=M_1=0 & J=0\\
\hline
B_R & M_2>1,\ M_1\geq 0 & J>\left|(M_2-M_1)/3-1\right|\\
\hline
B_1 & M_2>M_1+3 & J=(M_2-M_1)/3-1\\
\hline
B_2 & M_2>M_1+3 & J=(M_2-M_1)/3-2\\
\hline
B_3 & M_2=M_1+3 & J=(M_2-M_1)/3-1\\
\hline
B_4 & M_1\geq M_2>1 & J=-(M_2-M_1)/3+1\\
\hline
B_5 &M_1\geq M_2>1 & J=-(M_2-M_1)/3\\
\hline
B_6 & M_1\geq M_2=1 & J>-(M_2-M_1)/3+1\\
\hline
B_7 & M_1\geq M_2=1 & J=-(M_2-M_1)/3+1\\
\hline
B_8 & M_1\geq M_2=1 & J=-(M_2-M_1)/3\\
\hline
B_9 & M_1\geq M_2=0 & J>-(M_2-M_1)/3+1\\
\hline
B_{10} & M_1\geq M_2=0  & J=-(M_2-M_1)/3+1\\
\hline
B_{11} & M_1\geq M_2=0 & J=-(M_2-M_1)/3\\
\hline
\end{array}
\]
\end{footnotesize}
\caption{Series of $G'$ representations}
\label{defseries}
\end{table}
\begin{table}
\centering
\begin{footnotesize}
\[
\begin{array}{||c||c|c|c|c|c|c|c|c|c|c|c|c||} \hline
J^c&0&0&0&1/2&1/2&1/2&1/2&1/2&1/2&1&1&1\\
Z'&0&-2i&2i&3i&-3i&i&-i&5i&-5i&0&-2i&2i\\
Z''&0&-3i&3i&-3/2i&3/2i&-9/2i&9/2i&3/2i&-3/2i&0&-3i&3i\\
\mu& & & &\left({\rm a}\right)&\left({\rm b}\right)&\left({\rm a}\right)
&\left({\rm b}\right)&\left({\rm a}\right)&\left({\rm b}\right)
&\left({\rm c}\right)&\left({\rm c}\right)&\left({\rm c}\right)
\\\hline\hline
A^{\ 0}_R&*&*&*&*&*&*&*&*&*&*&*&*\\
\hline
A^{\ 0}_1&*&*& &*&*&*& & &*&*&*& \\
\hline
A^{*\ 0}_1&*& &*&*&*& &*&*& &*& &*\\
\hline
A^{\ 0}_2& &*& & & &*& & &*& &*& \\
\hline
A^{*\ 0}_2& & &*& & & &*&*& & & &*\\
\hline
A^{\ 0}_3&*&*&*& &*& &*& &*& & & \\
\hline
A^{*\ 0}_3&*&*&*&*& &*& &*& & & & \\
\hline
A^{\ 0}_4&*&*& & &*& & & &*& & & \\
\hline
A^{*\ 0}_4&*& &*&*& & & &*& & & & \\
\hline
A^{\ 0}_5& &*& & & & & & &*& & & \\
\hline
A^{*\ 0}_5& & &*& & & & &*& & & & \\
\hline
A^{\ 0}_6&*& & &*&*& & & & &*& & \\
\hline
A^{\ 0}_7&*&*&*& & & & & & & & & \\
\hline
A^{\ 0}_8&*& & & & & & & & & & & \\
\hline
\end{array}
\]
\end{footnotesize}
\caption{Harmonics content for the series of type $^0$}
\label{0series}
\end{table}
\begin{table}
\centering
\begin{footnotesize}
\[
\begin{array}{||c||c|c|c|c||}
\hline
J^c & 0 & 0 & 1/2 & 1/2\\
Z'  & 2i & 4i & -i & i\\
Z'' & -3i &0 & -3i/2 & 3i/2\\
\mu & & &\left({\rm b}\right)&\left({\rm b}\right)\\
\hline
\hline
A_R^{\ +} & * & * & * & *\\
\hline
A_1^{\ +} & * & * & * & * \\
\hline
A_1^{*+} & & * & & *\\
\hline
A_2^{\ +} & * & & * & \\
\hline
A_2^{*+} & & & & \\
\hline
A_3^{\ +} & * & * & * & *\\
\hline
A_3^{*+} & * & * & & \\
\hline
A_4^{\ +} & * & * & * & *\\
\hline
A_4^{*+} & & * & & \\
\hline
A_5^{\ +} & * & & * & \\
\hline
A_5^{*+} & & & & \\
\hline
A_6^{\ +} & & * & & * \\
\hline
A_7^{+} & * & * & & \\
\hline
A_8^{+} & & * & & \\
\hline
\end{array}
\]
\end{footnotesize}
\caption{Harmonics content for the series of type $^+$}
\label{+series}
\end{table}
\begin{table}
\centering
\begin{footnotesize}
\[
\begin{array}{||c||c|c|c|c||}
\hline
J^c & 0 & 0 & 1/2 & 1/2\\
Z' & -2i & -4i & i & -i\\
Z'' & 3i & 0 & 3i/2 & -3i/2\\
\mu & & &\left({\rm a}\right)&\left({\rm a}\right)\\
\hline
\hline
A_R^{\ -} & * & * & * & *\\
\hline
A_1^{\ -} & & * & & *\\
\hline
A_1^{*-} & * & * & * & *\\
\hline
A_2^{\ -} & & & &\\
\hline
A_2^{*-} & * & & * &\\
\hline
A_3^{\ -} & * & * & &\\
\hline
A_3^{*-} & * & * & * & *\\
\hline
A_4^{-} & & * & &\\
\hline
A_4^{*-} & * & * & * & *\\
\hline
A_5^{-} & & & &\\
\hline
A_5^{*-} & * & & * &\\
\hline
A_6^{-} & & * & & *\\
\hline
A_7^{-} & * & * & &\\
\hline
A_8^{-} & & * & &\\
\hline
\end{array}
\]
\end{footnotesize}
\caption{Harmonics content for the series of type $^-$}
\label{-series}
\end{table}
\begin{table}
\centering
\begin{footnotesize}
\[
\begin{array}{||c||c|c|c|c|c|c|c|c|c||} \hline
J^c&0&0&0&1/2&1/2&1/2&1&1&1\\
Z'&6i&8i&4i&3i&i&5i&0&-2i&2i\\
Z''&-3i&0&-6i&-3/2i&-9/2i&3/2i&0&-3i&3i\\
\mu & & & &\left({\rm b}\right)&\left({\rm b}\right)
&\left({\rm b}\right)&\left({\rm d}\right)
&\left({\rm d}\right)&\left({\rm d}\right)\\
\hline\hline
B^{\ ++}_R&*&*&*&*&*&*&*&*&*\\
\hline
B^{\ ++}_1&*& &*&*&*& &*&*& \\
\hline
B^{\ ++}_2& & &*& &*& & &*& \\
\hline
B^{\ ++}_3&*& & &*& & &*& & \\
\hline
B^{\ ++}_4&*&*& &*& &*&*& &*\\
\hline
B^{\ ++}_5& &*& & & &*& & &*\\
\hline
B^{\ ++}_6&*&*&*&*&*&*& & & \\
\hline
B^{\ ++}_7&*&*& &*& &*& & & \\
\hline
B^{\ ++}_8& &*& & & &*& & & \\
\hline
B^{\ ++}_9&*&*&*& & & & & & \\
\hline
B^{\ ++}_{10}&*&*& & & & & & & \\
\hline
B^{\ ++}_{11}& &*& & & & & & & \\
\hline
\end{array}
\]
\end{footnotesize}
\caption{Harmonics content for the
series of type $^{++}$}
\label{++series}
\end{table}
\begin{table}
\centering
\begin{footnotesize}
\[
\begin{array}{||c||c|c|c|c|c|c|c|c|c||} \hline
J^c&0&0&0&1/2&1/2&1/2&1&1&1\\
Z'&-6i&-8i&-4i&-3i&-i&-5i&0&2i&-2i\\
Z''&3i&0&6i&3/2i&9/2i&-3/2i&0&3i&-3i\\
\mu & & & &\left({\rm a}\right)&\left({\rm a}\right)
&\left({\rm a}\right)&\left({\rm e}\right)
&\left({\rm e}\right)&\left({\rm e}\right)\\
\hline\hline
B^{*\ --}_R&*&*&*&*&*&*&*&*&*\\
\hline
B^{*\ --}_1&*& &*&*&*& &*&*& \\
\hline
B^{*\ --}_2& & &*& &*& & &*& \\
\hline
B^{*\ --}_3&*& & &*& & &*& & \\
\hline
B^{*\ --}_4&*&*& &*& &*&*& &*\\
\hline
B^{*\ --}_5& &*& & & &*& & &*\\
\hline
B^{*\ --}_6&*&*&*&*&*&*& & & \\
\hline
B^{*\ --}_7&*&*& &*& &*& & & \\
\hline
B^{*\ --}_8& &*& & & &*& & & \\
\hline
B^{*\ --}_9&*&*&*& & & & & & \\
\hline
B^{*\ --}_{10}&*&*& & & & & & & \\
\hline
B^{*\ --}_{11}& &*& & & & & & & \\
\hline
\end{array}
\]
\end{footnotesize}
\caption{Harmonics content for the series
of type $^{--}$}
\label{--series}
\end{table}
\newpage
\section{Differential calculus via harmonic analysis}
The Kaluza Klein kinetic operators $\xbox_y^{[\l_1\l_2\l_3]}$
are covariant differential operators.
Eq. \eqn{dercomm} then implies that they are $SO(7)$-invariant:
\begin{equation}
\cL_{\Lambda}\xbox_y^{[\l_1\l_2\l_3]}=\xbox_y^{[\l_1\l_2\l_3]}
\cL_{\Lambda}.
\end{equation}
Using this fact and Schur's lemma we conclude that they act
irreducibly on the harmonics.
In other words they act as (finite dimensional) matrices on
the harmonic subspaces of fixed $G$-quantum numbers:
\begin{equation}
\xbox_y^{[\l_1\l_2\l_3]} \cH^{[M_1M_2J\,Y]}_{\zeta}(y)=
\cM\left([M_1M_2J\,Y]\right)_{\zeta}^{\ \zeta'}
\cH^{[M_1M_2J\,Y]}_{\zeta'}(y)
\end{equation}
where, for short, we have summarized with $\zeta$ the whole set
of indices labelling the $H$ representations of $\cH$.
\par
Let us now consider the explicit action of the covariant
derivative \eqn{covder} on the harmonics.
Following to the standard procedures of harmonic analysis
\cite{spectfer, univer},
$H$ is a subgroup of $SO(7)$.
The $SO(7)$--covariant derivative \eqn{covder} can then be decomposed as:
\begin{equation}\label{MM}
D = d + \Omega^Ht_H + \cB^{\g}\IM_{\g} \equiv D^H + \cB^{\g}\IM_{\g}\,,
\end{equation}
where $t_H$ are the generators of $H$ and 
$\IM_{\g}$ the part of the $SO(7)$-connection not belonging to $H$.
The $\IM$ generators in the fundamental $SO(7)$ representation
acquire the following form:
\begin{eqnarray*}
\IM_1=\left(\begin{array}{cc|c|cccc}
0&0&0&0&0&0&0\\
0&0&-2&0&0&0&0\\\hline
0&2&0&0&0&0&0\\\hline
0&0&0&0&0&0&0\\
0&0&0&0&0&0&0\\
0&0&0&0&0&0&0\\
0&0&0&0&0&0&0
\end{array}\right),
\qquad
\IM_2=\left(\begin{array}{cc|c|cccc}
0&0&2&0&0&0&0\\
0&0&0&0&0&0&0\\\hline
-2&0&0&0&0&0&0\\\hline
0&0&0&0&0&0&0\\
0&0&0&0&0&0&0\\
0&0&0&0&0&0&0\\
0&0&0&0&0&0&0
\end{array}\right),\\
\begin{array}{c}
\underbrace{\hspace{1.2 cm}}_{m}
\underbrace{\hspace{0.7 cm}}_{3}
\underbrace{\hspace{2.2 cm}}_{A}\\
\\
\\
\\
\\
\\
\\
\end{array}\qquad
\IM_3=\frac{1}{3}\left(\begin{array}{cc|c|cccc}
0&2&0&0&0&0&0\\
-2&0&0&0&0&0&0\\\hline
0&0&0&0&0&0&0\\\hline
0&0&0&0&0&0&0\\
0&0&0&0&0&0&0\\
0&0&0&0&0&0&0\\
0&0&0&0&0&0&0
\end{array}\right),\\
\IM_4=\left(\begin{array}{cc|c|cccc}
0&0&0&0&0&0&0\\
0&0&0&0&0&0&0\\\hline
0&0&0&0&-2&0&0\\\hline
0&0&0&0&0&0&0\\
0&0&2&0&0&0&0\\
0&0&0&0&0&0&0\\
0&0&0&0&0&0&0
\end{array}\right),\qquad
\IM_5=\left(\begin{array}{cc|c|cccc}
0&0&0&0&0&0&0\\
0&0&0&0&0&0&0\\\hline
0&0&0&2&0&0&0\\\hline
0&0&-2&0&0&0&0\\
0&0&0&0&0&0&0\\
0&0&0&0&0&0&0\\
0&0&0&0&0&0&0
\end{array}\right),\\
\IM_6=\left(\begin{array}{cc|c|cccc}
0&0&0&0&0&0&0\\
0&0&0&0&0&0&0\\\hline
0&0&0&0&0&0&-2\\\hline
0&0&0&0&0&0&0\\
0&0&0&0&0&0&0\\
0&0&0&0&0&0&0\\
0&0&2&0&0&0&0
\end{array}\right),\qquad
\IM_7=\left(\begin{array}{cc|c|cccc}
0&0&0&0&0&0&0\\
0&0&0&0&0&0&0\\\hline
0&0&0&0&0&2&0\\\hline
0&0&0&0&0&0&0\\
0&0&0&0&0&0&0\\
0&0&-2&0&0&0&0\\
0&0&0&0&0&0&0
\end{array}\right).
\end{eqnarray*}
The harmonics
transform as the inverse $L^{-1}(y)$ of the fundamental elements
of $G$  (see \cite{spectfer}).
Eq. \eqn{Omega} can be restated as
\begin{equation}
\Omega L^{-1} = L^{-1} {\rm d}L\,L^{-1} = -{\rm d}L^{-1}
\end{equation}
from which we deduce that
\begin{equation}
D^H\cH = (d+\Omega^Ht_H)\cH = -\Omega^{\a}t_{\a}\cH.
\end{equation}
By means of eq. \eqn{B} we can calculate the explicit components
of $D^H$, i.e. its projection along the vielbeins:
\begin{eqnarray}
D^H = \cB^{\a}D^H_{\a} = -\Omega^{\a}t_{\a},\nonumber\\
\left\{\begin{array}{ccl}
D_A & = & -\ft{4}{\sqrt{3}}i\l_A,\\
D_m & = & -\ft{4}{\sqrt{2}}i\s_m,\\
D_3 & = & -\ft{4}{3}Z,
\end{array}\right.
\end{eqnarray}
where the coset generators $t_{\a}$ act on $\cH$ as follows.
$\l_A$ acts on the $SU(3)$ part of the $G$ representation of the harmonic.
The fundamental representation of $\lambda_A$ is given by the
Gell--Mann matrices (see Appendix \ref{convenzioni}). 
On a generic Young tableau $\lambda_A$ acts as the tensor
representation. To give an example, consider a component 
of an harmonic with $SU(3)$ indices given by
$$
\begin{array}{l}
\begin{array}{|c|c|c|c|}
\hline
            1 & 1 & 3 & 3 \\
\hline
\end{array}\\
\begin{array}{|c|c|}
            2 & 3 \\
\hline
\end{array}
\end{array}
\,.
$$
Then $\l_4$, which exchanges $1$ with $3$, acts as follows:
\begin{eqnarray*}
\l_4\,
\begin{array}{l}
\begin{array}{|c|c|c|c|}
\hline
            1 & 1 & 3 & 3 \\
\hline
\end{array}\\
\begin{array}{|c|c|}
            2 & 3 \\
\hline
\end{array}
\end{array}=\hspace{10 cm}\\
=\begin{array}{l}
\begin{array}{|c|c|c|c|}
\hline
            3 & 1 & 3 & 3 \\
\hline
\end{array}\\
\begin{array}{|c|c|}
            2 & 3 \\
\hline
\end{array}
\end{array}+
\begin{array}{l}
\begin{array}{|c|c|c|c|}
\hline
            1 & 3 & 3 & 3 \\
\hline
\end{array}\\
\begin{array}{|c|c|}
            2 & 3 \\
\hline
\end{array}
\end{array}+
\begin{array}{l}
\begin{array}{|c|c|c|c|}
\hline
            1 & 1 & 3 & 3 \\
\hline
\end{array}\\
\begin{array}{|c|c|}
            2 & 1 \\
\hline
\end{array}
\end{array}+2\,
\begin{array}{l}
\begin{array}{|c|c|c|c|}
\hline
            1 & 1 & 1 & 3 \\
\hline
\end{array}\\
\begin{array}{|c|c|}
            2 & 3 \\
\hline
\end{array}
\end{array}=\\
=\begin{array}{l}
\begin{array}{|c|c|c|c|}
\hline
            3 & 1 & 3 & 3 \\
\hline
\end{array}\\
\begin{array}{|c|c|}
            2 & 3 \\
\hline
\end{array}
\end{array}+2\,
\begin{array}{l}
\begin{array}{|c|c|c|c|}
\hline
            1 & 1 & 1 & 3 \\
\hline
\end{array}\\
\begin{array}{|c|c|}
            2 & 3 \\
\hline
\end{array}
\end{array}.
\end{eqnarray*}
Similarly, $\s^m\ (m=\{1,2\})$ acts as the $m$-th Pauli matrix
on the fundamental representation of $SU(2)$, and as its $n$-th
tensor power on the $n$-boxes $SU(2)$ Young tableau:
$$
\s^1\,
\begin{array}{|c|c|c|c|c|}
\hline
            1 & 1 & 1 & 2 & 2\\
\hline
\end{array}= 3\,
\begin{array}{|c|c|c|c|c|}
\hline
            1 & 1 & 2 & 2 & 2 \\
\hline
\end{array}+ 2\,
\begin{array}{|c|c|c|c|c|}
\hline
            1 & 1 & 1 & 1 & 2\\
\hline
\end{array}\,.
$$
Finally, $Z$ acts trivially, multiplying $\cH$ by its
$Z$-charge:
\begin{eqnarray*}
Z\,\begin{array}{l}
\begin{array}{|c|c|c|c|}
\hline
            1 & 1 & 3 & 3 \\
\hline
\end{array}\\
\begin{array}{|c|c|}
            2 & 3 \\
\hline
\end{array}
\end{array}\otimes\,
\begin{array}{|c|c|c|c|c|}
\hline
            1 & 1 & 1 & 2 & 2 \\
\hline
\end{array}\ =\hspace{8cm}\\
=\left(-\frac{3}{2}i-\frac{1}{2}i+iY\right)\,\begin{array}{l}
\begin{array}{|c|c|c|c|}
\hline
            1 & 1 & 3 & 3 \\
\hline
\end{array}\\
\begin{array}{|c|c|}
            2 & 3 \\
\hline
\end{array}
\end{array}\otimes\,
\begin{array}{|c|c|c|c|c|}
\hline
            1 & 1 & 1 & 2 & 2 \\
\hline
\end{array}\,.
\end{eqnarray*}
In the course of the calculations, one often encounters
the $H$-covariant Laplace-Beltrami operator on $G/H$:
\begin{equation}\label{HLaplace}
\eta^{\a\b}D^H_{\,\a}D^H_{\,\b}=\frac{16}{3}\l_A\l_A+
\frac{16}{2}\s^m\s^m-\frac{16}{9}Z^2
\end{equation}
The eigenvalues of the first operator, $\l_A\l_A$, are listed
in the following table:
$$
\begin{array}{|c|c||c|}\hline
J^c & \mu & \l^A\l^A\quad{\rm eigenvalues}\\
\hline\hline
0 & - & 4(M_1+M_2+M_1M_2)\\
1/2 & (a) &2(4M_1+2M_1M_2-3)\\
1/2 & (b) &2(4M_2+2M_1M_2-3)\\
1 & (c) &4(M_1+M_2+M_1M_2-2)\\
1 & (d) &4(3M_2-M_1+M_1M_2-5)\\
1 & (e) &4(3M_1-M_2+M_1M_2-5)\\\hline
\end{array}
$$
while the eigenvalues of $\s^m\s^m$ depend on the
\emph{flavor isospin} quantum numbers $J$ and $J^3$:
\begin{eqnarray}
\s^m\s^m &
\begin{array}{|c|c|c|c|c|c|}
\hline
             1 & \cdots & 1 & 2 & \cdots & 2 \\
\hline
\end{array} & = 4\left[J(J+1)-J_3^2\right]
\begin{array}{|c|c|c|c|c|c|}
\hline
             1 & \cdots & 1 & 2 & \cdots & 2 \\
\hline
\end{array}\,, \\
& \underbrace{\hskip 2 cm}_{m_1}
\underbrace{\hskip 2 cm}_{m_2} & \nonumber
\end{eqnarray}
where $2J=m_1+m_2$ and $2J^3=m_1-m_2$.
The complete Kaluza Klein mass operator heavily depends
on the kind of field it acts on and will be analyzed in
details in the next sections.
\subsection{The zero-form}
The only representation into which the $[0,0,0]$ (i.e. the scalar)
of $SO(7)$ breaks under $H$, is obviously the $H$-scalar representation.
The question now is: which $G$-irreducible representations
 contain the $H$-scalar?
From equations (\ref{z1}),(\ref{z2}) we see that $Z'=Z''=0$ implies
\begin{equation}
2J_3=Y={2\over 3}\left(M_2-M_1\right).
\end{equation}
This means that
\begin{itemize}
\item $M_2-M_1\in 3\ZZ$
\item $J\in\IN$
\item $J\ge\left|{1\over 3}\left(M_2-M_1\right)\right|$
\item $Y={2\over 3}\left(M_2-M_1\right)$.
\end{itemize}
We will denote the scalar as
\begin{equation}
{\cal Y} = [0|{\rm I}] \equiv {\sum}'_{\left[M_1M_2J\,Y\right]} \cH_{[000]}^{[M_1M_2 J Y]}(y)
\cdot S_{[M_1M_2 Y J]}(x).
\end{equation}
The Kaluza Klein mass operator for the zero-form ${\cal Y}$ is given by
\begin{equation}
\xbox^{[000]}{\cal Y} \equiv D_{\b}D^{\b}{\cal Y} = D^H_{\b}D^{H\b}{\cal Y}.
\end{equation}
For the scalar, there are no $\IM$--connection terms.
So, by means of eq. \eqn{HLaplace}, the computation of its
eigenvalues, on the $G$--representations as listed above, is immediate:
\begin{eqnarray}
\xbox^{[000]}{\cal Y} \equiv M_{(0)^3}{\cal Y} &=& 
\left[\ft{64}{3}(M_1+M_2+M_1M_2)+ 32 J(J+1) +
\ft{32}{9}(M_2-M_1)^2\right]{\cal Y}=\nonumber\\
&=&H_0{\cal Y}
\end{eqnarray}
where $H_0$ is the same quantity defined in eq. (\ref{defH0}).
\par
As we see from the Kaluza Klein expansion  (\ref{kkexpansion}),
the eigenvalues of the zero--form harmonic allow us to determine (see \cite{univer}) 
the masses of the $AdS_4$ graviton field $h$ and the scalar fields $S,\Sigma$.
\subsection{The one-form}
The decomposition under $H$ of the vector representation of 
$SO(7)$ is the following
\footnote{When we write a pair of complex conjugate 
representations we assume a conjugation relation between 
them. For example, by writing $[0,-2i,-3i]\oplus[0,2i,3i]$
we intend a complex representation of complex dimension one or
real dimension two.}:
\begin{equation}
[1,0,0] \to [0,0,0] \oplus [0,-2i,-3i] \oplus [0,2i,3i] \oplus
[\ft{1}{2},3i,-\ft{3}{2}i] \oplus [\ft{1}{2},-3i,\ft{3}{2}i].
\end{equation}
Concretely, the decomposition of the one--form
in $H$--irreducible fragments is done as follows (see also \cite{spec321}):
\begin{eqnarray}
{\cal Y}^A & = & \l^A_{3i}\langle 1 | {\rm I } \rangle_i+
\l^A_{i3}\langle 1 | {\rm I } \rangle_i^*\\
{\cal Y}^m & = & \s^m_{21}\langle 1 | {\rm I } \rangle_{.}+
\s^m_{12}\langle 1 | {\rm I } \rangle_{.}^*\\
{\cal Y}^3 & = & [ 1 , {\rm I} ]
\end{eqnarray}
These $H$--irreducible fragments can be expanded as in (\ref{expansion})
\footnote{Using the same conventions as in 
\cite{spectfer,univer,castdauriafre,spec321}, the reader might notice that
there appears a sign $(-1)^{J - J_3}$ upon taking the
complex conjugate of the fragments $\langle \dots | \dots \rangle_x$.
In order to reduce the notation we have absorbed this sign
in the $x$--space fields $\tilde W\langle \dots , \dots \rangle$.
This will be done for all the
complex conjugates henceforth.  } (summation over
the $G$-quantum numbers is intended):
\begin{eqnarray}
&& {\rm For\ type}\quad ^0:\nonumber\\
&&\nonumber\\
\langle 1 | {\rm I } \rangle_i &=& {\cal H}^{[1/2,3i,-{3i}/{2}](a)}_i \cdot
                    W\langle \ft12 , {\rm I} \rangle \,,\nonumber\\
\langle 1 | {\rm I } \rangle_i^* &=& \varepsilon^{ij}
{\cal H}^{[1/2,-3i,{3i}/{2}](b)}_j \cdot
                    \tilde{W}\langle \ft12 , {\rm I} \rangle \,,\nonumber\\
\langle 1 | {\rm I } \rangle_\cdot &=&
{\cal H}^{[0,-2i,-3i]} \cdot
                 W\langle 0 , {\rm I} \rangle \,,\nonumber\\
\langle 1 | {\rm I } \rangle_\cdot^* &=&
{\cal H}^{[0,2i,3i]} \cdot
                    \tilde{W}\langle 0 , {\rm I} \rangle\,,\nonumber\\
{}[1 | {\rm I } ]_\cdot &=&
{\cal H}^{[0,0,0]} \cdot W[ 0 , {\rm I} ] \,,\label{1f0serexpansion}
\end{eqnarray}
\begin{eqnarray*}
 &&{\rm For\ type}\quad ^{++} :\nonumber\\
 &&\nonumber\\
 \langle 1 | {\rm I } \rangle_i &=& {\cal H}^{[1/2,3i,-3/2i](b)}_i \cdot
W\langle \ft12 , {\rm II} \rangle\,,\\
\nonumber \\
&&{\rm For\ type}\quad ^{--}:\nonumber\\
 &&\nonumber\\
 \langle 1 | {\rm I } \rangle_i^* &=& -\varepsilon^{ij}{\cal H}^{[1/2,-3i,3/2i](a)}_j \cdot
\tilde{W}\langle \ft12 , {\rm II} \rangle\,.
\end{eqnarray*}
As we see, there are five different $AdS_4$ fields ($W,\tilde{W}$)
in the case of the $^0$ series, and one field in the case of the
$^{++}$ and $^{--}$ series.
So, for the regular $^0$ series the Laplace Beltrami operator acts 
on the $AdS_4$ fields as a $5\times 5$ matrix.
For the {\sl exceptional} series it acts as a matrix of lower dimension.
\par
The Laplace Beltrami operator for the transverse one-form field ${\cal Y}^{\a}$,
is given by
\begin{equation}
\xbox^{[100]}{\cal Y}^{\a} \equiv M_{(1)(0)^2}{\cal Y}^{\a} = 2D_{\b}D^{[\b}{\cal Y}^{\a]} =
(D^{\b}D_{\b}+24){\cal Y}^{\a}\ ,
\end{equation}
where transversality of ${\cal Y}^{\a}$ means that $D_{\a}{\cal Y}^{\a}=0$.
From the decomposition $D_{\a}=D^{H}_{\a}+\IM_{\a}$ we obtain:
\begin{equation}
\xbox^{[100]}{\cal Y}^{\a} =
(D^{H\b}D^{H}_{\b}+24){\cal Y}^{\a}+\eta^{\g\d}\left(2(\IM_{\g})^{\a}_{\ \b}
D^H_{\d}+(\IM_{\g})^{\a}_{\ \e}(\IM_{\d})^{\e}_{\ \b}\right) {\cal Y}^{\b}\,.
\end{equation}
The matrix of this operator on the $AdS_4$ fields is given by
\begin{footnotesize}
\begin{eqnarray}\label{onematrix}
\begin{array}{|c||c|c|c|c|c|}
\hline
 M_{(1)(0)^2}
  & W\langle \ft12,{\rm I}  \rangle
  & \tilde {W}\langle\ft12,{\rm I}  \rangle
  & W\langle 0,{\rm I} \rangle
  & \tilde {W}\langle 0,{\rm I}  \rangle
  & W[0,{\rm I}]\\
\hline
\hline
 W\langle \ft12,{\rm I}\rangle & H_0\!-\!\frac{32(M_2-M_1)}{3} &
 0 & 0 & 0 & \frac{16M_1}{\sqrt{3}}\\
 \tilde{W}\langle\ft12 ,{\rm I}\rangle & 0 & H_0\!+\!\frac{32(M_2-M_1)}{3} &
 0 & 0 & \frac{16M_2}{\sqrt{3}}\\
 W\langle 0,{\rm I}\rangle & 0 & 0 & H_0\!+\!\frac{32(M_2-M_1)}{3} &
 0 & -\frac{8(2J+Y)}{\sqrt{2}}\\
 \tilde{W}\langle 0,{\rm I}\rangle & 0 & 0 & 0 & H_0\!-\!\frac{32(M_2-M_1)}{3} &
 \frac{8(2J-Y)}{\sqrt{2}}\\
 W[0,{\rm I}] & \frac{32(2+M_2)}{\sqrt{3}} & \frac{32(2+M_1)}{\sqrt{3}} &
 -\frac{16(2+2J-Y)}{\sqrt{2}} & \frac{16(2+2J+Y)}{\sqrt{2}} & H_0\!+\!48\\
\hline
\end{array}.
\end{eqnarray}
\end{footnotesize}
Its eigenvalues are:
\begin{eqnarray}\label{eigenAoneform}
\lambda_1 &=& H_0 + \ft{32}{3}(M_2-M_1)\,,
\nonumber \\
\lambda_2 &=& H_0 - \ft{32}{3}(M_2-M_1)\,,
\nonumber \\
\lambda_3 &=& H_0\,,
\\
\lambda_4 &=& H_0 + 24 + 4 \sqrt{H_0 + 36}\,,
\nonumber \\
\lambda_5 &=& H_0 + 24 - 4 \sqrt{H_0 + 36}\,.
\nonumber
\end{eqnarray}
Actually, what we have just calculated are the eigenvalues of 
\begin{equation}\label{actualM100}
M_{(1)(0)^2}{\cal Y}^\alpha + D^{\a}D_{\b}{\cal Y}^\b\,.
\end{equation}
It coincides with $M_{(1)(0)^2}$ when acting on a transverse one-form.
But on a generic ${\cal Y}^{\a}$, which possibly contains a longitudinal
term, the second part of \eqn{actualM100}, $D^{\a}D_{\b}{\cal Y}^{\b}$,
is not inert.
Indeed, let us suppose
$$
{\cal Y}^{\a}=D^{\a}{\cal Y}
$$
for some scalar function ${\cal Y}$.
Then
\begin{equation}
D^{\a}D_{\b}{\cal Y}^{\b}=D^{\a}D_{\b}D^{\b}{\cal Y}=D^{\a}M_{(0)^3}
{\cal Y}=M_{(0)^3}{\cal Y}^{\a}.
\end{equation}
So, our actual operator \eqn{onematrix} contains the eigenvalues
of $M_{(0)^3}$, which are \emph{longitudinal} (hence \emph{non-physical})
for the one--form.
This fact is true also for the two-form.
\par
The eigenvalue $\lambda_3$ in \eqn{eigenAoneform} is the longitudinal one,
equal to the zero--form eigenvalue $H_0$.
The other four, instead, are transverse physical eigenvalues.
\par
The matrices corresponding to the exceptional series are easily
obtained from \eqn{onematrix} by removing the rows and the
columns of the fields that disappear in the expansions (\ref{1f0serexpansion}), 
as we read from table \ref{0series}.
We list the mass eigenvalues of each series:
\begin{footnotesize}
\begin{eqnarray}
\begin{array}{|c||c|}
\hline
\, A_R \, & \, \lambda_1, \lambda_2, \lambda_3, \lambda_4, \lambda_5 \, \cr
\hline
A_1 & \lambda_1,  \lambda_3, \lambda_4, \lambda_5 \cr
\hline
A_1^* & \lambda_2,  \lambda_3, \lambda_4, \lambda_5 \cr
\hline
A_2 & \lambda_1  \cr
\hline
A_2^* & \lambda_2  \cr
\hline
A_3 & \lambda_1,  \lambda_3, \lambda_4, \lambda_5 \cr
\hline
A_3^* & \lambda_2,  \lambda_3, \lambda_4, \lambda_5 \cr
\hline
A_4 & \lambda_1, \lambda_3, \lambda_4  \cr
\hline
A_4^* & \lambda_2, \lambda_3, \lambda_4  \cr
\hline
A_5 & \lambda_1 \cr
\hline
A_5^* & \lambda_2 \cr
\hline
A_6 & \lambda_3, \lambda_4, \lambda_5 \cr
\hline
A_7 & \lambda_3, \lambda_4, \lambda_5 \cr
\hline
A_8 & \lambda_4 \cr
\hline
\end{array}
\label{eigenvaluesAoneform}
\end{eqnarray}
\end{footnotesize}
For the series of type $^{++}$ the operator $M_{(1)(0)^2}$ 
acts as a $1\times 1$ matrix on the $AdS_4$ fields and has eigenvalue:
\begin{equation}
H_0+\ft{32}{3}(M_2-M_1)
\label{eigenBoneform}
\end{equation}
for the series $B_R, B_1, B_3, B_4, B_6$ and $B_7$.
For the type $^{--}$-series the eigenvalue is the conjugate one
($M_2 \leftrightarrow M_1$) in the conjugate series.
\par
We can use the eigenvalues of the one--form harmonic to determine (see \cite{univer}) 
the masses of the $AdS_4$vector field $A,W$.
\subsection{The two-form}
Under the action of $H=SU(2)\times U(1)'\times U(1)''$
the 21 components of the $SO(7)$ two-form transform into the
completely reducible representation:
\begin{eqnarray}
[1,1,0] & \to & [1,0,0] \oplus [0,0,0] \oplus [0,0,0]
\oplus [0,6i,-3i] \oplus [0,-6i,3i] \oplus\nonumber\\
 && \oplus [1/2,i,-9/2i] \oplus [1/2,-i,9/2i] \oplus [1/2,5i,3/2i]
\oplus [1/2,-5i,-3/2i] \nonumber\\
 && \oplus [1/2,3i,-3/2i] \oplus [1/2,-3i,3/2i] \oplus [0,-2i,-3i]
\oplus [0,2i,3i]\,.
\end{eqnarray}
The  decomposition of the two--form in $H$--irreducible fragments is as follows:
\begin{eqnarray*}
{\cal Y}^{AB} & = & -i\l^{[A}_{\ i3}\l^{B]}_{\ 3i}\ [2|{\rm I}]_{.}-
i\l^{[A}_{\ i3}\,\l^{B]}_{\ 3j}\,\varepsilon^{ik}[2|{\rm I}]_{jk}+\\
 && \l^{[A}_{\ i3}\l^{B]}_{\ 3j}\,\varepsilon^{ik}\big<2|{\rm I}\big>_{jk}+
\l^{[A}_{\ 3i}\l^{B]}_{\ j3}\,\varepsilon^{ik}\big<2|{\rm I}\big>^*_{jk}+\\
 && \l^{[A}_{\ 3i}\l^{B]}_{\ 3j}\,\varepsilon^{ij}\big<2|{\rm I}\big>_{.}+
\l^{[A}_{\ i3}\l^{B]}_{\ j3}\,\varepsilon^{ij}\big<2|{\rm I}\big>^*_{.}\\
{\cal Y}^{Am} & = & \l^A_{\ 3i}\,\s^m_{\ 21}\big<2|{\rm II}\big>_i
+\l^A_{\ i3}\,\s^m_{\ 12}\big<2|{\rm II}\big>^*_i+\\
 && \l^A_{\ 3i}\s^m_{\ 12}\big<2|{\rm III}\big>_i+
\l^A_{\ i3}\s^m_{\ 21}\big<2|{\rm III}\big>^*_i\\
{\cal Y}^{mn} & = & \varepsilon^{mn}[2|{\rm II}]_{.}\\
{\cal Y}^{m3} & = & \s^m_{\ 21}\big<2|{\rm II}\big>_{.}+\s^m_{\ 12}\big<2|{\rm II}\big>^*_{.}\\
{\cal Y}^{A3} & = & \l^A_{\ 3i}\big<2|{\rm I}\big>_i+\l^A_{\ i3}\big<2|{\rm I}\big>^*_i\,,
\end{eqnarray*}
where:
\begin{eqnarray*}
[2|{\rm I}]_{.} & = & \cH^{[0,0,0]}\ Z [0,{\rm I}|\r]\\
\,[2|{\rm II}]_{.} & = & \cH^{[0,0,0]}\ Z [0,{\rm II}|\r]\\
\big<2|{\rm I}\big>_{.} & = & \cH^{[0,6i,-3i]}\ Z\big<0,{\rm I}|\r\big>\\
\big<2|{\rm I}\big>^*_{.} & = & \cH^{[0,-6i,3i]}\ \tilde{Z}
\big<0,{\rm I}|\r\big>\\
\big<2|{\rm II}\big>_{.} & = & \cH^{[0,-2i,-3i]}\ Z
\big<0,{\rm II}|\r\big>\\
\big<2|{\rm II}\big>^*_{.} & = & \cH^{[0,2i,3i]}\ \tilde{Z}
\big<0,{\rm II}|\r\big>\\
\big<2|{\rm I}\big>_i & = & \cH^{[1/2,3i,-3/2i](a)}_i\ Z
\big<1/2,{\rm I}|\r\big>\\
\big<2|{\rm I}\big>^*_i & = & \varepsilon^{ij}\cH^{[1/2,-3i,3/2i](b)}_j\ \tilde{Z}
\big<1/2,{\rm I}|\r\big>\\
\big<2|{\rm II}\big>_i & = & \cH^{[1/2,i,-9/2i](a)}_i\ Z
\big<1/2,{\rm II}|\r\big>\\
\big<2|{\rm II}\big>^*_i & = & \varepsilon^{ij}\cH^{[1/2,-i,9/2i](b)}_j\ \tilde{Z}
\big<1/2,{\rm II}|\r\big>\\
\big<2|{\rm III}\big>_i & = & \cH^{[1/2,5i,3/2i](a)}_i\ Z
\big<1/2,{\rm III}|\r\big>\\
\big<2|{\rm III}\big>^*_i & = & \varepsilon^{ij}\cH^{[1/2,-5i,-3/2i](b)}_j\ \tilde{Z}
\big<1/2,{\rm III}|\r\big>\\
\,[2|{\rm I}]_{ij} & = & \cH^{[1,0,0](c)}_{ij}\ Z[1,{\rm I}|\r]\\
\big<2|{\rm I}\big>_{ij} & = & \cH^{[1,0,0](d)}_{ij}\ Z
\big<1,{\rm I}|\r\big>\\
\big<2|{\rm I}\big>^*_{ij} & = & \varepsilon^{ik}\e_{jl}\cH^{[1,0,0](e)}_{kl}\ \tilde{Z}
\big<1,{\rm I}|\r\big>
\end{eqnarray*}
The Laplace Beltrami operator for the transverse two-form field ${\cal Y}^{\a\b}$,
is given by
\begin{equation}
\xbox^{[110]}{\cal Y}^{[\a\b]} \equiv M_{(1)^2(0)}{\cal Y}^{[\a\b]} =
3D_{\g}D^{[\g}{\cal Y}^{\a\b]} = (D^{\g}D_{\g}+48){\cal Y}^{[\a\b]}-
4\cR^{[\a\ \b]}_{\ [\g\ \d]}{\cal Y}^{[\g\d]}
\end{equation}
From the decomposition $D_{\a}{\cal Y}^{\b\g}=D^H_{\a}{\cal Y}^{\b\g}
+(\IM_{\a})^{\b}_{\ \d}{\cal Y}^{\d\g}+(\IM_{\a})^{\g}_{\ \d}{\cal Y}^{\b\d}$
we obtain:
\begin{eqnarray}
\xbox^{[110]}{\cal Y}^{[\a\b]} = \left\{48\,\d^{[\a\ \b]}_{\ [\g\ \d]}-
4\cR^{[\a\ \b]}_{\ [\g\ \d]} + \right. & \nonumber\\
 & \nonumber\\
2\eta^{\mu\nu}(\IM_{\mu})^{[\a}_{\ \,[\g}(\IM_{\nu})^{\b]}_{\ \,\d]}+
2\eta^{\mu\nu}(\IM_{\mu}\IM_{\nu})_{\ [\g}^{[\a}\d^{\b]}_{\ \d]}\, + & \!\!\!\left.
4\eta^{\mu\nu}(\IM_{\mu})_{\ [\g}^{[\a}\d^{\b]}_{\ \d]}D^H_{\nu}
\right\} {\cal Y}^{[\g\d]}\,.
\end{eqnarray}
For the regular $G$ representations of type $^0$ this operators acts on 
$AdS_4$ fields as the following $11\times 11$ matrix:\\
Column one to three:
\begin{footnotesize}
\begin{eqnarray*}
\begin{array}{|c||c|c|c|}
\hline
  M_{(1)^2(0)}
    & Z[0,{\rm I}]
    & Z[0,{\rm II}]
    & Z[1,{\rm I}]
\cr
\hline
\hline
   Z[0,{\rm I}] &
   H_0\!+\!32 & \!-\!16 &
   \ft{16}{\sqrt{3}}i(M_2\!+\!2)
   \cr
   Z[0,{\rm II}] &
   \!-32 & H_0\!+\!16 & 0
   \cr
   Z[1,{\rm I}] &
   0 & 0 & H_0
   \cr
   Z \langle \ft12, {\rm I} \rangle &
   \!-\ft{16}{\sqrt{3}}iM_1 & 0 & \ft{16}{\sqrt{3}}i(M_1\!+\!2)
   \cr
   \tilde{Z} \langle \ft12, {\rm I} \rangle &
   \ft{16}{\sqrt{3}}iM_2 & 0 & \!-\ft{16}{\sqrt{3}}i(M_2\!\!+\!\!2)
   \cr
   Z \langle 0, {\rm II} \rangle &
   0 & \ft{16}{3\sqrt{2}}i(M_2\!-\!M_1\!+\!3J) & 0
   \cr
   \tilde{Z} \langle 0, {\rm II}\rangle &
   0 & \!\!-\ft{16}{3\sqrt{2}}i(M_2\!\!-\!\!M_1\!\!-\!\!3J) & 0
   \cr
   Z \langle 0, {\rm I} \rangle &
   0 & 0 & 0
   \cr
   \tilde{Z} \langle 0, {\rm II}  \rangle &
   0 & 0 & 0
   \cr
   Z \langle \ft{1}{2}, {\rm III}  \rangle &
   0 & 0 & 0
   \cr
   \tilde{Z} \langle \ft{1}{2}, {\rm III}  \rangle &
   0 & 0 & 0
   \cr
\hline
\end{array}
\end{eqnarray*}
\end{footnotesize}
Column four to seven:
\begin{footnotesize}
\begin{eqnarray*}
\begin{array}{|c||c|c|c|c|}
\hline
  M_{(1)^2(0)}
    & Z \langle \ft12, {\rm I} \rangle
    & \tilde{Z} \langle \ft12, {\rm I} \rangle
    & Z \langle 0, {\rm II} \rangle
    & \tilde{Z} \langle 0 , {\rm II} \rangle
\cr
\hline
\hline
   Z[0,{\rm I}] & \!-\ft{16}{\sqrt{3}}i(M_1\!+\!2)& 0
   & 0 & 0
   \cr
   Z[0,{\rm II}]  & 0 & 0
   & \ft{32}{2\sqrt{2}}i(M_2\!\!-\!\!M_1\!\!-\!\!3J)
   & -\ft{32}{2\sqrt{2}}i(M_2\!\!-\!\!M_1\!\!-\!\!3J)
   \cr
   Z[1,{\rm I}]  & \!-\ft{16}{\sqrt{3}}iM_2 & \ft{16}{\sqrt{3}}iM_1
   & 0 & 0
   \cr
   Z \langle \ft12, {\rm I} \rangle &
   \!H_0\!\!+\!\!32\!\!-\!\!\ft{32}{3}(M_2\!\!-\!\!M_1)  & 0
   & 0 & 0
   \cr
   \tilde{Z} \langle \ft12, {\rm I} \rangle  &
   0 & \!H_0\!\!+\!\!32\!\!+\!\!\ft{32}{3}(M_2\!\!-\!\!M_1)
   & 0 & 0
   \cr
   Z \langle 0, {\rm II} \rangle & 0 & 0
   & \!H_0\!\!+\!\!32\!\!+\!\!\ft{32}{3}(M_2\!\!-\!\!M_1)
   & 0
   \cr
   \tilde{Z} \langle 0, {\rm II}\rangle & 0 & 0
   & 0 &\! H_0\!\!+\!\!32\!\!-\!\!\ft{32}{3}(M_2\!\!-\!\!M_1)
   \cr
   Z \langle 0, {\rm I} \rangle & \!-\ft{8}{3\sqrt{2}}(M_2\!-\!M_1\!+\!3J) & 0
   & -\ft{16}{\sqrt{3}}M_1 & 0
   \cr
   \tilde{Z} \langle 0, {\rm II}  \rangle
    & 0 & \!-\ft{16}{3\sqrt{2}}(M_2\!-\!M_1\!-\!3J)
    & 0 & -\ft{16}{\sqrt{3}} M_2
   \cr
   Z \langle \ft{1}{2}, {\rm III}  \rangle+
   & \!-\ft{8}{3\sqrt{2}}(M_2\!-\!M_1\!-\!3J)& 0
   & 0 & -\ft{16}{\sqrt{3}} M_1
   \cr
   \tilde{Z} \langle \ft{1}{2}, {\rm III}  \rangle
   & 0 & \!-\ft{16}{3\sqrt{2}}(M_2\!-\!M_1+\!3J)
   & -\ft{16}{\sqrt{3}}M_2
   & 0
   \cr
\hline
\end{array}
\end{eqnarray*}
\end{footnotesize}
\par
Column eight to eleven:
\begin{footnotesize}
\begin{eqnarray*}
\begin{array}{|c||c|c|c|c|}
\hline
    & Z\langle \ft12,{\rm II}\rangle
    & \tilde{Z} \langle \ft12, {\rm II} \rangle
    & Z\langle \ft12, {\rm III} \rangle
    & \tilde{Z} \langle \ft12, {\rm III} \rangle
\cr
\hline
\hline
   Z[0,{\rm I}] &
   0 & 0 & 0 & 0
   \cr
   Z[0,{\rm II}]  & 0 & 0 & 0 & 0\cr
   Z[1,{\rm I}] & 0 & 0 & 0 & 0
   \cr
   Z \langle \ft12, {\rm I} \rangle
    & \ft{32}{3\sqrt{2}}(M_2\!\!-\!\!M_1\!\!-\!\!3J) & 0 &
   \ft{32}{3\sqrt{2}}(M_2\!\!-\!\!M_1\!\!+\!\!3J) & 0
   \cr
   \tilde{Z} \langle \ft12, {\rm I} \rangle
   & 0 & \ft{32}{3\sqrt{2}}(M_2\!\!-\!\!M_1\!\!+\!\!3J) & 0 &
   \ft{32}{3\sqrt{2}}(M_2\!\!-\!\!M_1\!\!-\!\!3J)
   \cr
   Z\langle 0,{\rm II}\rangle &
   \!\!-\ft{32}{\sqrt{3}}(M_2\!\!+\!\!2) & 0 & 0 &
   \!\!-\ft{32}{\sqrt{3}}(M_1\!\!+\!\!2)
   \cr
   \tilde{Z}\langle 0,{\rm II}\rangle & 0 &
   \!\!-\ft{32}{\sqrt{3}}(M_1\!\!+\!\!2) &
   \!\!-\ft{32}{\sqrt{3}}(M_2\!\!+\!\!2) & 0
   \cr
   Z\langle \ft12,{\rm II}\rangle & H_0 & 0 & 0 & 0
   \cr
   \tilde{Z} \langle \ft12, {\rm II} \rangle & 0 & H_0 & 0 & 0
   \cr
   Z\langle \ft12, {\rm III} \rangle & 0 & 0 &
   H_0\!\!-\ft{64}{3}(M_2\!\!-\!\!M_1) & 0
   \cr
   \tilde{Z} \langle \ft12, {\rm III} \rangle & 0 & 0 & 0 &
   H_0\!\!+\!\!\ft{64}{3}(M_2\!\!-\!\!M_1)
   \cr
\hline
\end{array}
\end{eqnarray*}
\end{footnotesize}
This matrix has the following eigenvalues:
\begin{eqnarray}
\lambda_1 &=& H_0 + \ft{32}{3}(M_2-M_1)\,,
\nonumber \\
\lambda_2 &=& H_0 - \ft{32}{3}(M_2-M_1)\,,
\nonumber \\
\lambda_3 &=& H_0\,,
\nonumber \\
\lambda_4 &=& H_0 + 24 + 4 \sqrt{H_0 + 36}\,,
\nonumber \\
\lambda_5 &=& H_0 + 24 - 4 \sqrt{H_0 + 36}\,,
\nonumber \\
\lambda_6 &=& H_0 + \ft{32}{3}(M_2-M_1) + 16 + 4 \sqrt{H_0 + \ft{32}{3}(M_2-M_1) + 16}
\,, \nonumber \\
\lambda_7 &=& H_0 + \ft{32}{3}(M_2-M_1) + 16 - 4 \sqrt{H_0 + \ft{32}{3}(M_2-M_1) + 16}
\,, \nonumber \\
\lambda_8 &=& H_0 - \ft{32}{3}(M_2-M_1) + 16 + 4 \sqrt{H_0 - \ft{32}{3}(M_2-M_1) + 16}
\,, \nonumber \\
\lambda_9 &=& H_0 - \ft{32}{3}(M_2-M_1) + 16 - 4 \sqrt{H_0 - \ft{32}{3}(M_2-M_1) + 16}
\,, \nonumber \\
\lambda_{10} &=& \lambda_{11} = H_0 + 32 \,.
\nonumber \\
\label{eigenAtwoform}
\end{eqnarray}
The eigenvalues $\lambda_1,\lambda_2,\lambda_4,\lambda_5$, equal
to the one--form physical ones, are the longitudinal eigenvalues.
The other seven are the physical two-form eigenvalues.
\par
As in the case of the one--form, by removing rows and columns we find the matrix
of each exceptional series, and the corresponding eigenvalues:
\begin{footnotesize}
\begin{eqnarray}
\begin{array}{|c||c|}
\hline
\, A_R \, & \, \lambda_1, \lambda_2, \lambda_3, \lambda_4, \lambda_5 , \lambda_6,
\lambda_7, \lambda_8, \lambda_9, \lambda_{10}, \lambda_{11} \,\cr
\hline
A_1 & \, \lambda_1, \lambda_3, \lambda_4, \lambda_5 , \lambda_6,
\lambda_7, \lambda_{10}, \lambda_{11} \, \cr
\hline
A_1^* &  \, \lambda_2, \lambda_3, \lambda_4, \lambda_5 , \lambda_8,
\lambda_9, \lambda_{10}, \lambda_{11} \,  \cr
\hline
A_2 & \, \lambda_1,  \lambda_6,
\lambda_7 \,  \cr
\hline
A_2^* & \, \lambda_2,  \lambda_8,
\lambda_9 \, \cr
\hline
A_3 & \, \lambda_1, \lambda_4, \lambda_5 , \lambda_6,
\lambda_7, \lambda_{10}, \lambda_{11} \, \cr
\hline
A_3^* & \, \lambda_2, \lambda_4, \lambda_5 , \lambda_8,
\lambda_9, \lambda_{10}, \lambda_{11} \, \cr
\hline
A_4 &  \, \lambda_1, \lambda_4, \lambda_6,
\lambda_7, \lambda_{10} \, \cr
\hline
A_4^* & \, \lambda_2, \lambda_4, \lambda_8,
\lambda_9, \lambda_{10} \, \cr
\hline
A_5 & \, \lambda_1, \lambda_6 \,\cr
\hline
A_5^* & \, \lambda_2, \lambda_8 \, \cr
\hline
A_6 & \, \lambda_3, \lambda_4, \lambda_5, \lambda_{10}, \lambda_{11}\,  \cr
\hline
A_7 &  \, \lambda_4, \lambda_5, \lambda_{10}, \lambda_{11} \, \cr
\hline
A_8 & \, \lambda_3, \lambda_4 \, \cr
\hline
\end{array}
\label{eigenvaluesAtwoform}
\end{eqnarray}
\end{footnotesize}
The two--form operator matrix in the representations $^{++}$ is the following
$5\times 5$ matrix:\\
Columns one to two:
\begin{footnotesize}
\begin{eqnarray*}
\begin{array}{|c||c|c|}
\hline
  M_{(1)^2(0)}
    & Z\langle0,{\rm I}\rangle
    & Z\langle{1\over 2},{\rm I}\rangle
\cr
\hline
\hline
   Z\langle 0,{\rm I}\rangle &
   H_0-{32\over 3}\left(M_2-M_1\right) & 
   {16\over\sqrt{3}}\left(M_1+2\right) 
   \cr
   Z\langle{1\over 2},{\rm I}\rangle &
   \ft{32}{\sqrt{3}}M_2 & H_0+32
   \cr
   Z\langle{1\over 2},{\rm II}\rangle &
   0 &  -\ft{16}{3\sqrt{2}}\left(M_2-M_1+3J-3\right)
   \cr
   Z\langle{1\over 2},{\rm III}\rangle &
   0 & -\ft{16}{3\sqrt{2}}\left(M_2-M_1-3J-3\right) 
   \cr
  Z\langle1,{\rm I}\rangle &
    0 & \ft{32}{\sqrt{3}}\left(M_2-1\right) 
   \cr
\hline
\end{array}
\end{eqnarray*}
\end{footnotesize}
Columns three to five:
\begin{footnotesize}
\begin{eqnarray*}
\begin{array}{|c||c|c|c|}
\hline
  M_{(1)^2(0)}
    & Z\langle{1\over 2},{\rm II}\rangle
    & Z\langle{1\over 2},{\rm III}\rangle
    & Z\langle1,{\rm I}\rangle
\cr
\hline
\hline
   Z\langle0,{\rm I}\rangle &
   0&0&0
   \cr
   Z\langle{1\over 2},{\rm I}\rangle &
   \ft{32}{3\sqrt{2}}\left(M_2-M_1-3J-6\right) &
   \ft{32}{3\sqrt{2}}\left(M_2-M_1+3J\right) &
   {16\over\sqrt{3}}\left(M_1+3\right)
   \cr
   Z\langle{1\over 2},{\rm II}\rangle &
   H_0+{32\over 3}\left(M_2-M_1\right)-32 & 0 & 0
   \cr
   Z\langle{1\over 2},{\rm III}\rangle &
   0 & H_0-{32\over 3}\left(M_2-M_1\right)+32 & 0
   \cr
  Z\langle1,{\rm I}\rangle &
    0 & 0 & H_0+{32\over 3}\left(M_2-M_1\right)-32
   \cr
\hline
\end{array}
\end{eqnarray*}
\end{footnotesize}
It has eigenvalues
\begin{eqnarray}
\lambda_1 &=& H_0 + \ft{32}{3}(M_2 - M_1)
\,, \nonumber \\
\lambda_2 &=& H_0 + \ft{32}{3}(M_2 - M_1) + 16 + 4  \sqrt{H_0+ \ft{32}{3}(M_2-M_1)+16}
\,, \nonumber \\
\lambda_3 &=& H_0 + \ft{32}{3}(M_2 - M_1) + 16 - 4 \sqrt{H_0+ \ft{32}{3}(M_2-M_1)+16}
\,, \nonumber \\
\lambda_4 &=& H_0 + 32
\,, \nonumber \\
\lambda_5 &=& H_0 + \ft{64}{3}(M_2-M_1) -32
\,.
\label{eigenBtwoform}
\end{eqnarray}
The eigenvalue $\lambda_1$, equal to the physical eigenvalue of the
($^{++}$) one--form, is longitudinal.
The other four are the physical eigenvalues.
\begin{footnotesize}
\begin{eqnarray}
\begin{array}{|c||c|}
\hline
\, B_R \, & \, \lambda_1, \lambda_2, \lambda_3, \lambda_4, \lambda_5 \, \cr
\hline
\, B_1 \, & \, \lambda_1, \lambda_2,  \lambda_3, \lambda_5 \, \cr
\hline
\, B_2 \, & \, \lambda_5 \, \cr
\hline
\, B_3 \, & \, \lambda_1, \lambda_2, \lambda_3 \, \cr
\hline
\, B_4 \, & \, \lambda_1, \lambda_2, \lambda_3, \lambda_4 \,  \cr
\hline
\, B_5 \, & \,  \lambda_4 \, \cr
\hline
\, B_6 \, & \,  \lambda_1, \lambda_2, \lambda_3, \lambda_4 \, \cr
\hline
\, B_7 \, & \, \lambda_1, \lambda_2, \lambda_4 \, \cr
\hline
\, B_8 \, & \, \lambda_4 \, \cr
\hline
\, B_9 \, & \, \lambda_4 \, \cr
\hline
\, B_{10} \, & \, \lambda_4 \, \cr
\hline
\end{array}
\label{eigenvaluesBtwoform}
\end{eqnarray}
\end{footnotesize}
For the $^{--}$ representations, the eigenvalues are the conjugates
($M_2 \leftrightarrow M_1$) of the ones in (\ref{eigenvaluesBtwoform}).
\par
We can use the eigenvalues of the two--form harmonic to determine (see \cite{univer}) 
the masses of the $AdS_4$ vector field $Z$.
\subsection{The three-form}
The $H$ decomposition of the three--form in $H$--irreducible fragments has been done
in \cite{spec321}: 
\begin{eqnarray*}
{\cal Y}^{ABC} & = & \varepsilon^{ABCD}\,\{\l^D_{3i}\langle 3| {\rm I}
\rangle_i+\l^D_{i3}\langle 3| {\rm I}\rangle_i^* \}\,\,,\\
{\cal Y}^{ABm} & = & \l^A_{i3}\l^B_{j3}\varepsilon^{ij}\{\s^m_{21}
\langle 3|{\rm II}\rangle_{.}+\s^m_{12}\langle 3|{\rm III}\rangle_{.}\}+
 \l^A_{3i}\l^B_{3j}\varepsilon^{ij}\{\s^m_{12}\langle 3|{\rm II}\rangle_{.}^*+
\s^m_{21}\langle 3|{\rm III}\rangle_{.}^*\}\\
 && +i\l^{[A}_{i3}\l^{B]}_{3i}\{\s^m_{21}\langle 3|{\rm IV}\rangle_{.}+
\s^m_{12}\langle 3|{\rm IV}\rangle_{.}^*\}+\l^{[A}_{i3}\l^{B]}_{3j}
\{\s^m_{21}\varepsilon^{ik}\langle 3|{\rm I}\rangle_{kj}-
\s^m_{12}\varepsilon^{jk}\langle 3|{\rm I}\rangle_{ik}^*\}\,\,,\\
{\cal Y}^{AB3} & = & \l^A_{3i}\l^B_{3j}\varepsilon^{ij}\langle 3| {\rm I}\rangle_{.}+
i\l^A_{i3}\l^B_{j3}\varepsilon^{ij}\langle 3| {\rm I}\rangle_{.}^*+
i\l^{[A}_{i3}\l^{B]}_{3i}[3|{\rm I}]_{.}
 +\l^{[A}_{i3}\l^{B]}_{3j}\varepsilon^{ik}[3|{\rm I}]_{kj}\,\,,\\
{\cal Y}^{Amn} & = & \varepsilon^{mn}\{ \l^A_{3i}\langle 3| {\rm II}
\rangle_i+\l^A_{i3}\langle 3| {\rm II}\rangle_i^* \}\,\,,\\
{\cal Y}^{Am3} & = & \l^A_{3i}\{\s^m_{12}\langle 3| {\rm IV}\rangle_i+
\s^m_{21}\langle 3| {\rm III}\rangle_i \}+
\l^A_{i3}\{ \s^m_{21}\langle 3| {\rm IV}\rangle_i^*+
\s^m_{12}\langle 3| {\rm III}\rangle_i^* \}\,,\\
{\cal Y}^{mn3} & = & \varepsilon^{mn}[3| {\rm II}]_{.}\,\,,
\end{eqnarray*}
where the fragments of type $^0$ are:
\begin{eqnarray*}
\langle 3 | {\rm I } \rangle_{ij} &=& {\cal H}^{[1,-2i,-{3i}](c)}_{ij} \cdot
                    \pi\langle 1 , {\rm I} \rangle \,,
\\
\langle 3 | {\rm I } \rangle_{ij}^* &=& -\varepsilon^{ik} \varepsilon^{jl}
            {\cal H}^{[1,2i,3i](c)}_{kl} \cdot
                    \tilde{\pi}\langle 1 , {\rm I} \rangle \,,
\\
{}[ 3 | {\rm I } ]_{ij} &=& {\cal H}^{[1,0,0](c)}_{ij} \cdot
                    \pi[ 1 , {\rm I} ] \,,
\\
\langle 3 | {\rm I } \rangle_i &=& {\cal H}^{[1/2,3i,-{3i}/{2}](a)}_i \cdot
                    \pi\langle \ft12 , {\rm I} \rangle \,,
\\
\langle 3 | {\rm I } \rangle_i^* &=& \varepsilon^{ij}
          {\cal H}^{[1/2,-3i,{3i}/{2}](b)}_j \cdot
                    \tilde {\pi}\langle \ft12 , {\rm I} \rangle \,,
\\
\langle 3 | {\rm II } \rangle_i &=& {\cal H}^{[1/2,3i,-{3i}/{2}](a)}_i \cdot
                    \pi\langle \ft12 , {\rm II} \rangle \,,
\\
\langle 3 | {\rm II } \rangle_i^* &=& \varepsilon^{ij}
          {\cal H}^{[1/2,-3i,{3i}/{2}](b)}_j \cdot
                    \tilde {\pi}\langle \ft12 , {\rm II} \rangle \,,
\\
\langle 3 | {\rm III } \rangle_i &=& {\cal H}^{[1/2,i,-{9i}/{2}](a)}_i \cdot
                    \pi\langle \ft12 , {\rm III} \rangle \,,
\\
\langle 3 | {\rm III } \rangle_i^* &=& \varepsilon^{ij}
          {\cal H}^{[1/2,-i,{9i}/{2}](b)}_j \cdot
                    \tilde {\pi}\langle \ft12 , {\rm III} \rangle \,,
\\
\langle 3 | {\rm IV } \rangle_i &=& {\cal H}^{[1/2,5i,{3i}/{2}](a)}_i \cdot
                    \pi\langle \ft12 , {\rm IV} \rangle \,,
\\
\langle 3 | {\rm IV } \rangle_i^* &=& \varepsilon^{ij}
          {\cal H}^{[1/2,-5i,-{3i}/{2}](b)}_j \cdot
                    \tilde {\pi}\langle \ft12 , {\rm IV} \rangle \,,
\\
\langle 3 | {\rm IV } \rangle_\cdot &=&
{\cal H}^{[0,-2i,-3i]} \cdot
                 \pi\langle 0 , {\rm IV} \rangle \,,
\\
\langle 3 | {\rm IV } \rangle_\cdot^* &=&
{\cal H}^{[0,2i,3i]} \cdot
                    \tilde{\pi}\langle 0 , {\rm IV} \rangle \,,
\\
{}[3 | {\rm I } ]_\cdot &=&
{\cal H}^{[0,0,0]} \cdot
                    \pi[ 0 , {\rm I} ] \,,
\\
{}[3 | {\rm II } ]_\cdot &=&
{\cal H}^{[0,0,0]} \cdot
                    \pi[ 0 , {\rm II} ] \,,
\end{eqnarray*}
while the fragments of type $^{++}$ are:
\begin{eqnarray*}
\langle 3 | {\rm I } \rangle_{ij} &=& {\cal H}^{[1,-2i,-{3i}](d)}_{ij} \cdot
                    \pi\langle 1 , {\rm I} \rangle \,,
\\
\langle 3 | {\rm I } \rangle_{ij}^* &=& \varepsilon^{ik} \varepsilon^{jl}
            {\cal H}^{[1,2i,3i](d)}_{kl} \cdot
                    \tilde{\pi}\langle 1 , {\rm I} \rangle \,,
\\
{}[ 3 | {\rm I } ]_{ij} &=& {\cal H}^{[1,0,0](d)}_{ij} \cdot
                    \pi[ 1 , {\rm I} ] \,,
\\
\langle 3 | {\rm I } \rangle_i &=& {\cal H}^{[1/2,3i,-{3i}/{2}](b)}_i \cdot
                    \pi\langle \ft12 , {\rm I} \rangle \,,
\\
\langle 3 | {\rm II } \rangle_i &=& {\cal H}^{[1/2,3i,-{3i}/{2}](b)}_i \cdot
                    \pi\langle \ft12 , {\rm II} \rangle \,,
\\
\langle 3 | {\rm III } \rangle_i &=& {\cal H}^{[1/2,i,-{9i}/{2}](b)}_i \cdot
                    \pi\langle \ft12 , {\rm III} \rangle \,,
\\
\langle 3 | {\rm IV } \rangle_i &=& {\cal H}^{[1/2,5i,{3i}/{2}](b)}_i \cdot
                    \pi\langle \ft12 , {\rm IV} \rangle \,,
\\
\langle 3 | {\rm I } \rangle_\cdot &=&
{\cal H}^{[0,6i,-3i]} \cdot
                 \pi\langle 0 , {\rm I} \rangle \,,
\\
\langle 3 | {\rm II } \rangle_\cdot^* &=&
{\cal H}^{[0,8i,0]} \cdot
                    \tilde{\pi}\langle 0 , {\rm II} \rangle \,,
\\
\langle 3 | {\rm III } \rangle_\cdot^* &=&
{\cal H}^{[0,4i,-6i]} \cdot
                    \tilde{\pi}\langle 0 , {\rm III} \rangle \,.
\end{eqnarray*}
The fragments that are present the type $^{--}$ series are the complex
conjugates of the fragments above.
The Laplace Beltrami operator for the transverse three-form
${\cal Y}^{[\a\b\g]}$, is a first-order differential operator, given by
\begin{eqnarray}
\xbox^{[111]}{\cal Y}^{[\a\b\g]} \equiv M_{(1)^3}{\cal Y}^{[\a\b\g]} =
\ft{1}{24}\e^{\a\b\g\d}_{\ \ \ \ \mu\nu\r}
D_{\d}{\cal Y}^{\mu\nu\r} =\hspace{6 cm}\nonumber\\
=\ft{1}{24}\e^{\a\b\g\d}_{\ \ \ \ \ \mu\nu\r}\left[
D^H_{\d}{\cal Y}^{\mu\nu\r}+(\IM_{\d})^{\mu}_{\ \s}{\cal Y}^{\s\nu\r}+
(\IM_{\d})^{\nu}_{\ \s}{\cal Y}^{\mu\s\r}+
(\IM_{\d})^{\r}_{\ \s}{\cal Y}^{\mu\nu\s}\right].
\end{eqnarray}
For  the regular series of type $^0$  this operator acts on the $AdS_4$ fields as a
$15\times 15$ matrix:\\
Column one to five:
\begin{footnotesize}
\begin{eqnarray}
\begin{array}{|c||c|c|c|c|c|}
\hline
   M_{(1)^3} &
        \pi\langle 1,{\rm I} \rangle &
        \tilde{\pi}\langle 1,{\rm I} \rangle &
        \pi [1,{\rm I}] &
        \pi \langle \ft12, {\rm I} \rangle &
        \tilde{\pi}\langle \ft12, {\rm I}  \rangle \cr
\hline
\hline
    \pi\langle 1,{\rm I} \rangle         &
    {Y} & 0 &
  {\frac{-\left( 2\,J + {Y} \right) }{2\,{\sqrt{2}}}} & 0 & 0
   \cr
     \tilde{\pi}\langle 1,{\rm I} \rangle      &
 0 & -{Y} & {\frac{-2\,J + {Y}}{2\,{\sqrt{2}}}}
   & 0 & 0 \cr
     \pi [1,{\rm I}]   &
     {\frac{-2 - 2\,J + {Y}}{{\sqrt{2}}}} &
  -{\frac{2 + 2\,J + {Y}}{{\sqrt{2}}}} & 1 & 0 & 0 \cr
     \pi \langle \ft12, {\rm I} \rangle   &
 0 & 0
   & 0 & 0 & 0 \cr
         \tilde{\pi}\langle \ft12, {\rm I}  \rangle     &
 0 & 0 & 0 & 0 & 0 \cr
     \pi \langle \ft12, {\rm II} \rangle          &
 0 & 0 &
  {\frac{{\frac{i}{2}}\,\left( 2 + {M_1} \right) }{{\sqrt{3}}}} &
  -i\,{Y} & 0 \cr
         \tilde{\pi}\langle \ft12, {\rm II}  \rangle      &
 0 & 0 &
  {\frac{{\frac{i}{2}}\,\left( 2 + {M_2} \right) }{{\sqrt{3}}}} & 0
   & -i\,{Y} \cr
     \pi \langle \ft12, {\rm III} \rangle                &
 {\frac{2 + {M_1}}{2\,{\sqrt{3}}}} & 0
   & 0 & {\frac{-\left( 2\,J + {Y} \right) }{2\,{\sqrt{2}}}} & 0
   \cr
        \tilde{\pi}\langle \ft12, {\rm III}  \rangle              &
 0 & {\frac{2 + {M_2}}{2\,{\sqrt{3}}}} & 0 & 0 &
  {\frac{2\,J - {Y}}{2\,{\sqrt{2}}}} \cr
     \pi \langle \ft12, {\rm IV} \rangle                   &
 0 &
  {\frac{2 + {M_1}}{2\,{\sqrt{3}}}} & 0 &
  {\frac{2\,J - {Y}}{2\,{\sqrt{2}}}} & 0 \cr
        \tilde{\pi}\langle \ft12, {\rm IV}  \rangle            &
  {\frac{2 + {M_2}}{2\,{\sqrt{3}}}} & 0 & 0 & 0 &
  {\frac{-\left( 2\,J + {Y} \right) }{2\,{\sqrt{2}}}} \cr
     \pi \langle 0 , {\rm IV} \rangle                    &
 0 & 0
   & 0 & 0 & 0 \cr
        \tilde{\pi}\langle 0 , {\rm IV}  \rangle              &
 0 & 0 & 0 & 0 & 0 \cr
 \pi[0,{\rm I}|\rho]           &
 0 & 0 & 0 & 0 & 0 \cr
\pi[0,{\rm II}|\rho]               &
 0
   & 0 & 0 & {\frac{2\,i\,\left( 2 + {M_2} \right) }{{\sqrt{3}}}}
   & {\frac{-2\,i\,\left( 2 + {M_1} \right) }{{\sqrt{3}}}} \cr
\hline
\end{array}
\end{eqnarray}
\end{footnotesize}
Column six to ten:
\begin{footnotesize}
\begin{eqnarray}
\begin{array}{|c||c|c|c|c|c|}
\hline
        &  \pi\langle \ft12, {\rm II}  \rangle
        & \tilde{\pi}\langle \ft12, {\rm II} \rangle
        & \pi\langle \ft12, {\rm III}  \rangle
        & \tilde{\pi}\langle \ft12, {\rm III} \rangle
        & \pi\langle \ft12, {\rm IV} \rangle
 \cr
\hline
\hline
    \pi\langle 1,{\rm I} \rangle            &
   0 & 0 & {\frac{2\,{M_2}}{{\sqrt{3}}}} & 0 & 0 \cr
    \tilde{\pi} \langle 1,{\rm I} \rangle          &
   0 & 0
   & 0 & {\frac{2\,{M_1}}{{\sqrt{3}}}} &
  {\frac{2\,{M_2}}{{\sqrt{3}}}} \cr
    \pi [1,{\rm I}]            &
  {\frac{-2\,i\,{M_2}}{{\sqrt{3}}}} &
  {\frac{-2\,i\,{M_1}}{{\sqrt{3}}}} & 0 & 0 & 0 \cr
\pi\langle \ft12, {\rm I} \rangle                &
  i\,{Y} & 0 & {\frac{-2 - 2\,J + {Y}}{{\sqrt{2}}}}
   & 0 & {\frac{2 + 2\,J + {Y}}{{\sqrt{2}}}} \cr
\tilde{\pi}\langle \ft12, {\rm I} \rangle                 &
  0 & i\,{Y} & 0 & {\frac{2 + 2\,J + {Y}}{{\sqrt{2}}}} &
  0 \cr
\pi\langle \ft12, {\rm II}  \rangle                &
   0 & 0 & 0 & 0 & 0 \cr
 \tilde{\pi}\langle \ft12, {\rm II} \rangle               &
   0 & 0 & 0 & 0 & 0 \cr
 \pi\langle \ft12, {\rm III}  \rangle                &
    0 & 0 & 1 & 0
   & 0 \cr
\tilde{\pi}\langle \ft12, {\rm III} \rangle                &
    0 & 0 & 0 & 1 & 0 \cr
  \pi\langle \ft12, {\rm IV} \rangle              &
    0 & 0 & 0 & 0 & -1 \cr
  \tilde{\pi}\langle \ft12, {\rm IV} \rangle                &
   0 & 0 & 0 & 0 & 0 \cr
    \pi \langle 0, {\rm IV} \rangle            &
    0 & 0 & {\frac{-i\,\left( 2 + {M_2} \right) }
    {{\sqrt{3}}}} & 0 & 0 \cr
     \tilde{\pi} \langle 0, {\rm IV} \rangle           &
   0 & 0 & 0 &
  {\frac{i\,\left( 2 + {M_1} \right) }{{\sqrt{3}}}} &
  {\frac{i\,\left( 2 + {M_2} \right) }{{\sqrt{3}}}} \cr
    \pi[0,{\rm I}]             &
  -{\frac{2 + {M_2}}{{\sqrt{3}}}} &
  -{\frac{2 + {M_1}}{{\sqrt{3}}}} & 0 & 0 & 0 \cr
   \pi[0,{\rm II}]             &
   0 & 0 & 0 & 0
   & 0 \cr
\hline
  \end{array}
\end{eqnarray}
\end{footnotesize}
Column eleven to fifteen:
\begin{footnotesize}
\begin{eqnarray}
\begin{array}{|c||c|c|c|c|c|}
\hline
     & \tilde{\pi}\langle \ft12, {\rm IV} \rangle
     & \pi \langle 0, {\rm IV} \rangle
     & \tilde{\pi} \langle 0, {\rm IV} \rangle
     & \pi[0,{\rm I}]
     & \pi[0,{\rm II}]
\cr
\hline
\hline
 \pi\langle 1,{\rm I} \rangle         &
   {\frac{2\,{M_1}}{{\sqrt{3}}}} & 0 & 0 & 0 & 0 \cr
  \tilde{\pi}\langle 1,{\rm I} \rangle        &
   0 & 0
   & 0 & 0 & 0 \cr
\pi [1,{\rm I}]          &
   0 & 0 & 0 & 0 & 0 \cr
 \pi\langle \ft12, {\rm I} \rangle         &
  0 & 0 & 0 & 0 &
  {\frac{-i\,{M_1}}{{\sqrt{3}}}} \cr
 \tilde{\pi} \langle \ft12, {\rm I} \rangle         &
  {\frac{-2 - 2\,J + {Y}}{{\sqrt{2}}}} & 0 & 0 & 0 &
  {\frac{i\,{M_2}}{{\sqrt{3}}}} \cr
 \pi\langle \ft12, {\rm II}  \rangle         &
  0 & 0 & 0 &
  -{\frac{{M_1}}{{\sqrt{3}}}} & 0 \cr
   \tilde{\pi} \langle \ft12, {\rm II} \rangle       &
  0 & 0 & 0 &
  -{\frac{{M_2}}{{\sqrt{3}}}} & 0 \cr
  \pi\langle \ft12, {\rm III}  \rangle        &
  0 &
  {\frac{i\,{M_1}}{{\sqrt{3}}}} & 0 & 0 & 0 \cr
     \tilde{\pi}\langle \ft12, {\rm III} \rangle     &
  0 & 0 &
  {\frac{-i\,{M_2}}{{\sqrt{3}}}} & 0 & 0 \cr
  \pi\langle \ft12, {\rm IV} \rangle        &
  0 & 0 &
  {\frac{-i\,{M_1}}{{\sqrt{3}}}} & 0 & 0 \cr
 \tilde{\pi}\langle \ft12, {\rm IV} \rangle          &
   -1 &
  {\frac{i\,{M_2}}{{\sqrt{3}}}} & 0 & 0 & 0 \cr
     \pi \langle 0, {\rm IV} \rangle     &
  {\frac{-i\,\left( 2 + {M_1} \right) }{{\sqrt{3}}}} &
  -{Y} & 0 & {\frac{2\,J + {Y}}{2\,{\sqrt{2}}}} & 0
   \cr
    \tilde{\pi} \langle 0, {\rm IV} \rangle      &
   0 & 0 & {Y} &
  {\frac{-2\,J + {Y}}{2\,{\sqrt{2}}}} & 0 \cr
  \pi[0,{\rm I}]        &
   0 &
  {\frac{2 + 2\,J - {Y}}{{\sqrt{2}}}} &
  -{\frac{2 + 2\,J + {Y}}{{\sqrt{2}}}} & -1 & -1 \cr
  \pi[0,{\rm II}]        &
   0 & 0 & 0
   & -2 & 0 \cr
\hline
\end{array}
\end{eqnarray}
\end{footnotesize}
This matrix has the following eigenvalues:
\begin{eqnarray}
\lambda_1 &=& \ft14 \sqrt{H_0 + \ft{32}{3}(M_2-M_1)+16} \,,
\nonumber \\
\lambda_2 &=& \ft14 \sqrt{H_0 - \ft{32}{3}(M_2-M_1)+16} \,,
\nonumber \\
\lambda_3 &=& - \ft14 \sqrt{H_0 + \ft{32}{3}(M_2-M_1)+16} \,,
\nonumber \\
\lambda_4 &=& - \ft14 \sqrt{H_0 - \ft{32}{3}(M_2-M_1)+16} \,,
\nonumber
\\
\lambda_5 &=& \ft14 \sqrt{H_0 + 36} - \ft12 \,,
\nonumber \\
\lambda_6 &=& -\ft14 \sqrt{H_0 + 36} - \ft12 \,,
\nonumber \\
\lambda_7 &=& -\ft14 \sqrt{H_0 + 4} + \ft12 \,,
\nonumber \\
\lambda_8 &=& \ft14 \sqrt{H_0 + 4} + \ft12 \,,
\nonumber \\
\lambda_9 &=& \dots = \lambda_{15} = 0 \,.
\label{eigenAthreeform}
\end{eqnarray}
We note that seven eigenvalues are $0$.
They correspond to the longitudinal three-forms
(${\cal Y}^{(3)}=D\wedge {\cal Y}^{(2)}$), which are annihilated by
$\xbox^{[111]}$ ($=\,^{*}D\wedge$).
\par
As in the cases of the one--form and of the two--form, by removing
rows and columns we find the matrix for each exceptional series, and
the corresponding eigenvalues:
\begin{footnotesize}
\begin{eqnarray}
\begin{array}{|c||c|}
\hline
\, A_R \, & \, \lambda_1, \lambda_2, \lambda_3, \lambda_4, \lambda_5, \lambda_6,
               \lambda_7, \lambda_8  \, \cr
\hline
\, A_1 \, & \, \lambda_1,  \lambda_3,  \lambda_5, \lambda_6,
               \lambda_7, \lambda_8  \cr
\hline
\, A_1^* \, & \,  \lambda_2, \lambda_4, \lambda_5, \lambda_6,
               \lambda_7, \lambda_8  \cr
\hline
\, A_2 \, & \, \lambda_1,  \lambda_3 \cr
\hline
\, A_2^* \, & \, \lambda_2,  \lambda_4 \cr
\hline
\, A_3 \, & \, \lambda_1,  \lambda_3, \lambda_5, \lambda_6
               \cr
\hline
\, A_3^* \, & \, \lambda_2,  \lambda_4, \lambda_5, \lambda_6
               \cr
\hline
\, A_4 \, & \, \lambda_1, \lambda_3,  \lambda_6
               \, \cr
\hline
\, A_4^* \, & \, \lambda_2, \lambda_4,  \lambda_6
               \, \cr
\hline
\, A_5 \, & \, \lambda_1, \lambda_3 \cr
\hline
\, A_5^* \, & \, \lambda_2, \lambda_4 \cr
\hline
\, A_6 \, & \,  \lambda_5, \lambda_6,
               \lambda_7, \lambda_8  \cr
\hline
\, A_7 \, & \, \lambda_5, \lambda_6
               \, \cr
\hline
\, A_8 \, & \, \lambda_6,
               \lambda_8  \cr
\hline
\end{array}
\label{eigenvaluesAthreeform}
\end{eqnarray}
\end{footnotesize}
The two--form operator matrix for the regular series of type $^{++}$ is
the following $10\times 10$ matrix:\\
Column one to five:
\begin{footnotesize}
\begin{eqnarray}
\begin{array}{|c||c|c|c|c|c|}
\hline
    & \pi \langle 1,{\rm I} \rangle
    & \tilde {\pi} \langle 1,{\rm I} \rangle
    & \pi[1,{\rm I}]
    & \pi \langle \ft12, {\rm I} \rangle
    & \pi \langle \ft12, {\rm II} \rangle
\cr
\hline
\hline
  \pi \langle 1,{\rm I} \rangle                 &
  {Y} & 0 &
  {\frac{-\left( 2\,J + {Y} \right) }{2\,{\sqrt{2}}}} & 0 & 0
   \cr
    \tilde {\pi} \langle 1,{\rm I} \rangle               &
 0 & -{Y} & {\frac{-\left( -2\,J + {Y} \right) }
    {2\,{\sqrt{2}}}} & 0 & 0 \cr
  \pi[1,{\rm I}]                 &
  {\frac{-2 - 2\,J + {Y}}{{\sqrt{2}}}} &
  {\frac{2 + 2\,J + {Y}}{{\sqrt{2}}}} & 1 & 0 &
  {\frac{-2\,i\,\left( -1 + {M_2} \right) }{{\sqrt{3}}}} \cr
   \pi \langle \ft12, {\rm I}   \rangle                 &
   0 & 0
   & 0 & 0 & i\,{Y} \cr
         \pi \langle \ft12, {\rm II}  \rangle          &
  0 & 0 &
  {\frac{i\,\left( 3 + {M_1} \right) }{{\sqrt{3}}}} &
  -i\,{Y} & 0 \cr
   \pi \langle \ft12, {\rm III}   \rangle                 &
 {\frac{3 + {M_1}}{{\sqrt{3}}}} & 0 &
  0 & {\frac{-\left( 2\,J + {Y} \right) }{2\,{\sqrt{2}}}} & 0
   \cr
      \pi \langle \ft12, {\rm IV}    \rangle              &
 0 & -{\frac{3 + {M_1}}{{\sqrt{3}}}} & 0 &
  {\frac{2\,J - {Y}}{2\,{\sqrt{2}}}} & 0 \cr
     \pi \langle 0, {\rm I} \rangle              &
  0 & 0 & 0 & 0 &
  {\frac{i\,\left( 2 + {M_1} \right) }{{\sqrt{3}}}} \cr
   \tilde{\pi} \langle 0, {\rm II}  \rangle                &
 0 & 0 & 0
   & 0 & 0 \cr
       \tilde{\pi} \langle 0, {\rm III}  \rangle             &
 0 & 0 & 0 & 0 & 0 \cr
\hline
\end{array}
\end{eqnarray}
\end{footnotesize}
Column six to ten:
\begin{footnotesize}
\begin{eqnarray}
\begin{array}{|c||c|c|c|c|c|}
\hline
    &\pi \langle \ft12, {\rm III}  \rangle
    &\pi \langle \ft12, {\rm IV} \rangle
    &\pi \langle 0, {\rm I} \rangle
    & \tilde{\pi} \langle 0, {\rm II}  \rangle
    &\tilde{\pi} \langle 0, {\rm III}  \rangle
\cr
\hline
\hline
   \pi \langle 1,{\rm I} \rangle             &
   {\frac{2\,\left( -1 + {M_2} \right) }{{\sqrt{3}}}} & 0 & 0
   & 0 & 0 \cr
         \tilde {\pi \langle 1,{\rm I} \rangle}       &
   0 & {\frac{-2\,\left( -1 + {M_2} \right) }
    {{\sqrt{3}}}} & 0 & 0 & 0 \cr
   \pi[1,{\rm I}]             &
   0 & 0 & 0 & 0 & 0 \cr
     \pi \langle \ft12, {\rm I}  \rangle            &
  {\frac{-2 - 2\,J + {Y}}{{\sqrt{2}}}} &
  {\frac{2 + 2\,J + {Y}}{{\sqrt{2}}}} & 0 & 0 & 0 \cr
       \pi \langle \ft12, {\rm II}  \rangle         &
 0 & 0 &
  {\frac{-2\,i\,{M_2}}{{\sqrt{3}}}} & 0 & 0 \cr
   \pi \langle \ft12, {\rm III}  \rangle             &
 1 & 0 & 0 & 0
   & {\frac{-2\,{M_2}}{{\sqrt{3}}}} \cr
   \pi \langle \ft12, {\rm IV} \rangle             &
 0 & -1 & 0 &
  {\frac{2\,{M_2}}{{\sqrt{3}}}} & 0 \cr
    \pi \langle 0, {\rm I} \rangle            &
 0 & 0 & -1 &
  -{\frac{2 + 2\,J + {Y}}{{\sqrt{2}}}} &
  {\frac{2 + 2\,J - {Y}}{{\sqrt{2}}}} \cr
   \tilde{\pi} \langle 0, {\rm II}  \rangle             &
 0 &
  {\frac{2 + {M_1}}{{\sqrt{3}}}} &
  {\frac{-2\,J + {Y}}{2\,{\sqrt{2}}}} & {Y} & 0 \cr
       \tilde{\pi} \langle 0, {\rm III}  \rangle          &
  -{\frac{2 + {M_1}}{{\sqrt{3}}}} & 0 &
  {\frac{2\,J + {Y}}{2\,{\sqrt{2}}}} & 0 & -{Y} \cr
\hline
\end{array}
\end{eqnarray}
\end{footnotesize}
It has eigenvalues:
\begin{eqnarray}
\lambda_1 &=& \ft14 \sqrt{H_0 + \ft{32}{3}(M_2 - M_1) +16} \,,
\nonumber \\
\lambda_2 &=& - \ft14 \sqrt{H_0 + \ft{32}{3}(M_2 - M_1) +16} \,,
\nonumber \\
\lambda_3 &=& \ft14 \sqrt{H_0 + 36} - \ft12 \,,
\nonumber \\
\lambda_4 &=& - \ft14 \sqrt{H_0 + 36} - \ft12 \,,
\nonumber \\
\lambda_5 &=& - \ft14 \sqrt{H_0 + \ft{64}{3}(M_2 - M_1) - 28} + \ft12 \,,
\nonumber \\
\lambda_6 &=&  \ft14 \sqrt{H_0 + \ft{64}{3}(M_2 - M_1) - 28} + \ft12 \,,
\nonumber \\
\lambda_7 &=& \dots = \lambda_{10} = 0 \,.
\label{eigenBthreeform}
\end{eqnarray}
The complete table of eigenvalues for the type $^{++}$ series is:
\begin{footnotesize}
\begin{eqnarray}
\begin{array}{|c||c|}
\hline
\, B_R \, & \, \lambda_1, \lambda_2, \lambda_3, \lambda_4, \lambda_5, \lambda_6 \cr
\hline
\, B_1 \, & \, \lambda_1, \lambda_2,  \lambda_5, \lambda_6 \cr
\hline
\, B_2 \, & \, \lambda_5, \lambda_6 \cr
\hline
\, B_3 \, & \, \lambda_1, \lambda_2 \cr
\hline
\, B_4 \, & \, \lambda_1, \lambda_2, \lambda_3, \lambda_4 \cr
\hline
\, B_5 \, & \, \lambda_3, \lambda_4 \cr
\hline
\, B_6 \, & \, \lambda_1, \lambda_2, \lambda_3, \lambda_4 \cr
\hline
\, B_7 \, & \, \lambda_2, \lambda_3, \lambda_4 \cr
\hline
\, B_8 \, & \,  \lambda_4 \cr
\hline
\, B_9 \, & \, \lambda_3, \lambda_4 \cr
\hline
\, B_{10} \, & \,  \lambda_4 \cr
\hline
\, B_{11} \, & \,  \lambda_4 \cr
\hline
\end{array}
\label{eigenvaluesBthreeform}
\end{eqnarray}
\end{footnotesize}
For the representations of the $^{--}$ series, the eigenvalues are the 
conjugates of the one in \eqn{eigenvaluesBthreeform}.
\subsection{The spinor}
The harmonic analysis of the eight--component Majorana spinor
has been completely worked out in \cite{spectfer}. We reformulate these results 
in our framework, in order to facilitate the matching of the spectrum with the ${\cal N}=2$ multiplets.
\par
The decomposition of the spinor
in its $H$--irreducible   components is
\begin{eqnarray}
\eta=
\left(
\matrix{
\langle \ft12 | {\rm I} \rangle_i \cr
\langle \ft12 | {\rm I} \rangle_\cdot \cr
\langle \ft12 | {\rm II} \rangle_\cdot \cr
-i\sigma_2 \langle \ft12 | {\rm I} \rangle_i^* \cr
\langle \ft12 | {\rm II} \rangle^*_\cdot \cr
-\langle \ft12 | {\rm I} \rangle^*_\cdot
}
\right)
\end{eqnarray}
where
\begin{eqnarray}
\langle \ft12 | {\rm I} \rangle_i
&=& {\cal H}_i^{[1/2, -i, -3i/2]\xi} \cdot \chi\langle \ft12, {\rm I} \rangle
\,, \nonumber \\
\langle \ft12 | {\rm I} \rangle_\cdot
 &=& {\cal H}^{[0, 2i, -3i]} \cdot \chi\langle 0, {\rm I} \rangle
\,, \nonumber \\
\langle \ft12 | {\rm II} \rangle_\cdot
 &=& {\cal H}^{[0, -4i, 0]} \cdot \chi\langle 0, {\rm II} \rangle
\,, \nonumber \\
\langle \ft12 | {\rm I} \rangle_i^*
 &=& \pm \varepsilon^{ij} {\cal H}_j^{[1/2, i, 3i/2]\xi} \cdot
\tilde \chi\langle \ft12, {\rm I} \rangle
\,, \nonumber \\
\langle \ft12 | {\rm I} \rangle^*_\cdot
 &=& {\cal H}^{[0, -2i, 3i]} \cdot \tilde \chi\langle 0, {\rm I} \rangle
\,, \nonumber \\
\langle \ft12 | {\rm II} \rangle^*_\cdot &=& {\cal H}^{[0, -4i, 0]} \cdot
\tilde \chi\langle 0, {\rm II} \rangle.
\end{eqnarray}
The fragments of type $^+$ are
\begin{eqnarray}
\langle \ft12 | {\rm I} \rangle_i &=& {\cal H}_i^{[1/2, -i, -3i/2](b)} \cdot \chi\langle \ft12, {\rm I} \rangle
\,, \nonumber \\
\langle \ft12 | {\rm I} \rangle_\cdot &=& {\cal H}^{[0, 2i, -3i]} \cdot \chi\langle 0, {\rm I} \rangle
\,, \nonumber \\
\langle \ft12 | {\rm I} \rangle_i^* &=&   \varepsilon^{ij} {\cal H}_j^{[1/2, i, 3i/2](b)} \cdot
\tilde \chi\langle \ft12, {\rm I} \rangle
\,, \nonumber \\
\langle \ft12 | {\rm II} \rangle_\cdot^* &=&
{\cal H}^{[0, 4i, 0]} \cdot \tilde \chi\langle 0, {\rm II} \rangle
\,.
\end{eqnarray}
\par
For the regular series $^+$ the spinor operator acts on the $AdS_4$ fields
as a $4\times 4$ matrix, whose eigenvalues are:
\begin{eqnarray}
\lambda_1 &=& - 6 + \sqrt{H_0 + 36} \,,
\nonumber \\
\lambda_2 &=& - 6 - \sqrt{H_0 + 36} \,,
\nonumber \\
\lambda_3 &=& - 8 + \sqrt{H_0 + 16 + \ft{32}{3}(M_2 - M_1)} \,,
\nonumber \\
\lambda_4 &=& - 8 - \sqrt{H_0 + 16 + \ft{32}{3}(M_2 - M_1)} \,.
\nonumber \\
\label{eigenA+spinor}
\end{eqnarray}
The eigenvalues for each exceptional series are
\begin{footnotesize}
\begin{eqnarray}
\begin{array}{|c||c|}
\hline
A_R^+, A_1^+, A_3^+, A_4^+ & \, \lambda_1, \lambda_2, \lambda_3, \lambda_4 \, \cr
\hline
A_2^+, A_5^+ & \, \lambda_3, \lambda_4 \, \cr
\hline
A_1^{+*}, A_6^+ & \, \lambda_1, \lambda_2 \, \cr
\hline
A_3^{+*}, A_7^+ & \, \lambda_1, \lambda_2 \, \cr
\hline
A_4^{+*}, A_8^+ & \, \lambda_1  \, \cr
\hline
\end{array}
\label{eigenvaluesA+spinor}
\end{eqnarray}
\end{footnotesize}
The fragments of type $^-$ are
\begin{eqnarray}
\langle \ft12 | {\rm I} \rangle_i &=& {\cal H}_i^{[1/2, -i, -3i/2](a)} \cdot \chi\langle \ft12, {\rm I} \rangle
\,, \nonumber \\
\langle \ft12 | {\rm II} \rangle_\cdot &=& {\cal H}^{[0, -4i, 0]} \cdot \chi\langle 0, {\rm II} \rangle
\,, \nonumber \\
\langle \ft12 | {\rm I} \rangle_i^* &=&  \varepsilon^{ij} {\cal H}_j^{[1/2, i, 3i/2](a)} \cdot
\tilde{ \chi\langle \ft12, {\rm I} \rangle }
\,, \nonumber \\
\langle \ft12 | {\rm I} \rangle_\cdot^* &=& {\cal H}^{[0, -2i, 3i]} \cdot \tilde{ \chi\langle 0, {\rm II} \rangle }
\,.
\end{eqnarray}
\par
For the regular series $^-$ the spinor operator acts on the $AdS_4$ fields
as a $4\times 4$ matrix, whose eigenvalues are:
\begin{eqnarray}
\lambda_1 &=& - 6 + \sqrt{H_0 + 36} \,,
\nonumber \\
\lambda_2 &=& - 6 - \sqrt{H_0 + 36} \,,
\nonumber \\
\lambda_3 &=& - 8 + \sqrt{H_0 + 16 - \ft{32}{3}(M_2 - M_1)} \,,
\nonumber \\
\lambda_4 &=& - 8 - \sqrt{H_0 + 16 - \ft{32}{3}(M_2 - M_1)} \,.
\nonumber \\
\label{eigenA-spinor}
\end{eqnarray}
The eigenvalues for each exceptional series are
\begin{footnotesize}
\begin{eqnarray}
\begin{array}{|c||c|}
\hline
A_R^-, A_1^{-*}, A_3^{-*}, A_4^{-*}
& \, \lambda_1, \lambda_2, \lambda_3, \lambda_4 \, \cr
\hline
A_2^{-*}, A_5^{-*} & \, \lambda_3, \lambda_4 \, \cr
\hline
A_1^-, A_6^- & \, \lambda_1, \lambda_2 \, \cr
\hline
A_3^-, A_7^- & \, \lambda_1, \lambda_2 \, \cr
\hline
A_4^-, A_8^- & \, \lambda_1  \, \cr
\hline
\end{array}
\label{eigenvaluesA-spinor}
\end{eqnarray}
\end{footnotesize}
\section{Matching the spectrum with the $Osp(2|4)$ multiplets}
As already mentioned, the structure of the long multiplets that arise
from ${\cal N}=2$ compactifications of eleven--dimensional supergravity
in $AdS_4$ has already been presented in \cite{multanna}.
The structure and the $G'$ representations of the long graviton multiplet, the long gravitino
multiplets and the massless multiplets are known since the eighties 
\cite{multanna}. However this is not the case for the long vector multiplets 
and for the short multiplets, due to the fact that the crucial knowledge of the
eigenvalues of the operators (\ref{operatori}) was lacking.
In this section we match the complete spectrum that we have  calculated
 with the multiplets as they are already known from old literature.
Furthermore, while doing this we also find the short ${\cal N}=2$ multiplets
as truncations of the long multiplets that were already
known.
\par
In order to achieve this we use a procedure of {\it exhaustion},
i.e. one starts with one of the
four different types of multiplets for which
all the masses of a certain field component 
are most easily retrieved
(this is for instance the case for the graviton field of the graviton multiplet)
and using the mass relations of \cite{univer} (see also appendix \ref{formule}),
one calculates all the masses
of the other types of fields present in the multiplet.
One uses also the information that all the fields in  a 
multiplet are in the same irreducible 
$G'=SU\left(3\right)\times SU\left(2\right)$ representation
and that their hypercharges are related according to the group
theoretical structure of the multiplets shown in tables
\ref{longgraviton}, \ref{longgravitino}, \ref{longvector}, 
\ref{shortgraviton}, \ref{shortgravitino}, \ref{shortvector}, 
\ref{hyper}, \ref{masslessgraviton}, \ref{masslessvector}.
So one knows in which $G$ representation  to find
the other fields of the multiplet, whose masses have been determined. 
Then, upon using the relations (\ref{massform}), these masses are 
compared with the eigenvalues
of the invariant operators on the spinor, the one--form,
the two--form or the three--form depending on the type of field one is 
considering. The upshot of this is
that some of these eigenvalues yield all the masses
obtained from the mass relations.
However, the remaining eigenvalues signal the
existence of some extra masses which then pertain to other
fields that are to be found in other multiplets.
In this way one establishes the existence of
new unknown multiplets and determines their structure
by filling out their field content.
After repeatedly applying this procedure one
will have filled out all the existing multiplets
in the spectrum.
\par
We should remark here that we did not calculate the eigenvalues 
of the Lichnerowicz  and Rarita--Schwinger operators 
$M_{(2)(0)^2}$ and $M_{(3/2)(1/2)^2}$.
However we succeeded in finding the complete
multiplet structure without making use of this.
The $AdS_4$ fields whose spectrum is determined by  
$M_{(2)(0)^2}$ and $M_{(3/2)(1/2)^2}$ are the scalar field $\phi$
and the transverse spinor field $\lambda_T$ (see (\ref{kkexpansion}).
We can fill the multiplets without knowing the spectrum of 
these two fields with the
help of the ${\cal N}=2\rightarrow {\cal N}=1$ decompositions
(see pictures \ref{N1longgraviton},\ref{N1longgravitino},\ref{N1longvector},
\ref{N1shortgraviton},\ref{N1shortgravitino},\ref{N1shortvector},\ref{N1hyper}).
Every theory with ${\cal N}=2$ supersymmetry also has 
${\cal N}=1$ supersymmetry, so every ${\cal N}=2$ multiplet has to be
decomposable in ${\cal N}=1$ multiplets. If we know every field of a multiplet
except for $\phi$ and $\lambda_T$, we can deduce which $\phi$ and $\lambda_T$
are present by trying to organize the ${\cal N}=2$ multiplet  in ${\cal N}=1$ multiplets. 
There is no ambiguity, because no ${\cal N}=1$ multiplet is built using $\phi$ and 
$\lambda_T$ fields alone.
In particular, a Wess Zumino multiplet with one $\lambda_T$ and two $\phi$ 's is
not allowed, since it has to contain both a scalar and a pseudoscalar.
\par
In practice one starts with the graviton multiplet
since the masses of the graviton field in the different
representations are immediate to derive, being the eigenvalues
of the scalar operator $M_{(0)^3}$.
By means of the above procedure,
one {\it exhausts} all the spin--$\ft32$ fields
 in the graviton multiplet comparing
the masses of the spin--$\ft32$ fields in the graviton multiplet
with the eigenvalues of the operator $M_{(1/2)^3}$.
The spin--$\ft32$ fields that provide the remaining
eigenvalues of the operator $M_{(1/2)^3}$,
can only be the highest--spin component gravitino fields
of the gravitino multiplet and hence we know all the masses of the gravitini
in the gravitino multiplet.  At this stage we can repeat the same procedure.
We use the eigenvalues of the one--form operator
$M_{(1)(0)^2}$ to identify the vector field $A$ and $W$
and we use the eigenvalues of the two--form operator
$M_{(1)^2 (0)}$ to identify the vector fields $Z$ in the graviton
and the gravitino multiplet. The remaining vector fields constitute the
highest--component vector fields of the vector multiplet.
Then we determine the masses of the longitudinal spinors, provided by the
eigenvalues of the operator $M_{(1/2)^3}$, and we find the longitudinal
spinors of the gravitino and vector multiplet.
The remaining longitudinal spinors  belong to hypermultiplets.
At the end we determine the masses of the scalars $S,\Sigma$, that are
provided by the eigenvalues of $M_{(0)^3}$, and of the pseudoscalar $\pi$,
provided by the eigenvalues of the three--form operator $M_{(1)^3}$.
At this point, the  matching of the spectrum with the multiplets will be complete.
\par
Since we are in particular interested in multiplet shortening, it is of
utmost importance
to pay attention to what happens with the eigenvalues in the exceptional series.
As it is clear from tables (\ref{eigenvaluesAoneform}), (\ref{eigenvaluesAthreeform}),
(\ref{eigenvaluesBthreeform}) of the eigenvalues, there are always less eigenvalues
present when the operators act on the harmonics in the exceptional
series. This is reflected into the fact that certain field components are
not present in the multiplets, thus multiplet shortening.
\par
In the next sections we give a detailed discussion of the matching
of the multiplets. Doing so we 
show that the information
that we collected about the invariant operators on the
zero form, the one--form, the two--form, the three--form and the spinor
is in perfect agreement with the group theoretical information
that was already known in \cite{multanna}.
\par
\subsection{The graviton multiplet}
As pointed out above, the graviton multiplet is the appropriate multiplet
to start with. In particular
we look at the spin--two graviton field.
The mass of the graviton is given by the eigenvalue of the scalar operator
(see eq.s (\ref{massform}):
\begin{equation}
m_h^2 = M_{(0)^3} \equiv H_0 \,.
\end{equation}
Using table (\ref{0series}) we find that its
harmonics can sit in
 all the $G$ representations of the series
\begin{eqnarray}
A_R^0, A_1^0, A_1^{*0}, A_3^0, A_3^{*0}, A_4^0, A_4^{*0}, A_6^0, A_7^0, A_8^0 \,,
\label{gravitonseries}
\end{eqnarray}
Remember that the superscripts $^0$ mean that the hypercharge is 
$Y={2\over 3}\left(M_2-M_1\right)$.
\par
Using the group--theoretical
information of the long graviton multiplet (see table \ref{longgraviton})
we find the energy and hypercharge $\left(E_0,y_0\right)$ 
of the graviton multiplet
\footnote{Remember that  $E_0, y_0$ denote the energy and
hypercharge of the Clifford vacuum of the multiplet}
\begin{eqnarray}
E_0 &=& \ft14 \sqrt{H_0+36}+\ft12\,\nonumber\\
y_0 &=&  {2\over 3}\left(M_2-M_1\right)\,,
\end{eqnarray}
and using table \ref{longgraviton} \cite{multanna} we find the energies and
hypercharges of all the fields in the multiplet. In particular, we see that the
gravitini are in $U(1)_R$ representations  $^+,~^-$, the $A,W$ vectors in
$U(1)_R$ representations $^0$, the $Z$  vectors in $U(1)_R$ representations $^0$, 
$^{++}$, $^{--}$.
From the mass of the graviton we deduce, using the mass relations
in appendix \ref{formule},
the masses of the gravitini and 
vectors present in the graviton multiplet,
\begin{eqnarray}
m_{\chi^\pm} &=& - 6 \pm \sqrt{H_0 + 36} \,,
\label{massgravitini} \\
m_A^2 &=& H_0 + 48 - 8 \sqrt{H_0 + 36} \,,
\nonumber \\
m_W^2 &=& H_0 + 48 + 8 \sqrt{H_0 + 36} \,,
\nonumber \\
m_Z^2 &=& H_0 + 32 \,.
\label{massesbosonsgraviton}
\end{eqnarray}
From equations (\ref{massform}), we predict the presence
of  the eigenvalues
$M_{(1/2)^3}=m_{\chi^\pm}$ for the spinor. Indeed, looking at
(\ref{eigenA+spinor}), we see that the two eigenvalues
$\lambda_1$ and $\lambda_2$
come from spin--$\ft32$ fields that belong to the graviton multiplet.
To find out whether there are some short graviton multiplets
present in the spectrum, we
now use table (\ref{eigenvaluesA+spinor}).
The absence of these eigenvalues $\lambda_1$ or
$\lambda_2$ in some of the exceptional series
implies the existence of a short graviton multiplet in that
particular $G'$ series.
Let us look at it more closely. For instance, for $A_2^+$ and $A_5^+$,
there is none of the eigenvalues $\lambda_1$ or $\lambda_2$.
This would imply a graviton multiplet without gravitino fields. But fortunately,
the series
$A_2$ and $A_5$ do not contain representations of $G^\prime$ in which there is a
graviton field, see (\ref{gravitonseries}). Considering the rest of
table (\ref{eigenvaluesA+spinor}) and also table (\ref{eigenvaluesA-spinor}),
we find three types of graviton multiplets:
a long graviton multiplet and two types of short graviton multiplets.
The long graviton multiplet
contains four spinors $\chi$: $\chi^{+}$ with hypercharge
$y_0 \pm 1$ and $\chi^{-}$ with hypercharge $y_0 \pm 1$.
They  are found in the $G^\prime$ representations of
$A_R, A_1, A_1^*, A_3, A_3^*, A_6, A_7$.
Then there is a short graviton multiplet in the
series $A_4$ and $A_4^*$. From
tables (\ref{eigenvaluesA+spinor}) and (\ref{eigenvaluesA-spinor}),
one sees that they contain the two $\chi^{+}$
with hypercharge $y_0 \pm 1$, but only one $\chi^{-}$,
i.e. for $A_4$ we have one $\chi^{-}$ with $y_0-1$, and for $A_4^*$ we have
one $\chi^{-}$ with $y_0+1$. We also find the massless multiplet
in $A_8$ for which none of the spin--$\ft32$ fields $\chi^{-}$ are present.
\par
At this stage, we know that the spin--$\ft32$ fields that
correspond to the eigenvalues $\lambda_1$ and $\lambda_2$ in
(\ref{eigenA+spinor}) and (\ref{eigenA-spinor}) sit
in the graviton multiplets. However, there are also
spin--$\ft32$ fields that yield the eigenvalues
$\lambda_3$ and $\lambda_4$ in
(\ref{eigenA+spinor}) and (\ref{eigenA-spinor}).
They can only be gravitini of the gravitino multiplets
in the spectrum.
So now we know the highest components of gravitino multiplets,
their energies, hypercharges and $G'$ representations.
But before we continue with the gravitino multiplet, let us look at the
vectors of the graviton multiplet.
\par
Let us consider $A$ and $W$ first.
We know that, if present, they should be in the series (\ref{gravitonseries}).
Using equations (\ref{massform}) we see that their $M_{(1)(0)^2}$
eigenvalues would then be
\begin{eqnarray}
M_{(1)(0)^2}^A &=& H_0 + 24 + 4 \sqrt{H_0 + 36} \,,
\nonumber \\
M_{(1)(0)^2}^W &=& H_0 + 24 - 4 \sqrt{H_0 + 36} \,.
\end{eqnarray}
Indeed, these eigenvalues are present, namely for
$A$ we find $\lambda_4$ and for
$W$ we find $\lambda_5$
of eq. (\ref{eigenAoneform}).
To determine whether, in the exceptional series, the
vector $A$ or the vector $W$ is present we use table
(\ref{eigenvaluesAoneform}). The absence of one of the vectors
will imply shortening of the graviton multiplet.
Studying the spin $3/2$ fields, we have found that there are
long graviton multiplets in the series $A_R, A_1, A_1^*, A_3, A_3^*, A_6, A_7$ 
and short graviton multiplets in the series $A_4, A_4^*$ . This is
confirmed here: in the former 
series both the $A$ and $W$ fields are present, 
in the latter only the field $A$ is present.
For the massless multiplet
of $A_8$ we also see that only the vector $A$ is present.
\par
Let us look at the vector $Z$ in the graviton multiplet.
We know that the $Z$ vectors should be in the same $G'$ representations
of the graviton:
\begin{equation}
A_R, A_1, A_1^*, A_3, A_3^*, A_4, A_4^*, A_6, A_7, A_8
\end{equation}
and that two $Z$ vectors should be in the series $^0$, one in the series $^{++}$ 
and one in the series $^{--}$.
For the operator $M_{(1)^2(0)}$ on the two--form
we predict, using eq.s (\ref{massform}), the presence the eigenvalue,
\begin{eqnarray}
M_{(1)^2(0)}^Z &=& H_0 +32
\,.
\end{eqnarray}
Indeed, it corresponds to $\lambda_{10}$ and $\lambda_{11}$ in (\ref{eigenAtwoform})
for the series $^0$, and $\lambda_4$ in (\ref{eigenBtwoform})
for series $^{++}$ (and $^{--}$, which are the series of
the conjugate representations of $^{++}$ ($M_2 \leftrightarrow M_1$)).
So we see that for the long graviton multiplets all the vectors $Z$ are present.
Using the fact that
\begin{eqnarray}
B_R \cup B_4 \cup B_5 \cup B_6 \cup B_7 \cup B_8 \cup B_9 \cup B_{10}
&=&
A_R \cup A_1 \cup A_1^* \cup A_3 \cup A_3^* \cup A_4 \cup A_6 \cup A_7,
\nonumber \\
B_R^* \cup B_4^* \cup B_5^* \cup B_6^* \cup B_7^* \cup B_8^* \cup B_9^* \cup B_{10}^*
&=&
A_R \cup A_1 \cup A_1^* \cup A_3 \cup A_3^* \cup A_4^* \cup A_6 \cup A_7,
\label{overlapgraviton}
\nonumber \\
\end{eqnarray}
and tables (\ref{eigenvaluesAtwoform}) and(\ref{eigenvaluesBtwoform})
we find that
for the short graviton multiplets of $A_4$ we have two $Z$'s, one with
hypercharge $y$ and one with hypercharge $y-2$; for the short graviton multiplets
of $A_4^*$ we have two $Z$'s, one with hypercharge $y$ and one with hypercharge $y+2$;
for the massless graviton multiplet we have no vectors $Z$.
\par
To determine which $\lambda_T$ fields and  scalar fields
$\phi$ are present, we use the ${\cal N}=2\rightarrow {\cal N}=1$
decomposition of the multiplets. We already know where 
$\lambda_T$ and $\phi$ are located in the long graviton multiplet
from table \ref{longgraviton}, \cite{multanna}.
From figure \ref{N1longgraviton} we see that the long ${\cal N}=2$ graviton multiplet
is made by four ${\cal N}=1$ massive multiplets: one graviton, two gravitino
and a vector multiplet. Harmonic analysis teaches us that in the short
graviton multiplet there are three gravitino fields and three vector fields.
The only possible structure of the short graviton multiplet is then the one
displayed in figure \ref{N1shortgraviton}.
\par
The  multiplet that we have found in the representation of
series $A_8$ is in fact the massless graviton multiplet.
In this case the field $A$ becomes the graviphoton.
The final structure of the short graviton multiplet
and the massless graviton multiplet
is displayed in tables \ref{shortgraviton} and 
\ref{masslessgraviton} respectively.
\subsection{The gravitino multiplet}
\par
As already previously explained, 
we know the $M_{(1/ 2)^3}$ eigenvalues 
and the $G$ representations of the spin--$\ft32$ in the
gravitino multiplet from the matching of the graviton multiplet. Their masses
are given by equations (\ref{massform}),
\begin{eqnarray}
m_{\chi^{+}}&=& - 8 + \sqrt{H_0 + 16 + \ft{32}{3}(M_2 - M_1)} = \lambda_3
\nonumber \\
m_{\chi^{-}}&=& - 8 - \sqrt{H_0 + 16 + \ft{32}{3}(M_2 - M_1)} = \lambda_4
\label{masseschiC+}
\end{eqnarray}
for series of type $^+$ and
\begin{eqnarray}
m_{\chi^{+}}&=& - 8 + \sqrt{H_0 + 16 - \ft{32}{3}(M_2 - M_1)} = \lambda_3
\nonumber \\
m_{\chi^{-}}&=& - 8 - \sqrt{H_0 + 16 - \ft{32}{3}(M_2 - M_1)} = \lambda_4
\label{gravitinisA-}
\end{eqnarray}
for series of type $^-$. Each of the above four different eigenvalues
gives rise to gravitino multiplets of different types
and/or in different $G'$ representations. Now we look at tables (\ref{eigenvaluesA+spinor})
and (\ref{eigenvaluesA-spinor}) and see that we have gravitino multiplets
for the series $A^\pm_R$ and $A^\pm_1$. We  consider the
gravitino multiplets in the series of type $^+$ only. The
gravitino multiplets in the series of type $^-$ coming from (\ref{gravitinisA-})
can be obtained be taking the conjugates of the gravitino multiplets
in the series of type $^+$.
\par
We start with $\chi^+$ in the series of type $^+$.
The energy and hypercharge $(E_0,~y_0)$ of the gravitino multiplets are given by,
\begin{eqnarray}
E_0 &=& \ft14 \sqrt{H_0 +16 +\ft{32}{3}(M_2 - M_1)} - \ft12 \,.\nonumber\\
y_0  &=& {2\over 3}\left(M_2-M_1\right)-1
\end{eqnarray}
Let us look at the vectors in the gravitino multiplets.
As we know from group theory (see table \ref{longgravitino}, \cite{multanna}) we
should find a vector
with hypercharge $y_0+1$ and energy $E_0+\ft12$, in the series $^0$.
However
group theory does not tell us whether it is the vector
$A$ or the vector $W$. But since we 
know that in series of type $^+$ we have
$m_{\chi^+} \geq -8$, we can use
the mass relations of appendix \ref{formule} to derive,
\begin{eqnarray}
m_A^2 &=& H_0 +\ft{32}{3}(M_2-M_1) + 48 - 12 \sqrt{H_0 + \ft{32}{3}(M_2-M_1)+ 16}
\end{eqnarray}
or
\begin{eqnarray}
m_W^2 &=& m_\chi^2 + 2 m_\chi + 192 \,.
\end{eqnarray}
We see from table \ref{longgravitino} that it is the $A$ vector 
which is present in the $\chi^+$ gravitino multiplet and not $W$.
Hence, comparing with the formula (\ref{massform})
in order to find $A$, we expect the following eigenvalue
\begin{equation}
M_{(1)(0)^2}^A = H_0 + \ft{32}{3}(M_2 - M_1) \,
\label{lambdaone}
\end{equation}
for the $M_{(1)(0)^2}$ operator. Looking at table (\ref{eigenAoneform})
we see that it is indeed present: $\lambda_1$.
Looking at table (\ref{eigenvaluesAoneform}) we see that it appears
in the series $A_R^0, A_1^0, A_2^0, A_3^0, A_4^0, A_5^0$.
We also find a vector $A$ with hypercharge $y_0-1$ in series $^{++}$. Indeed,
using
\begin{equation}
B_R \cup B_1 \cup B_3 \cup B_4 \cup B_6 \cup B_7
=
A_R \cup A_1 \cup A_2 \cup A_3 \cup A_4 \cup A_5 \,,
\end{equation}
we see that (\ref{lambdaone}) is an eigenvalue of the one--form operator
$M_{(1)(0)^2}$ in series $^{++}$ as given in (\ref{eigenBoneform}).
Both the spin--$1$ fields $A$ with $y_0-1$ and $y_0+1$
of the gravitino multiplet for
$\chi^+$ are present and there are no other left with eigenvalue
(\ref{lambdaone}).
For the vector $Z$ sector, we expect the presence of two states
with mass (using the table \ref{longgraviton})
\begin{equation}
m_Z^2 = H_0 + 16 + \ft{32}{3} (M_2 - M_1)
         -4 \sqrt{H_0 +16 +\ft{32}{3}(M_2 - M_1)},
\label{massz32}
\end{equation}
one in the $G$ representations of type $^0$, the other in the 
representations $^{++}$ or $^{--}$ (depending on the $G$ representation
of the gravitino).
The mass (\ref{massz32}) 
corresponds to $\lambda_7$ in (\ref{eigenAtwoform}) and $\lambda_3$
in (\ref{eigenBtwoform}). From this we see that $Z$ is present except for
series $A_5$, and series $B_7$. The series $A_5$ and $B_7$ have no overlap.
So we conclude that we have long gravitino multiplets except if the
multiplet sits in a representation of $A_5$ or $B_7$.
For the gravitino multiplet with $\chi^+$ in the series $^+$, we now look
at the mass of the scalar $\pi$,
\begin{equation}
m_\pi^2 = 16 \, (\ft14\sqrt{H_0+16+\ft{32}{3}}-1) (\ft14\sqrt{H_0+16+\ft{32}{3}}-2)
\end{equation}
From eq.s (\ref{massform})) we predict the eigenvalue
\begin{eqnarray}
M_{(1)^3}^\pi = \ft14\sqrt{H_0+16+\ft{32}{3}(M_2-M_1)}
\end{eqnarray}
which we do find as $\lambda_1$ in (\ref{eigenAthreeform})
in series $A_R^0, A_1^0, A_2^0, A_3^0, A_4^0$
(see (\ref{eigenvaluesAthreeform}))
and  as  $\lambda_1$ in (\ref{eigenBthreeform}) in the series
$B_R^{++}, B_1^{++}, $ $ B_3^{++}, B_4^{++}, B_6^{++}$ (see
(\ref{eigenvaluesBthreeform})).
So none of the fields $\pi$ with $y_0-1$ and $y_0+1$ is
present in the short gravitino multiplets with $\chi^+$ in the series of type
$^+$ .
Let us now consider the spin--$\ft12$ field $\lambda_L^+$.
Looking at the expansion (\ref{kkexpansion}),
we see that $\lambda_L$ appears in the expansion of the spinor.
So we can check whether it is present in the
gravitino multiplet with $\chi^+$ in the series $^+$.
We know its mass from group theory \cite{multanna},
\begin{eqnarray}
m_{\lambda_L^+}=-8+\sqrt{H_0+16+\ft{32}{3}(M_2-M_1)}\,,
\end{eqnarray}
so, from eq.s (\ref{massform}) we expect the
eigenvalue
\begin{eqnarray}
M_{(1/2)^3}^{\lambda_L^+}=-8-\sqrt{H_0+16+\ft{32}{3}(M_2-M_1)}
\end{eqnarray}
which we do find  as $\lambda_4$ in (\ref{eigenA+spinor})
in $A_R^+, A_1^+, A_2^+, A_3^+, A_4^+,  A_5^+$ (see
(\ref{eigenvaluesA+spinor})).
So the field $\lambda_L^+$ is present in both long and short
gravitino multiplets with hypercharge $y_0$. In fact
it has to be there since it provides the Clifford vacuum
of the representation.
For the short gravitino multiplets
we have found which of  the fields $\phi$ and $\lambda_T$ are 
present by using the ${\cal N}=2\rightarrow {\cal N}=1$
decomposition (see figures \ref{N1longgravitino},\ref{N1shortgravitino}) and 
by calculating the norms of the states using
creation and annihilation operators \cite{frenico,multanna}. 
The result is displayed in table \ref{shortgravitino}.
\par
Let us consider $\chi^-$ for the series of type $^+$.
It has mass $m_{\chi^-}$ from (\ref{masseschiC+}).
The energy and hypercharge $(E_0,~y_0)$ of the multiplet are
\begin{eqnarray}
E_0 &=& \ft14\sqrt{H_0+16+\ft{32}{3}(M_2-M_1)}+\ft32\,.\nonumber\\
y_0 &=& {2\over 3}\left(M_2-M_1\right)-1\,.
\end{eqnarray}
We now have $m_{\chi}\leq -8$.
So, using the mass relations for $W$ we find
\begin{eqnarray}
m_W^2 &=& H_0 +\ft{32}{3}(M_2-M_1) + 48 + 12 \sqrt{H_0 + \ft{32}{3}(M_2-M_1)+ 16}\,.
\end{eqnarray}
Thus in this case it is $W$ that is present and not $A$.
We find the same eigenvalue (\ref{lambdaone}), so we conclude that $W$
is present in all types of gravitino multiplets with $\chi^-$ in series of type $^+$.
For $Z$ we have
\begin{eqnarray}
m_Z^2 = H_0 +16 +\ft{32}{3}(M_2-M_1)+4\sqrt{H_0+16+\ft{32}{3}(M_2-M_1)}
\end{eqnarray}
which, according to eq.s (\ref{massform}),
 has to be an eigenvalue of the two--form mass operator. Indeed,
for series of type $^0$ it corresponds to $\lambda_6$, which is present in series
$A_R, A_1, A_2, A_3, A_4, A_5$ (see (\ref{eigenvaluesAtwoform})).
Notice that these  are the same series of representations as the ones
in which we found $\chi^+$. For the series $^{++}$ we find $\lambda_2$, which is
present in the series $B_R, B_1, B_3, B_4, B_6, B_7$
(see (\ref{eigenvaluesBtwoform})), which are again the same series
of representations
as for $\chi^+$. The fields $\pi$ present have mass,
\begin{equation}
m_\pi^2 = 16 \, (-\ft14\sqrt{H_0+16+\ft{32}{3}(M_2-M_1)}-1)
(-\ft14\sqrt{H_0+16+\ft{32}{3}(M_2-M_1)}-2)
\end{equation}
So we predict the eigenvalue
\begin{equation}
M_{(1)^3}^{\pi}= - \ft14 \sqrt{H_0+16+\ft{32}{3}(M_2-M_1)}
\end{equation}
Indeed it is $\lambda_3$ in (\ref{eigenAthreeform}),
present in the series
$A_R, A_1, A_2, A_3, A_4, A_5$
(\ref{eigenvaluesAthreeform}) and
$\lambda_2$ in (\ref{eigenvaluesBthreeform}), present
in the series $B_R, B_1, B_3, B_4,$ $ B_6, B_7$
(\ref{eigenBthreeform}).
We conclude that all the gravitino multiplets with $\chi^-$
are long gravitino multiplets.
\subsection{The vector multiplet}
\par
What are the vector field we have been left with?
They have to be the highest components of the vector multiplets.
Well, we have a multiplet with highest component vector $A$ with
eigenvalue $\lambda_5$ in (\ref{eigenAoneform}). We have
a vector multiplet with highest vector component $W$ with
eigenvalue $\lambda_4$ in (\ref{eigenAoneform}). We have
some vector multiplets with highest vector component $Z$
with eigenvalues $\lambda_3$ in (\ref{eigenAtwoform}),
$\lambda_5$ in (\ref{eigenBtwoform}) and
$\lambda_5^*$ in the series $^{--}$. All  these eigenvalues give
rise to the existence of different types of
vector multiplets in different representations of $G^\prime$.
\par
Let us start with $A$. We call this the $A$--vector multiplet.
It has eigenvalue $\lambda_5$ in (\ref{eigenAoneform}).
Its energy and hypercharge are
\begin{eqnarray}
E_0 &=& \ft14 \sqrt{H_0 + 36} - \ft32 \,\nonumber\\
y_0 &= & {2\over 3}\left(M_2-M_1\right)
\end{eqnarray}
and the mass of the field component $A$ is
\begin{equation}
m_A^2 = H_0 + 96 - 16\,\sqrt{H_0 + 36} \,.
\label{massAvector}
\end{equation}
This eigenvalue is present in  the series
$A_R^0, A_1^0, A_1^{*0}, A_3^0, A_3^{*0}, A_6^0, A_7^0$. We now figure
out for which of these there is shortening. From the table in
\cite{multanna} we see that
$\pi$ has the same mass as $A$ (\ref{massAvector}), and using
eq.s (\ref{massform}) we conclude that we should find the eigenvalues
\begin{eqnarray}
M_{(1)^3}^\pi = \ft14 \sqrt{H_0 + 36} -\ft12 \,,
\end{eqnarray}
which is present: $\lambda_5$ in
$A_R^0, A_1^0, A_1^{0*}, A_3^0, A_3^{0*}, A_6^0, A_7^0$
(\ref{eigenAthreeform}) (\ref{eigenvaluesAthreeform}).
It is also present as $\lambda_3$ in
$B_R^{++}, B_4^{++}, B_5^{++}, B_6^{++}, B_7^{++}, B_9^{++}$
(\ref{eigenBthreeform}) (\ref{eigenvaluesBthreeform}). Considering
(\ref{overlapgraviton}) this seems strange at
first sight. However, what happens is that here we discover a
scalar $\pi$ in the series $A_4$ of a
hypermultiplet. We can see this as follows.
Suppose the eigenvalue were also present in series $B_8$
and series $B_{10}$. Then the eigenvalue $\lambda_3$ would
appear in the representations of $B$ that are on the right--hand side
of (\ref{overlapgraviton}). So we would find the field $\pi$ in the
$G^\prime$ representations
$A_R, A_1, A_1^*, A_3, A_3^*, A_6, A_7$ and in $A_4$, with $Y=\ft23(M_2-M_1)-2$.
The series $A_4$ and $B_8$ and $B_{10}$ have no overlap.
Consequently,
the $\pi$ in $A_4$ can not belong to the $A$--vector multiplet and
thus has to be a scalar of a hypermultiplet. Similarly, we find $\pi$
in $B_R^{--*}, B_4^{--*} B_5^{--*}, B_6^{--*},$ $B_7^{--*}, 
B_8^{--*}, B_9^{--*}, B_{10}^{--*}$. With the
same reasoning, we conclude that $\pi$ in $A_4^*$ with $Y=\ft23(M_2-M_1)+2$
has to be a scalar of some hypermultiplet.
However, $\lambda_3$ does not sit in the series $B_8, B_8^*, B_{10}, B_{10}^*$.
So we conclude that that we get shortening in these series. Now we
get different types of short vector multiplets. This is due to fact the
$B_8$ and $B_8^*$ have overlap, namely if $M_1=M_2=1, J=0$ and
that also $B_{10}$ and $B_{10}^*$ have overlap, namely
for the representation $M_1=M_2=0, J=1$.
For the representations in the series $B_8$ and $B_{10}$
with $M_1 > M_2 = 1$,
we find that the field $\pi$ with hypercharge $y-2$
in the long vector multiplet decouples. The representations
\begin{eqnarray}
&M_1=M_2, &J=1\nonumber\\
&M_1=M_2=1, &J=0\label{shvecreps}
\end{eqnarray}
yield massless  vector multiplets.
They contain the vectors that gauge $SU(2)$ and $SU(3)$
respectively.
\par
Let us now figure out whether we can learn something about the
presence of $\phi$, $S$ and $\Sigma$ in the $A$--vector multiplet. The table in
\cite{multanna} gives the mass,
\begin{eqnarray}
m_{\phi,S/\Sigma}^2 = 16 \, E_0 (E_0 + 1) = H_0 + 48 - 4 \sqrt{H_0 + 36} \,.
\end{eqnarray}
Looking at eq.s (\ref{massform}), we see that the entry in the table can not be
$S$ or $\Sigma$, but has to be $\phi$. If we look at the other
$\phi, S/\Sigma$ in the table with  mass
\begin{eqnarray}
m_{\phi,S/\Sigma}^2 = 16 \, (E_0 - 2)(E_0 - 1)
                    = H_0 + 176 - 24 \sqrt{H_0 + 36} \,,
\end{eqnarray}
we see that it is the mass for the field $S$.
So at this place in the table we
find the field $S$. The field $S$ is found in the series
$A_R^0, A_1^0, A_1^{0*}, A_3^0, A_3^{0*}, 
A_4^0, A_4^{0*}, A_6^0, A_7^0, A_8^0$. So
it is always present in the $A$--vector multiplets. Besides, we
get some extra $S$--fields that are to be put in the hypermultiplets
in the series $A_4, A_4^*, A_8$.
\par
To conclude the discussion of the $A$ vector multiplet, there is
shortening of $A$--vector multiplets in series
$B_8, B_{8}^*$ and $B_{10}, B_{10}^*$.
In the representation (\ref{shvecreps}) there are massless vector
multiplets, in the other $B_8,B_8^*,B_{10},B_{10}^*$ representations
there are short vector multiplets.
The $\phi$ and $\lambda_T$ contents of the short vector multiplets
can be determined by using the ${\cal N}=2\rightarrow {\cal N}=1$
decomposition (see pictures \ref{N1longvector},\ref{N1shortvector}).
The structure of the long vector multiplet and
the short vector multiplet
is displayed in table \ref{longvector} and \ref{shortvector}
respectively.
\par
Let us now consider the vector multiplet with highest vector component $W$.
We will call this the $W$-- vector multiplet.
We expect eigenvalue $\lambda_5$
in (\ref{eigenAoneform}) and (\ref{eigenvaluesAoneform}),
which we find in series
$A_R^0, A_1^0, A_1^{0*}, A_3^0, A_3^{0*},
A_4^0, A_4^{0*}, A_6^0, A_7^0, A_8^0$.
This multiplet has energy and hypercharge,
\begin{eqnarray}
E_0 &=& \ft14 \sqrt{H_0 + 36} +\ft52 \,,\nonumber \\
y_0 &=& {2\over 3}\left(M_2-M_1\right),
\end{eqnarray}
the $W$ field has mass
\begin{equation}
m_W^2 = H_0 + 96 + 16 \sqrt{H_0 +36} \,.
\end{equation}
For the fields $\pi$, we expect to find the eigenvalues $\lambda_6$ in
series $A_R^0, A_1^0, A_1^{*0}, A_3^0, A_3^{*0}, A_4^0, A_4^{*0}, A_6^0,$ $ A_7^0, A_8^0$
(\ref{eigenAthreeform})(\ref{eigenvaluesAthreeform}), and
$\lambda_4$ in series
$B_R^{++}, B_4^{++}, B_5^{++}, B_6^{++}, B_7^{++}, B_8^{++}, B_9^{++}, 
B_{10}^{++}, B_{11}^{++}$ 
(\ref{eigenBthreeform}) (\ref{eigenvaluesBthreeform}),
and $\lambda_4^*$ in series
$B_R^{--}, B_4^{--}, B_5^{--}, B_6^{--}, B_7^{--}, B_8^{--}, B_9^{--}, 
B_{10}^{--}, B_{11}^{--}$. 
Using
\begin{eqnarray}
B_{11} &=& A_4 \cup A_8 \,, \nonumber \\
B_{11}^* &=& A_4^* \cup A_8 \,,
\end{eqnarray}
and (\ref{overlapgraviton}), we see that all these $^0,~^{++},$ and $^{--}$ series coincide.
Thus all the
fields $\pi$ in the table of \cite{multanna} are
always present and we find no fields $\pi$ that have to be put in
other multiplets. So the $W$--vector multiplet is always long.
Which of the fields $\phi, S/\Sigma$ are present? Let us look at
$\phi, S/\Sigma$ with mass
\begin{eqnarray}
m_{\phi, S/\Sigma}^2 = 16 \, E_0 (E_0 + 1) = H_0 + 176 + 24 \sqrt{H_0 + 36}
\,.
\end{eqnarray}
From eq.s (\ref{massform}) we see
that it is the field $\Sigma$ that is  present in the series
$A_R^0, A_1^0, A_1^{0*},$ $ A_3^0, A_3^{0*}, A_4^0,$ $ A_4^{0*},
A_6^0,$ $ A_7^0, A_8^0$.
So this confirms that there is no shortening and we do not
find any extra fields $\Sigma$ that are to be put in the
hypermultiplets.
Let us look at
$\phi, S/\Sigma$ with mass
\begin{eqnarray}
m_{\phi, S/\Sigma}^2 = 16 \, (E_0 - 2) (E_0 - 1) = H_0 + 48 + 8 \sqrt{H_0 + 36}
\,.
\end{eqnarray}
This can only be the field $\phi$.
So we conclude that the $W$--vector multiplets are always long vector multiplets.
And there are no scalar left that have to be put in hypermultiplets.
Its structure is displayed in table \ref{longvector}.
\par
Let us now look at the $Z$--vector multiplet with eigenvalue $\lambda_3$
in series $A_R, A_1, A_1^*, A_6, A_8$ (\ref{eigenAtwoform})
(\ref{eigenvaluesAtwoform}). 
The multiplet has energy and hypercharge
\begin{eqnarray}
E_0 &=& \ft14\sqrt{H_0 + 4} + \ft12 \,,\nonumber \\
y_0 &=& {2\over 3}\left(M_2-M_1\right),
\end{eqnarray}
the field $Z$ has mass
\begin{equation}
m_Z^2 = H_0 \,.
\end{equation}
What about the two fields $\pi$? Let us look at $\pi$ with mass
\begin{eqnarray}
m_\pi^2 = 16\, E_0 (E_0 + 1)
        = H_0 + 16 +  \sqrt{H_0 + 4}
\end{eqnarray}
From eq.s (\ref{massform}) we expect there to be
$\lambda_7$ in (\ref{eigenAthreeform}). Indeed, it is present in series
$A_R^0, A_1^0, A_1^{*0}, A_6^0$. So we get shortening in the singlet representation 
$A_8$. For $\pi$ with mass
\begin{eqnarray}
m_\pi^2 = 16\, (E_0 -2)(E_0 - 1) \,,
\end{eqnarray}
we find $\lambda_8$ in series $A_R,^0 A_1^0, A_1^{0*}, A_6^0, A_8^0$.
So finally, we conclude that for this type of $Z$--vector multiplet
(with $\lambda_3$ in (\ref{eigenAtwoform})) there is
shortening in series $A_8$, which yields the
massless Betti multiplet.
The structure of the long $Z$--vector multiplet and the
massless Betti multiplet is displayed in table \ref{longvector}
and \ref{masslessvector} respectively.
\par
Let us now look at the $Z$--vector multiplet with $\lambda_5$ in
(\ref{eigenBtwoform}). It appears in series $B_R, B_1, B_2$
(\ref{eigenvaluesBtwoform}). 
The multiplet has energy and hypercharge
\begin{eqnarray}
E_0 &=& \ft14 \sqrt{H_0 + \ft{64}{3}(M_2- M_1) -28} +\ft12\, \nonumber\\
y_0 &=& {2\over 3}\left(M_2-M_1\right)-2,
\end{eqnarray}
the field $Z$ has mass
\begin{equation}
m_Z^2 = H_0 + \ft{64}{3}(M_2-M_1) -32 \,.
\end{equation}
What about the presence of the fields $\pi$?
For $\pi$ with mass
\begin{eqnarray}
m_\pi^2 = 16 \, (E_0-2)(E_0-1) \,,
\end{eqnarray}
we expect the eigenvalue $\lambda_5$ in (\ref{eigenBthreeform}), which
is found in the series $B_R^{++}, B_1^{++}, B_2^{++}$ (\ref{eigenvaluesBthreeform}).
For $\pi$ with mass
\begin{eqnarray}
m_\pi^2 = 16 \, E_0(E_0+1) \,,
\end{eqnarray}
we expect $\lambda_6$
in (\ref{eigenBthreeform}), which
is found in the series $B_R^{++}, B_1^{++}, B_2^{++}$ (\ref{eigenvaluesBthreeform}).
So we conclude that for the $Z$--vector multiplet (with vector $Z$ with
eigenvalue $\lambda_5$ in (\ref{eigenBtwoform})), there is never
shortening. We do not find extra scalars that are to be put
in hypermultiplets either.
The structure of this long $Z$ vector multiplet
is displayed in table \ref{longvector}.
\par
For the $Z$--vector multiplet with $\lambda_5^*$ in series
$B_R^*, B_1^*, B_2^*$, one just takes the conjugate of the previous results.
\subsection{The hypermultiplet}
After having put the scalars $\pi$ in the right places in the
graviton, the gravitino and the vector multiplet, we are only left with scalars
$\pi$ in series $A_4^0$ and $A_4^{0*}$ and $S$ in series $A_4,^0 A_4^{0*}, A_8^0$.
\par
So for each representation of $A_4$ we find a hypermultiplet
with energy
\begin{eqnarray}
E_0 = \ft14 \sqrt{H_0 + 36} -\ft32
\end{eqnarray}
containing
the field $\pi$ with hypercharge $Y=\ft23(M_2-M_1)-2$ and mass
\begin{eqnarray}
m_\pi^2= H_0 + 96 - 16\,\sqrt{H_0+36}
\end{eqnarray}
and the field $S$ with $Y=\ft23(M_2 - M_1)$ and mass
\begin{eqnarray}
m_S^2 = H_0 +176 - 24\, \sqrt{H_0+36}
\end{eqnarray}
The scalars
of this hypermultiplet are complete if we add the scalars $\pi$ and $S$
of $A_4^*$, which are in fact the complex conjugates of the scalars in
$A_4$. From the eigenvalues of the operator $M_{(1/ 2)^3}$
we find the $\lambda_L$ necessary to fill all the hypermultiplets.
The structure of the hypermultiplets is displayed in
the table \ref{hyper}. 
\par
In order to correctly match the fields with the multiplets, it is
important to note that in the singlet $G$ representation $M_1=M_2=J=Y=0$
the scalar $S$ is absent. This is due to the fact that, from the Kaluza Klein expansion of
the eleven-dimensional field $h_{mn}\left(x,y\right)$, the scalar $S$ 
appears in the expressions $(6-\sqrt{M_{(0)^3}+36})S^I\left(x\right)$ and
${\cal D}_{(m}{\cal D}_{n)}(2+\sqrt{M_{(0)^3}+36})S^I\left(x\right)$. 
The coefficient of the former, $6-\sqrt{M_{(0)^3}+36}$, disappears
in the singlet representation. The latter become a pure gauge term,
due to the freedom of coordinate reparametrization, being the graviton
in the singlet $G$ representation the massless graviton.
\vskip 1cm
At this point we have done the complete matching of the multiplets 
with the spectrum of Laplace Beltrami operators. It is reassuring that all the fields
we have found have been organized in ${\cal N}=2~AdS_4$ multiplets.
An important result is that we have established the existence of short multiplets.
From the expressions of the energies and hypercharges
$(E_0,y_0)$ we have found, we can easily derive that
\begin{itemize}
\item for all the long multiplets
\[
E_0>\left|y_0\right|+s_0+1
\]
\item for all the short graviton, gravitino and vector multiplets
\[
E_0=\left|y_0\right|+s_0+1
\]
\item for all the hypermultiplets
\[
E_0=\left|y_0\right|\ge {1\over 2}
\]
\item for all the massless multiplets
\[
E_0=s_0+1~~y_0=0.
\]
\end{itemize}
This confirms the algebraic lore on multiplet shortening from the literature
\cite{frenico,multanna}.
\par
\section{Outlook and plan for future development}
In this paper we have have constructed the complete Kaluza Klein
spectrum of the $AdS_4 \times M^{111}$ compactification, organizing
it into $Osp(2\vert 4)$ supermultiplets. As stressed in the
introduction this must be regarded both as the completion of an
outstanding problem and as a first essential step along a further
research plan that can be outlined as follows:
\begin{itemize}
  \item{ Construction of the three ${\cal N}=2$ multiplet spectra:
  \begin{enumerate}
  \item   $Osp(2\vert 4) \times SU(3) \times SU(2) $ for $M^{111}$
  \item   $Osp(2\vert 4) \times SO(5) $ for $V_{5,2}$
  \item    $Osp(2\vert 4) \times SU(2)^3 $ for $Q^{111}$
\end{enumerate}}
  \item Construction of the single ${\cal N}=3$ multiplet spectrum,
  corresponding to the $N^{010}$ space. In this case the Kaluza Klein
  states are organized in $Osp(3\vert 4) \times SU(3)$ multiplets.
  \item Search of the appropriate K\"ahlerian conifold that leads to the
   three dimensional conformal field theory corresponding to each of
   the above cases.
   \item Test of the $AdS/CFT$ conjecture at the level of both the
   long and the short multiplets.
   \item Determination of the effective low energy ${\cal N}=2$
   lagrangians, namely search of the appropriate Special K\"ahler
   manifold and of its geometrical interpretation as moduli space of
   an appropriate structure.
   \item As for the ${\cal N}=3$ theory, here supersymmetry is
   already strong enough to predict the structure of the effective
   Lagrangian. As shown in \cite{n3} the $3\times 8$ complex scalars
   of the $N^{010}$ compactification will fill the coset manifold:
\begin{equation}
  \frac{SU(3,8)}{SU(3) \times SU(8) \times U(1)}
\label{su38}
\end{equation}
\end{itemize}
The final goal of our research plan is to provide new non trivial
checks of the $AdS/CFT$ correspondence in cases where the spectrum of
primary conformal operators is not too much restricted by
supersymmetry. In particular we are interested in exploring
properties of non trivial three--dimensional conformal field theories
by means of classical   supergravity on anti de Sitter
compactifications $AdS_4 \times X_7$ where $X_7$ are suitable
7-manifolds with interesting geometrical structures and to establish
the relation between the quantum aspects of the 3D field theory and
the geometry of $X_7$.
\appendix
\section{Conventions used in the calculations}
\par
\label{convenzioni}
The
 Gell--Mann matrices are:
\begin{eqnarray}
    \lambda_1    =
    \left(\begin{array}{ccc}
          0   &  1  & 0 \\
          1   &  0   & 0 \\
          0   &  0  & 0 \\
          \end{array}
    \right)\,, \quad
    \lambda_2    =
    \left(\begin{array}{ccc}
          0   & -i   & 0 \\
          i   &  0  & 0  \\
          0   &  0  & 0  \\
          \end{array}
    \right)\,, \quad
    \lambda_3    =
    \left(\begin{array}{ccc}
          1   & 0   & 0 \\
          0   & -1  & 0 \\
          0   & 0   & 0 \\
          \end{array}
    \right)\,,
\nonumber \\
    \lambda_4    =
    \left(\begin{array}{ccc}
           0  & 0   & 1 \\
           0  & 0   & 0 \\
           1  &  0  & 0 \\
          \end{array}
    \right)\,, \quad
    \lambda_5    =
    \left(\begin{array}{ccc}
           0  & 0   & -i \\
           0  & 0   &  0\\
           i  & 0   &  0\\
          \end{array}
    \right)\,,\quad
    \lambda_6    =
    \left(\begin{array}{ccc}
           0  & 0   & 0 \\
           0  & 0   & 1 \\
            0 & 1   & 0 \\
          \end{array}
    \right)\,, \quad
\nonumber \\
    \lambda_7    =
    \left(\begin{array}{ccc}
           0  & 0   & 0 \\
           0  & 0   & -i \\
            0 & i   & 0 \\
          \end{array}
    \right)\,,
    \lambda_8    = \ft{1}{\sqrt{3}}
    \left(\begin{array}{ccc}
           1  &  0  & 0 \\
           0  &  1  & 0 \\
           0  &  0  & -2 \\
          \end{array}
    \right)\,.
\nonumber \\
\end{eqnarray}
The Pauli matrices are:
\begin{equation}
\sigma_1 =
    \left(\begin{array}{cc}
           0  &  1  \\
           1  &  0  \\
          \end{array}
    \right) \,,
\sigma_2 =
    \left(\begin{array}{cc}
           0  &  -i  \\
           i  &  0  \\
          \end{array}
    \right) \,,
\sigma_3 =
    \left(\begin{array}{cc}
           1  &  0  \\
           0  &  -1  \\
          \end{array}
    \right) \,.
\end{equation}
Furthermore, in order to follow the notations of \cite{spectfer} we define
\begin{equation}
\sigma_{\ddot{3}}\equiv -\sigma_{3}.
\end{equation}
The generators of $G=SU\left(3\right)^c\times SU\left(2\right)^w\times U\left(1\right)$ are
\footnote{the subfix $_{\stackrel{.\!.\!.}{3}}$
has been given in order to follow the notations of
\cite{spectfer}}:
\begin{eqnarray}
SU\left(3\right)^c : &~~~&{i\over 2}\lambda_1,\dots,{i\over 2}\lambda_8\nonumber\\
SU\left(2\right)^w : &~~~&{i\over 2}\sigma_1,\dots,{i\over 2},\sigma_3\nonumber\\
U\left(1\right):         &~~~&i Y_{\stackrel{.\!.\!.}{3}}
 \nonumber
\end{eqnarray}
The orthogonal decomposition gives
\begin{equation}
\IG=\IH\oplus\IK
\end{equation}
where $\IH$ is a subalgebra of $\IG$, and $\IK$ is a representation of $\IH$.
\par
The generators of $H=SU\left(2\right)^c\times U\left(1\right)'\times U\left(1\right)''
$ are:
\begin{eqnarray}
SU\left(2\right)^c : &~~~&{i\over 2}\lambda_{\dot{m}}={i\over 2}\lambda_1,\dots,{i\over 2}
\lambda_3\nonumber\\
U\left(1\right)':         &~~~& Z'=\sqrt{3}i\lambda_8+i\sigma_{\ddot{3}}-4i
Y_{\stackrel{.\!.\!.}{3}}\nonumber\\
U\left(1\right)'':         &~~~& Z''=-{\sqrt{3}\over 2}i\lambda_8+{3\over 2}i\sigma_{
\ddot{3}}\nonumber
\end{eqnarray}
so the generators of the orthogonal space $\IK$ are
\begin{eqnarray}
{i\over 2}\lambda_A&=&{i\over 2}\lambda_4\dots{i\over 2}\lambda_7,\nonumber\\
\sigma_{m}&=&{i\over 2}\sigma_1,{i\over 2}\sigma_2\nonumber\\
Z&=&{\sqrt{3}\over 2}i\lambda_8+{1\over 2}i\sigma_{\ddot{3}}+iY_{\stackrel{.\!.\!.}{3}}\,.
\end{eqnarray}
\par
Due to this decomposition we divide the indices into six groups:
\begin{eqnarray}
&\dot{m},\dot{n}=1,2,3,\nonumber\\
&{\stackrel{.\!.\!.}{3}},\nonumber\\
&m,n=1,2,\nonumber\\
&\ddot{3},\\
&A,B,C=4,5,6,7,\nonumber\\
&8\nonumber
\end{eqnarray}
\par
Other indices used in this paper are:
\begin{eqnarray}
\Sigma,\Lambda: &~~~&{\rm indices~of~the~adjoint~representation~of~}G\nonumber\\
\alpha,\beta:&~~~&{\rm indices~of~the~vector~representation~of~}SO\left(7\right)\nonumber\\
i,j:&~~~&{\rm indices~of~the~vector~representation~of~}SU\left(2\right)^c
\end{eqnarray}
\par
Out conventions for the $\varepsilon$ tensors are the following:
\begin{equation}
\begin{array}{lll}
SU\left(2\right)^W: & \varepsilon^{mn} & \varepsilon^{12}=-1 \\
SU\left(3\right)^c:  & \varepsilon^{\dot{m}\dot{n}\dot{r}} &
 \varepsilon^{\dot{1}\dot{2}\dot{3}}=1 \\
SU\left(2\right)^c:  & \varepsilon^{\dot{m}\dot{n}} &
\varepsilon^{\dot{1}\dot{2}}=1 \\
SO\left(7\right)^c:  & \varepsilon^{\alpha\beta\gamma\delta\mu\nu\rho} &
\varepsilon^{1234567}=-1 
\end{array}
\end{equation}
\par
\section{Main formulas of harmonic analysis}
\label{formule}
The fields of the four dimensional Kaluza Klein theory are defined in  the expansion
the eleven dimensional fields in $M^{111}$ harmonics \cite{castdauriafre}, \cite{univer}:
\begin{eqnarray}
h_{mn}\left(x,y\right)
         &=& \Big( h_{mn}^I
\left(x\right)
                 - \frac{3}{M_{(0)^3}+32}{\cal D}_{(m}{\cal D}_{n)}
                 \left[(2+\sqrt{M_{(0)^3}+36}\,) S^I\left(x\right)\right.+ \nonumber\\
        & &  \left. (2-\sqrt{M_{(0)^3}+36}\,)
                 \Sigma^I
\left(x\right)\right]
                 + \ft54 \delta_{mn}
                   \left[(6-\sqrt{M_{(0)^3}+36}\,)S^I\left(x\right)+\right.\nonumber\\
        & &    \left. (6+\sqrt{M_{(0)^3}+36}\,)\Sigma^I
\left(x\right)\right]
            \Big) \, Y^I \left(y\right)\,,
\nonumber \\
h_{ma}\left(x,y\right)
            &=& \big[
                (\sqrt{M_{(1)(0)^2}+16}-4)A_m^I\left(x\right) +
                (\sqrt{M_{(1)(0)^2}+16}+4)W_m^I
\left(x\right)\big] \, Y^I_a\left(y\right) \,,
\nonumber \\
h_{ab}\left(x,y\right)
           &=& \phi^I\left(x\right) Y^I_{(ab)}\left(y\right) - \delta_{ab}
           \left[(6-\sqrt{M_{(0)^3}+36}) S^I\left(x\right)+\right. \nonumber \\
 & & \left.  (6+\sqrt{M_{(0)^3}+36})\Sigma^I
\left(x\right)\right] \, Y^I\left(y\right) \,,
\nonumber \\
a_{mnr}\left(x,y\right)
             &=& 2 \, \varepsilon_{mnrp} \,{\cal D}_p (S^I\left(x\right)+\Sigma^I\left(x\right))
              Y^I\left(y\right) \,,
\nonumber \\
a_{mna}\left(x,y\right)
             &=& \ft23 \, \varepsilon_{mnrs} \,
            ({\cal D}_r A_s^I\left(x\right) + {\cal D}_r W_s^I\left(x\right))\, Y_a^I\left(y\right) \,,
\nonumber \\
a_{mab}\left(x,y\right)
             &=& Z_m^I\left(x\right) Y^I_{[ab]}\left(y\right) \,,
\nonumber \\
a_{abc}\left(x,y\right)
            &=& \pi^I\left(x\right) Y^I_{[abc]}\left(y\right) \,,
\nonumber \\
\psi_m\left(x,y\right)
            &=& \Big(
                \chi_m^I\left(x\right) +\frac{\ft47 M_{(1/2)^3}+8}{M_{(1/2)^3}+8}
                  \big[D_m \lambda_L^I\left(x\right) \big]_{3/2}
-\nonumber\\
 &&    (6+\ft37 M_{(1/2)^3})\gamma_5\gamma_m \lambda_L^I
\left(x\right)
           \Big) \, \Xi^I\left(y\right) \,,
\nonumber \\
\psi_a &=& \lambda_T^I\left(x\right) \Xi_a^I\left(y\right) + \lambda_L^I\left(x\right)
                \big[ \nabla_a \Xi ^I\left(y\right) \big]_{3/2}
\,.\label{kkexpansion}
\end{eqnarray}
where we call $x$ the coordinates of the four dimensional space, and $y$ the coordinates
of the internal compact space.
\par
The convention for the names of the eigenvalues of the  transverse harmonics are the 
following:
\begin{equation}
\begin{array}{|c|c|}
\hline
{\rm Harmonic}&{\rm Eigenvalue}\\\hline
{\cal Y}^I & M_{\left(0\right)^3} \\
{\cal Y}^I_{a} & M_{\left(1\right)\left(0\right)^2}\\
{\cal Y}^I_{\left[ab\right]} & M_{\left(1\right)^2\left(0\right)}\\
{\cal Y}^I_{\left[abc\right]} & M_{\left(1\right)^3}\\
{\cal Y}^I_{\left(ab\right)} & M_{\left(2\right)\left(0\right)^2}\\
\Xi^I & M_{\left(1\over 2\right)^3} \\
\Xi^I_a & M_{\left(3\over 2\right)\left(1\over 2\right)^2}\\
\hline
\end{array}
\end{equation}
In the previous expansion, to each eigenvalue of a harmonic does correspond
an $AdS_4$ field of mass:
\begin{eqnarray}
m_h & = & M_{\left(0\right)^3}\,, \nonumber \\
m_{\Sigma}^2&=& M_{\left(0\right)^3} +176+24\sqrt{ M_{\left(0\right)^3}+36}\,,\nonumber\\
m_S^2 &=& M_{\left(0\right)^3} +176-24\sqrt{ M_{\left(0\right)^3}+36}\,,\nonumber\\
m_{\phi}^2 &=& M_{\left(2\right)\left(0\right)^2} \,,\nonumber\\
m_{\pi}^2 &=& 16\left( M_{\left(1\right)^3}-2\right)\left( M_{\left(1\right)^3}-1\right)
\,,\nonumber\\
m_W^2 &=& M_{\left(1\right)\left(0\right)^2} + 48 + 12 
\sqrt{M_{\left(1\right)\left(0\right)^2}+16}\,,\nonumber\\
m_A^2 &=& M_{\left(1\right)\left(0\right)^2} + 48 - 12 
\sqrt{M_{\left(1\right)\left(0\right)^2}+16}\,,\nonumber\\
m_Z^2 & = &  M_{\left(1\right)^2\left(0\right)} \,,\nonumber\\
m_{\lambda_L} & = & -\left( M_{\left(1\over 2\right)^3} +16\right)\,,
\nonumber\\
m_{\lambda_T} & = & M_{\left(3\over 2\right)\left(1\over 2\right)^2}+8\,,\nonumber\\
m_{\chi} & = & M_{\left(1\over 2\right)^3}\,.\label{massform}
\end{eqnarray}
\par
\begin{figure}
\label{molecola}
\centering
\begin{picture}(268,200)
\put (15,100){\circle{30}}
\put (12,97){\shortstack{\large{$2$}}}
\put (12,72){\shortstack{\large{$h$}}}
\put (60,100){\vector (-1,0){30}}
\put (75,100){\circle{30}}
\put (72,97){\shortstack{\large{$3\over 2$}}}
\put (72,72){\shortstack{\large{$\chi$}}}
\put (85.61,110.61){\vector (1,1){21.22}}
\put (85.61,89.39){\vector (1,-1){21.22}}
\put (117.43,142.43){\circle{30}}
\put (114.43,139.43){\shortstack{\large{$1^-$}}}
\put (114.43,114.43){\shortstack{\large{$Z$}}}
\put (117.43,57.57){\circle{30}}
\put (114.43,54.57){\shortstack{\large{$1^+$}}}
\put (104.43,29.57){\shortstack{\large{$A,W$}}}
\put (132.43,142.43){\vector (1,0){30}}
\put (132.43,57.57){\vector (1,0){30}}
\put (125.04,68.18){\vector (2,3){42}}
\put (177.43,142.43){\circle{30}}
\put (174.43,139.43){\shortstack{\large{$1\over 2$}}}
\put (174.43,114.43){\shortstack{\large{$\lambda_T$}}}
\put (177.43,57.57){\circle{30}}
\put (174.43,54.57){\shortstack{\large{$1\over 2$}}}
\put (174.43,29.57){\shortstack{\large{$\lambda_L$}}}
\put (192.43,142.43){\vector (1,0){30}}
\put (188.04,131.82){\vector (1,-1){32.5}}
\put (188.04,68.18){\vector (1,1){32.5}}
\put (192.43,57.57){\vector (1,0){30}}
\put (237.43,142.43){\circle{30}}
\put (234.43,139.43){\shortstack{\large{$0^+$}}}
\put (259.43,139.43){\shortstack{\large{$\phi$}}}
\put (237.43,100){\circle{30}}
\put (234.43,97){\shortstack{\large{$0^-$}}}
\put (259.43,97){\shortstack{\large{$\pi$}}}
\put (237.43,57.57){\circle{30}}
\put (234.43,54.57){\shortstack{\large{$0^+$}}}
\put (259.43,54.57){\shortstack{\large{$S,\Sigma$}}}
\end{picture}
\caption{Supersymmetry relations between the
Kaluza Klein fields: for every couple of fields linked
by an arrow there is a mass relation descending by
supersymmetry.}
\end{figure}
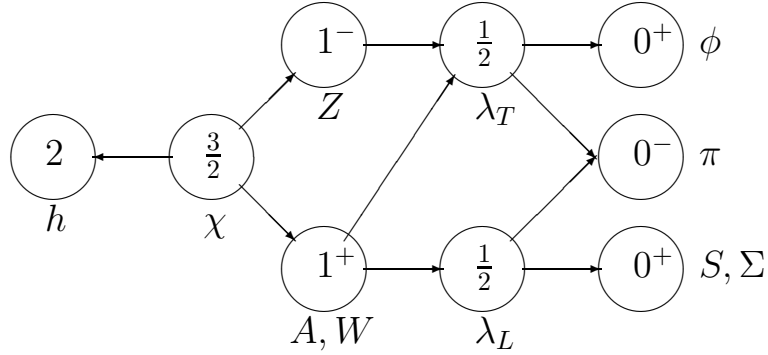

In order to determine the matching of
the spectrum with the ${\cal N}=2$ multiplets and the
structure of the ${\cal N}=2$ multiplets themselves
the following mass relations \cite{univer}
, \cite{castdauriafre},
due to the supersymmetry
relations between the various fields, have been used
\begin{eqnarray}
m_h^2 =& m_\chi (m_\chi + 12)\,,
&
\nonumber \\
m_A^2 =&
m_\chi (m_{\chi} + 4) &{\rm  \hskip .5 cm    if  \hskip .5 cm  } m_{\chi} \geq -8
\,,
\nonumber \\
m_A^2 =&
m_\chi^2+2m_\chi+192 &{\rm  \hskip .5 cm    if  \hskip .5 cm  } m_{\chi} \leq -8
\,,
\nonumber \\
m_W^2 =&
m_\chi^2+2m_\chi+192 &{\rm  \hskip .5 cm    if  \hskip .5 cm  } m_{\chi} \geq -8
\,,
\nonumber \\
m_W^2 =&
m_\chi (m_{\chi} + 4) &{\rm  \hskip .5 cm    if  \hskip .5 cm  } m_{\chi} \leq -8
\,,
\nonumber \\
m_Z^2 =& (m_\chi + 8)(m_\chi + 4)
\,,\nonumber \\
\label{massrelationchi}
\end{eqnarray}
\begin{eqnarray}
m_{\pi}^2=& m_{\lambda_T}(m_{\lambda_T}+4)
 & \,,\nonumber \\
m_{\phi}^2=& m_{\lambda_T}(m_{\lambda_T}-4)
 & \,,\nonumber \\
m_A^2 =& m_{\lambda_T}^2-20\, m_{\lambda_T} + 96
 &{\rm  \hskip .5 cm    if  \hskip .5 cm  } m_{\lambda_T} \geq 4
\,,\nonumber \\
m_A^2 =& m_{\lambda_T}(m_{\lambda_T}+4)
 &{\rm  \hskip .5 cm    if  \hskip .5 cm  } m_{\lambda_T} < 4
\,,\nonumber \\
m_W^2 =& m_{\lambda_T}(m_{\lambda_T}+4)
 &{\rm  \hskip .5 cm    if  \hskip .5 cm  } m_{\lambda_T} \geq 4
\,,\nonumber \\
m_W^2 =& m_{\lambda_T}^2-20\, m_{\lambda_T} + 96
 &{\rm  \hskip .5 cm    if  \hskip .5 cm  } m_{\lambda_T} < 4
\,,\nonumber \\
m_Z^2 =& m_{\lambda_T}(m_{\lambda_T}-4)\,,
 &
\label{massrelationlambdaT}
\end{eqnarray}
\begin{eqnarray}
m_{\pi}^2=& m_{\lambda_L}(m_{\lambda_L}+4)
 & \,,\nonumber \\
m_S^2 =& \left(m_{\lambda_L}+24\right)\left(m_{\lambda_L}+20\right)
 &{\rm  \hskip .5 cm    if  \hskip .5 cm  } m_{\lambda_L} < -10
\,,\nonumber \\
m_S^2 =& m_{\lambda_L}(m_{\lambda_L}-4)
 &{\rm  \hskip .5 cm    if  \hskip .5 cm  } m_{\lambda_L} \geq -10
\,,\nonumber \\
m_{\Sigma}^2 =& m_{\lambda_L}(m_{\lambda_L}-4)
 &{\rm  \hskip .5 cm    if  \hskip .5 cm  } m_{\lambda_L} <-10
\,,\nonumber \\
m_{\Sigma}^2 =& \left(m_{\lambda_L}+24\right)\left(m_{\lambda_L}+20\right)
 &{\rm  \hskip .5 cm    if  \hskip .5 cm  } m_{\lambda_L} \geq -10
\,, \nonumber\\
m_A^2 =& m_{\lambda_L}^2-2\, m_{\lambda_L} + 192
 &{\rm  \hskip .5 cm    if  \hskip .5 cm  } m_{\lambda_L} <-8
\,,\nonumber \\
m_A^2 =& m_{\lambda_L}(m_{\lambda_L}+4)
 &{\rm  \hskip .5 cm    if  \hskip .5 cm  } m_{\lambda_L} \geq -8
\,,\nonumber \\
m_W^2 =& m_{\lambda_L}(m_{\lambda_L}+4)
 &{\rm  \hskip .5 cm    if  \hskip .5 cm  } m_{\lambda_L} < -8
\,,\nonumber \\
m_W^2 =& m_{\lambda_L}^2-2\, m_{\lambda_L} + 192
 &{\rm  \hskip .5 cm    if  \hskip .5 cm  } m_{\lambda_L} \geq -8\,.
\label{massrelationlambdaL}
\end{eqnarray}
These supersymmetry relations are pictorially represented
in Figure \ref{molecola}.
\newpage
In the construction of the ${\cal N}=2$ multiplets, the following
equations (see \cite{multanna}), that give the masses of the
$Osp\left(4|2\right)$ fields in terms of their energies, has been
used
\begin{eqnarray}
m_{\left(0\right)}^2&=&16\left(E_{\left(0\right)}-2\right)\left(E_{\left(0\right)}-1\right)
\,,\nonumber\\
\left|m_{\left(1\over 2\right)}\right|&=&4E_{\left(1\over 2\right)}-6\,,\nonumber\\
m_{\left(1\right)}^2&=&16\left(E_{\left(1\right)}-2\right)\left(E_{\left(1\right)}-1\right)
\,,\nonumber\\
\left|m_{\left(3\over 2\right)}+4\right|&=&4E_{\left(3\over 2\right)}-6\label{massenergy} \,.
\end{eqnarray}


\begin{thebibliography}{99}
\bibitem{stiefel} A. Ceresole, G. Dall'Agata and  R. D'Auria,
 paper in preparation.
\bibitem{n010} L. Castellani, D. Fabbri, P. Fr\'e, L. Gualtieri, P.
Termonia, {\it M-theory on $AdS_4 \times N^{010}$: the complete
spectrum of $Osp(3\vert 4) \times SU(3)$ supermultiplets}, paper in
preparation.
\bibitem{castromwar} L. Castellani, L.J. Romans and N.P. Warner,{\it
A Classification of Compactifying solutions for D=11 Supergravity}
Nucl. Phys. {\bf B2421} (1984) 429
\bibitem{noi321} L. Castellani, R. D'Auria and P. Fr\'e {\it
$SU(3)\times SU(2) \times U(1)$ from D=11 supergravity} Nucl. Phys. {\bf B239} (1984)
60
\bibitem{spectfer} R. D'Auria and P. Fr\'e, {\it On the fermion
mass-spectrum of Kaluza Klein supergravity} Ann. of Physics. {\bf
157} (1984) 1.
\bibitem{univer} R. D'Auria and P. Fr\'e {\it Universal Bose-Fermi
mass--relations in Kaluza Klein supergravity and harmonic analysis on
coset manifolds with Killing spinors} Ann. of Physics {\bf 162}
(1985) 372.
\bibitem{spec321} R. D'Auria and P. Fr\'e {\it On the spectrum of the
${\cal N}=2$ $SU(3)\times SU(2) \times U(1)$ gauge theory from D=11
supergravity} Class. Quantum Grav. {\bf 1} (1984) 447.
\bibitem{multanna} A. Ceresole, P. Fr\'e and H. Nicolai, {\it
Multiplet structure and spectra of ${\cal N}=2$ compactifications}
Class. Quantum Grav. {\bf 2} (1985) 133.
\bibitem{g/hm2} L. Castellani, A. Ceresole, R. D'Auria, S. Ferrara,
P. Fr\'e and M. Trigiante {\it $G/H$ M-branes and $AdS_{p+2}$
geometries}Nucl. Phys. {\bf B527} (1998) 142, hep-th 9803039.
\bibitem{noim2} G. Dall'Agata, D. Fabbri, C. Fraser, P. Fr\'e, P. Termonia, M. Trigiante
{\it The Osp(8|4) singleton action from the supermembrane}
hep-th/9807115, to appear on Nucl. Phys.
\bibitem{tatar} K.Oh and R. Tatar {\it Three dimensional SCFT from M2
branes at conifold singularities} hep-th 9810244.
\bibitem{witkleb} I. Klebanov and E. Witten {\it Superconformal Field
Theory on Threebranes at a Calabi Yau Singularity} hep-th 9807080.
\bibitem{cre}E. Cremmer, B. Julia, {\it The $N=8$ supergravity
theory.1. the Lagrangian}, Phys.Lett. {\bf 80B} (1978) 48; {\it The $SO(8)$
supergravity}, Nucl.Phys. {\bf B159} (1979) 141.
\bibitem{ricpie11} R. D'Auria and P. Fr\'e, {\it Geometric d=11 supergravity and its hidden
supergroup} Nucl.Phys. {\bf B201} (1982) 101
\bibitem{Kkidea} Th. Kaluza: {\it Zum Unit\"atsproblem der Physik}
Sitzungsber. Preuss. Akad. Wiss. Phys. Math. K1 (1921) 966,\\
O. Klein {\it Quantum Theory and Five Dimensional Theory of
Relativity} Z. Phys. 37 (1926) 895.
\bibitem{freurub} P.G.O. Freund and M.A. Rubin {\it Dynamics of
dimensional reduction} Phys. Lett. {\bf B97} (1980) 233.
\bibitem{duffrev} {\em For an early review see}: M.J. Duff, B.E.W. Nilsson
and C.N. Pope {\it Kaluza Klein Supergravity}, Phys. Rep. {\bf 130}
(1986) 1.
\bibitem{round7a} M.J. Duff, C.N. Pope, {\it Kaluza Klein supergravity and the seven
sphere} ICTP/82/83-7, Lectures given at September School on Supergravity and Supersymmetry,
Trieste, Italy, Sep 6-18, 1982. Published in Trieste Workshop 1982:0183
(QC178:T7:1982).
\bibitem{squash7a} M.A. Awada, M.J. Duff, C.N. Pope {\it N=8 supergravity breaks down to
N=1.} Phys. Rev. Letters {\bf 50} (1983) 294.
\bibitem{englert} F. Englert {\it Spontaneous compactification of
11--dimensional supergravity} Phys. Lett. 119B (1982) 339.
\bibitem{biran} B. Biran, F. Englert, B. de Wit and H. Nicolai,
{\it Gauged ${\cal N}=8$ supergravity and its breaking from spontaneous
compactifications} {\it Phys. Lett.} {\bf B124}, (1983) 45
\bibitem{casher} A. Casher, F. Englert, H. Nicolai and M. Rooman
{\it The mass spectrum of Supergravity on the round seven sphere}
Nucl. Phys. {\bf B243} (1984) 173.
\bibitem{dewit2} B. de Wit and H. Nicolai,
{\it On the relation between $D=4$ and $D=11$ supergravity}
Nucl. Phys. {\bf 243} (1984), 91;  
B. de Wit and H. Nicolai and N. P. Warner,
{\it The embedding of gauged ${\cal N}=8$ supergravity into $D=11$ supergravity}
{\it Nucl. Phys.} {\bf B255}, (1985) 29.
B. de Wit and H. Nicolai, 
{\it The consistency of the $S^7$ truncation of $D=11$ supergravity};
{\it Nucl. Phys.} {\bf B281}, (1987) 211;
\bibitem{dewit1} B. de Wit and H. Nicolai,
{\it ${\cal N}=8$ supergravity} {\it Nucl. Phys.} {\bf B208}, (1982) 323.
\bibitem{osp48} R. D'Auria, P. Fre'
{\it Spontaneous generation of Osp(4/8) symmetry in the spontaneous compactification of d=11
supergravity} Phys. Lett. {\bf B121} (1983) 225.
\bibitem{gunawar} M. Gunaydin and N.P. Warner. {\it Unitary
Supermultiplets of Osp(8/4,R) and the spectrum of the $S^7$
compactification of 11--dimensional supergravity} Nucl. Phys. {\bf B272} (1986) 99
\bibitem{gunay2} M. Gunaydin, N. Marcus
{\it The spectrum of the $S^5$ compactification of the chiral N=2,
 D = 10 supergravity and the unitary supermultiplets of U(2, 2/4).}
 Class. Quantum Grav. {\bf 2} (1985) L11
\bibitem{maldapasto}J. Maldacena, {\it The Large N Limit of Superconformal Field
Theories and Supergravity} Adv.Theor.Math.Phys. 2 (1998) 231,  hep-th/9711200.
\bibitem{minwalla} S. Minwalla
{\it  Particles on $AdS_{4/7}$ and Primary Operators on $M_{2/5}$ Brane
Worldvolumes} JHEP 9810 (1998) 002,  hep-th/9803053
\bibitem{serfro1} S. Ferrara, C. Fronsdal
{\it  Conformal Maxwell theory as a singleton field theory on $AdS_5$,
IIB three-branes and duality}
Class.Quant.Grav. 15 (1998) 2153, hep-th/9712239.
\bibitem{serfro2} S. Ferrara, C. Fronsdal,{\it Gauge fields as composite boundary
excitations} Phys.Lett. B433 (1998) 19, hep-th/9802126
\bibitem{serfro3}  S. Ferrara, A. Zaffaroni
{\it Bulk Gauge Fields in AdS Supergravity and Supersingletons} hep-th/9807090
\bibitem{serlau} L. Andrianopoli, S. Ferrara
{\it Non chiral  primary superfields in the $AdS_{d+1}/CFT_d$
correspondence} Lett.Math.Phys. 46 (1998) 265, hep-th/9807150
\bibitem{serlau2}  L. Andrianopoli, S. Ferrara
 {\it On short and long SU(2,2/4) multiplets in the AdS/CFT
 correspondence} hep-th/9812067.
\bibitem{kkwitten} E. Witten, {\it Search for a realistic Kaluza Klein
Theory} Nucl. Phys. {\bf B186} (1981) 412
\bibitem{fretrst} P. Fr\'e {\it Lectures given at the 1984 Trieste Spring
School}, P. Van Nieuwenhuizen et al editors, World Scientific,
publisher
\bibitem{dafrepvn} R. D'Auria, P. Fr\'e and P. van Nieuwenhuizen
{\it N=2 matter coupled supergravity from compactification on a coset
with an extra Killing vector} Phys. Lett. {\bf B136B} (1984) 347
\bibitem{n3} L. Castellani, A. Ceresole, R. D'Auria, S. Ferrara, P. Fre', E. Maina
{\it  The complete N=3 matter coupled supergravity} Nucl. Phys. {\bf B268} (1986)
317
\bibitem{castdauriafre} L. Castellani, R. D'Auria, P. Fr\'e, {\it
Supergravity and Superstring theory: a geometric perspective} World
Scientific, Singapore 1990.
\bibitem{frenico} D. Z. Freedman, H. Nicolai, {\it Multiplet Shortening in $Osp(N \vert 4)$ }
Nucl. Phys. {\bf B237} (1984) 342
\bibitem{bosmass} L. Castellani, R. D'Auria, P. Fr\'e, K. Pilch, P. van Nieuwenhuizen, {\it
The bosonic mass formula for Freund--Rubin solutions of $D$=11 supergravity on general
coset manifolds} Class.Quant.Grav. 1 (1984) 339.
\end{thebibliography}
\end{document}